\documentclass[draft]{agujournal2019}
\usepackage{url} 
\usepackage{lineno}
\usepackage[inline]{trackchanges} 
\usepackage{soul}
\usepackage{amsmath}
\usepackage{multirow}
\usepackage{longtable}
\usepackage{tabularx}
\usepackage{array}
\usepackage{amsfonts}
\usepackage{apacite}
\usepackage{bm}
\usepackage{comment}
\newcolumntype{Y}{>{\raggedright\arraybackslash}X}
\newcommand{\longoverbrace}[2]{\overbrace{#1}^{\text{\hbox to 0cm{\hss #2 \hss}}}}  

\usepackage{booktabs}   

\newcommand{\stnote}[1]{\newline{\footnotesize #1}}

\usepackage{setspace}   
\newcolumntype{S}{>{\setstretch{0.95}\raggedright\arraybackslash}p{0.5cm}}
\newcolumntype{D}{>{\setstretch{0.95}\raggedright\arraybackslash}p{6 cm}}
\newcolumntype{M}{>{\setstretch{0.98}\raggedright\arraybackslash}p{7.5cm}}

\usepackage{pdflscape}
\usepackage{threeparttable}

\draftfalse

\journalname{Water Resources Research}

\begin{document}

%
%


\title{Extending the Joint Probability Method to Compound Flooding: Transition Zone Delineation, Flood Depth Attribution, and Design Event Selection}


%
%




\authors{ Mark S. Bartlett$^{1,2}$, Nathan Geldner$^{1,8}$, Hugh J. Roberts$^{1,2}$, Zach Cobell$^{1,2}$, Brett McMann$^1$,  Luis Partida$^1$, Ovel Diaz$^1$, David R. Johnson$^{5,6}$, Hanbeen Kim$^{3,4}$, Gabriele Villarini$^{3,4}$, Shubhra Misra$^{1,7}$, Muthukumar Narayanaswamy$^1$}

\affiliation{1}{The Water Institute, Baton Rouge, Louisiana, USA}
\affiliation{2}{ICEYE, Espoo, Finland}
\affiliation{3}{Department of Civil and Environmental Engineering,
Princeton University, Princeton, New Jersey USA}
\affiliation{4}{High Meadows Environmental Institute,
Princeton University, Princeton, New Jersey USA}
\affiliation{5}{Edwardson School of Industrial Engineering, Purdue University, West Lafayette, Indiana USA}
\affiliation{6}{Department of Political Science, Purdue University, West Lafayette, Indiana USA}
\affiliation{7}{Coastal Engineering and Adaptation and Solutions (CEAS)}
\affiliation{8}{The Barbara Geldner Foundation}





\correspondingauthor{Mark Bartlett}{Mark.Bartlett@gmail.com}
\correspondingauthor{Muthu Narayanaswamy}{mnarayanaswamy@thewaterinstitute.org}



\begin{keypoints}

\item The Joint Probability Method (JPM) is extended with stochastic rainfall fields and stochastic hydrologic drivers for compound flooding
\item Compound flooding increases annual exceedance probability flood depths through both nonlinear interactions and multiple flood-generation pathways

\item The framework enables statistical CFTZ delineation, probabilistic flood attribution, and response-based design storm selection
\end{keypoints}

%
%

%
%


\begin{abstract}

Quantifying the frequency of compound flood depths is a fundamental challenge in low-gradient coastal watersheds, where flood hazards arise from the nonlinear interaction of storm surge, rainfall, and riverine flooding. Existing approaches often characterize either the joint occurrence of flood drivers or the flood response for prescribed events, but they do not derive the long-term frequency distribution of compound flood depths from probabilistic descriptions of rainfall and antecedent hydrologic conditions. Traditionally, coastal flood frequency has been quantified using the Joint Probability Method (JPM), which represents storm surge probabilistically. Although recent studies have incorporated rainfall into JPM-based analyses, rainfall is treated as a deterministic function of JPM storm characteristics rather than as a conditional probability distribution. Here, we extend the JPM by coupling its event-scale stochastic description of storm characteristics with probabilistic rainfall realizations and stochastic antecedent hydrologic conditions, thereby enabling propagation of these stochastic processes through the flood response to derive the compound flood-depth distribution. The framework provides a statistical basis for delineating compound flood transition zones, probabilistically attributing flood depths to hydrologic and coastal processes, and selecting response-based design storms for specified annual exceedance probabilities (or return periods). Application to the Lake Maurepas basin, Louisiana, shows that the statistically defined compound flood transition zone is more than twice the area identified from event-based analyses and that compound interactions increase flood depths by up to 0.7 m. This extended JPM establishes a probabilistic foundation for compound flood hazard assessment and response-based design.


\end{abstract}


%
%

%


%
%
%
%


\section{Introduction}
Quantifying the frequency of flood depths is the primary objective in many flood studies \cite{maranzoni2023quantitative, nofal2022understanding, de2015flood}. This becomes considerably more complex in low-gradient coastal watersheds, where flood-depth frequency reflects not only storm surge and coastal dynamics but also their nonlinear compounding with runoff and riverine overflows \cite{zscheischler2018future, green2025comprehensive, bensi2020multi, habel2020sea, bilskie2018defining, santiago2019comprehensive}. Traditionally, coastal flood hazards in the United States have been quantified using the Joint Probability Method (JPM) and related optimal sampling variants, forming the probabilistic basis for FEMA flood hazard assessments \cite{FEMA2023Coastal, nadal2016statistical,nadal2022coastal,resio2009surge,toro2008joint,yang2019objective}. These methods account for coastal and oceanic drivers of flooding, but have not been extended to incorporate probabilistic representations of rainfall and antecedent hydrologic processes needed to quantify the frequency of compound flood depths, a hazard projected to become increasingly important under future climate conditions \cite{couasnon2020measuring,ghanbari2021climate,hsiao2021flood,nasr2021assessing, emanuel2008hurricanes}. Numerous studies have demonstrated that compound processes can amplify individual flood events \cite<e.g.>{bilskie2018defining,santiago2019comprehensive,bilskie2021enhancing}; however, how these nonlinear processes alter the long-term frequency of the flood depth response remains largely unexplored \cite{santamaria2026large, bevacqua2020more, gori2020tropical}.

The traditional JPM represents tropical cyclones as a marked Poisson process in which storms arrive randomly in time and each event is assigned a mark---a set of storm characteristics with an associated statistical weight---that parameterizes a coastal hydrodynamic simulation \cite{ganguli2013probabilistic, serafin2019s, demoel2011effect, gonzalez2019quantification, hinkel2021uncertainty,resio2009surge, vousdoukas2018understanding,ahmadisharaf2019coupled,bensi2020multi,kheradmand2018evaluation,marijnissen2019re,thompson2014deterministic,voortman2003risk,johnson2023coastal}. In contrast, stochastic ecohydrology considers the continuous hydrologic evolution across all the storm events over yearly to decadal timescales, deriving analytical probability distributions for soil moisture, runoff, and baseflow by representing rainfall as a marked Poisson process acting on continuous watershed dynamics \cite{bartlett2025stochasticE,botter2007basin,botter2007signatures,bartlett2015stochastic,basso2015emergence,basso2016physically,porporato2022ecohydrology,eagleson1978climate2}. Although developed independently, the two frameworks share the same stochastic foundation---a marked Poisson representation of storm arrivals---but at different temporal scales. This common stochastic foundation naturally suggests extending the JPM beyond event-scale coastal forcing by replacing deterministic representations of watershed forcing with probabilistic descriptions of rainfall and the continuous hydrologic evolution described by stochastic ecohydrology.

Despite this natural connection, most studies of compound flooding have focused either on individual flood events \cite<e.g.,>{han2024compound,gori2020assessing} or on the statistical dependence among flood drivers \cite<e.g.,>{wahl2015increasing,moftakhari2019linking,jane2022assessing,kim2023generation}. Event-based analyses reveal how surge and runoff interact during particular storms, while copula methods quantify the joint occurrence of coastal water levels and precipitation. Neither approach, however, provides a probabilistic description of antecedent hydrology or derives the long-term frequency distribution of flood depths. More recently, extensions of the JPM have incorporated rainfall into probabilistic compound flood assessments by parameterizing rainfall fields from storm characteristics \cite<e.g.,>{gori2020tropical,gori2022jpmosbq,bass2018surrogate}. However, rainfall remains a deterministic function of the JPM storm characteristics, and the continuously evolving antecedent hydrologic state is not represented probabilistically. Consequently, they do not derive the probability distribution of compound flood depths by propagating the joint stochastic descriptions of storm characteristics, rainfall, and antecedent watershed conditions through the flood response. Addressing compound flood hazard therefore requires a probabilistic framework that propagates these stochastic descriptions through the flood response to estimate flood-depth exceedance probabilities.

Because existing probabilistic methods focus on characterizing the drivers rather than the flood response, several important capabilities remain unavailable.  Compound flood transition zones are still identified primarily from individual events, revealing where a particular storm exhibits nonlinear amplification but not where multiple flood pathways systematically increase the probability of large flood depths across the full storm population \cite<e.g.,>{bilskie2018defining,bilskie2021enhancing,gori2020tropical,shen2019flood,han2024compound}. Flood depths likewise cannot be probabilistically decomposed into the relative contributions of their generating mechanisms. Consequently, for a given AEP flood depth, the probability that a specified fraction arises from hydrologic versus coastal processes cannot presently be quantified---an important consideration for planning and design decisions. The same limitation extends to design-event selection: multivariate methods \cite<e.g.,>{jane2020multivariate} identify combinations of extreme drivers, but the annual exceedance probability (AEP) associated with the selected driver combination typically diverges from the AEP of the flood depth for planning and engineering design \cite{serinaldi2015dismissing,volpi2014hydraulic}. Selecting representative storms---each with different attributions to the same target annual exceedance flood depth---requires conditioning on the flood depth response itself.

To quantify how compound flooding alters the long-term frequency of flood depths, we extend the Joint Probability Method by replacing deterministic mappings within the flood-generation process with probabilistic descriptions of rainfall and antecedent hydrology (Section \ref{sec:theory_depths}). The resulting probabilistic framework derives the compound flood-depth distribution from the joint distribution of coastal and hydrologic drivers, providing a statistical basis for delineating compound flood transition zones, probabilistically attributing flood depths to hydrologic and coastal processes, and selecting representative design storms for target AEPs (Section \ref{sec:theory_characterization}). The framework is demonstrated for the Lake Maurepas basin, Louisiana, where compound flood transition zones are delineated statistically, flood depths are probabilistically attributed to hydrologic and coastal processes, and response-based design storms are identified (Section \ref{sec:case_study}). This is followed by a discussion of the results and broader implications for compound flood hazard assessment and design practice (Sections \ref{sec:results}-\ref{sec:discussion}).

\begin{figure}
    \centering
    \includegraphics[width=6in]{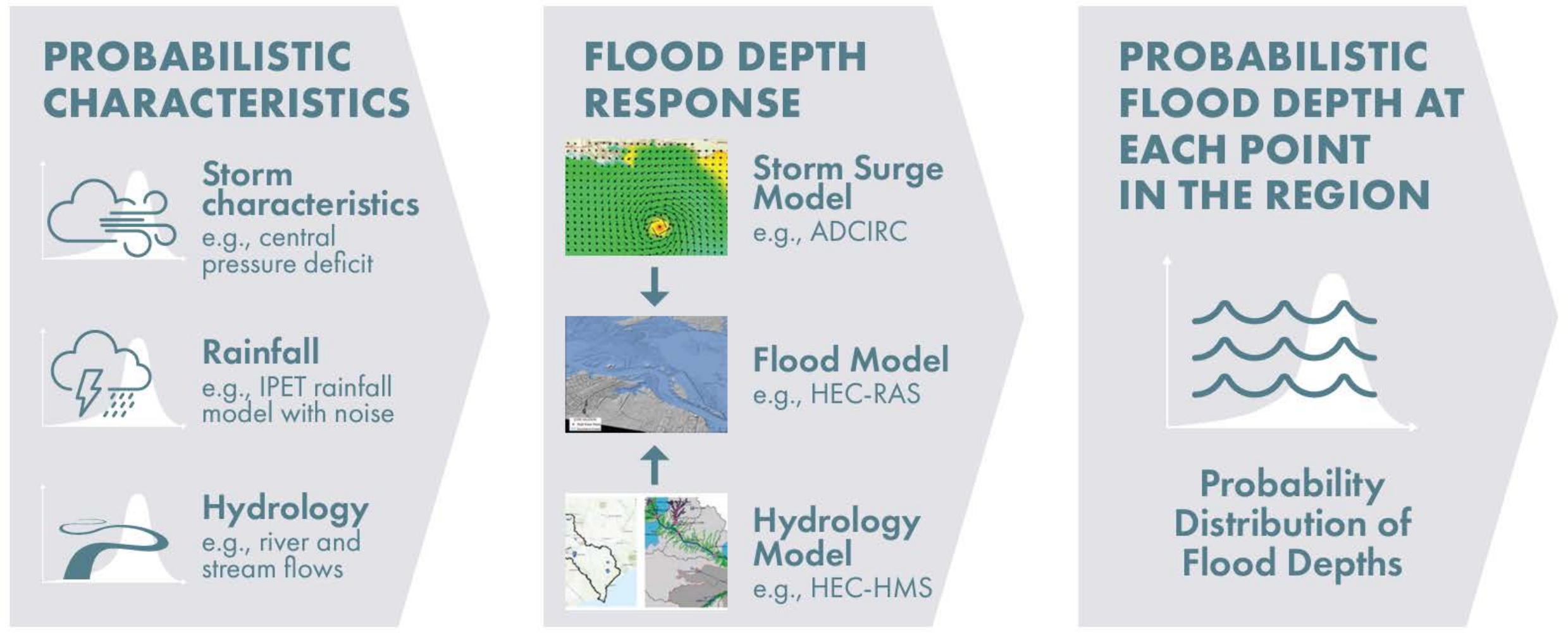}
    \caption{The extended Joint Probability Method processes the probabilistic characteristics of storm events (including precipitation, and hydrology) through a flood depth response (based on respective models) and derives a probability distribution of the flood depth response for any point over the study region.}
    \label{fig:1}
\end{figure}

\begin{table}
\begin{minipage}{\textwidth}
\linespread{.5}\selectfont
\caption{Theory variables and parameters$^a$ \label{tab:vars_params} }
\noindent
\begin{tabular}{c p{13cm}}
\hline
\noalign{\vskip 0.04in}
Symbol & Description \\
\hline
\noalign{\vskip 0.05in}
$H_{\%}$ & Percent of the compound flood depth at a point, $\eta_{\max}$, attributed to hydrologic drivers.  \\
$S_{\%}$ & Percent of the compound flood depth at a point, $\eta_{\max}$, attributed to storm surge.  \\
$\mathbf{x}_{(\cdot)}$ &Set of characteristics driving flooding for a generic storm type.\\
$\mathbf{x}_{TC}$         &Set of characteristics driving flooding for tropical cyclones, $ \{ \mathbf{x}_{JPM}, \mathbf{r}(t), \mathbf{\overline{r}}(t), \mathbf{s}, \mathbf{\overline{s}}, \mathbf{w},\mathbf{\overline{w}}, \mathbf{\overline{q}}_b \}$     \\
$\mathbf{x}_{NT}$         & Set of characteristics driving flooding for non-tropical storms,  $\{\kappa, \tau_l, \mathbf{u}(t), \mathbf{r}(t), \mathbf{\overline{r}}(t), \mathbf{s}, \mathbf{\overline{s}}, \mathbf{w}, \mathbf{\overline{w}},\mathbf{\overline{q}}_b\}$ \\
$\mathbf{x}_{JPM}$        & Set of JPM tropical cyclone characteristics, $\left\{x_l,c_p,\theta,R_{\max},v_f \right\}$  \\ 
$x_o$                     & Tropical cyclone landfall location relative to a CRL. \\
$x_l$                     & Tropical cyclone landfall location; see Fig. \ref{fig:2}. \\
$c_p$                     & Tropical cyclone central pressure deficit \\
$\theta$                  & The storm heading at landfall \\
$R_{\max}$                & The radius of maximum wind speed \\
$v_f$                     & The storm forward velocity \\
$\mathbf{x}_{Storm}$      & Storm parameters governing the coastal boundary condition; for the case study equal to $\mathbf{x}_{JPM}$ for tropical events and equal to $\{ \kappa, \tau_l, \mathbf{u}(t) \}$ for non-tropical events.  \\
$T_o$                     & Period of historical record observation \\
$\lambda$                 & Overall regional frequency of storms, $\lambda = \lambda_{TC}+ \lambda_{NT}$\\
$\lambda_{TC}$            & Time-averaged regional frequency of tropical storms, $\lambda_{TC} = \frac{1}{T_o}\int_0^{T_o}\lambda_{TC,t}(t)dt$\\
$\lambda_{NT}$            & Time-averaged regional frequency of non-tropical storms, $\lambda_{NT} = \frac{1}{T_o}\int_0^{T_o}\lambda_{NT,t}(t)dt$\\
$\lambda_{TC,t}(t)$       & Time varying regional frequency of tropical storms\\
$\lambda_{NT,t}(t)$       & Time varying regional frequency of non-tropical storm arrivals\\
$l_k$                     & Limits of the \% surge attribution, $S_{\%}$, defining a design storm scenario\\
$p(\mathbf{.})$           & General function for PDF\\
$P(\mathbf{.})$           & General function for CDF\\
$\mathbf{r}(t)$           & Vector representing the 2d rainfall field, which varies in time \\
$\overline{\mathbf{r}}(t)$  & Vector of spatial average rainfall, which varies in time; for the case represented by the IPET model \\
$\mathbf{\overline{q}}_b$   & Vector of river and stream baseflow; for the 23 basins of the case study, $\mathbf{\overline{q}}_b = \{q_{b,1}, q_{b,2},...,q_{b,23}\}$\\
$\mathbf{u}(t)$           & Wind field\\
$\kappa$                  & Non-tidal residual peaking factor\\
$\tau_l$                  & Lag time between peak river flow and the non-tidal residual\\
$\eta_{\max}$             & Maximum flood depth at a point per storm event\\
$\eta_{\max,H}$           & Maximum flood depth at a point per storm event attributed to hydrologic drivers\\
$\eta_{\max,S}$           & Maximum flood depth at a point per storm event attributed to storm surge drivers\\
$\boldsymbol{\eta}_s$     & Coastal boundary water depth (storm surge or non-tidal residual); for the case study discretized to eight boundary values $\boldsymbol{\eta}_s = \{\eta_{s,1},\eta_{s,2},\eta_{s,3},\eta_{s,4},\eta_{s,5},\eta_{s,6},\eta_{s,7},\eta_{s,8} \}$\\
$\delta(\cdot)$           & Dirac delta function\\
$\eta(t)$                 & Flood depths at each point over the storm duration\\
$\mathbf{q}$              & River inflows\\
$\mathbf{x}_S$            & Storm surge and winds, $\mathbf{x}_S(t) = \{\boldsymbol{\eta}_s(t), \mathbf{u}(t)\}$\\
$\boldsymbol{\mu}$        & Set of ensemble average soil moisture values; for the 23 case study basins $\boldsymbol{\mu} = \{\mu_1, \mu_2,...,\mu_{23}\}$ \\
$\Sigma$                  & Covariance matrix\\
$\boldsymbol{\sigma}$     & Set of ensemble soil moisture standard deviations; for the 23 case study basins $\boldsymbol{\sigma} = \{\sigma_1, \sigma_2,...,\sigma_{23}\}$ \\
$T_d$                     & Storm event duration\\
$\mathbf{s}$              & Vector of soil moisture values at each spatial location in the 2D watershed area\\
$\mathbf{\overline{s}}$   & Vector of basin average soil moisture (on a unit-area basis); for the 23 basins of the case study, $\mathbf{\overline{s}} = \{\overline{s}_1, \overline{s}_2,...,\overline{s}_{23}\}$\\
$\mathbf{w}$              & Vector of total storage capacity at each spatial location in the 2D watershed area\\
$\mathbf{\overline{w}}$   & Vector of basin average total storage capacity (on a unit-area basis); for the 23 basins of the case study, $\mathbf{\overline{w}} = \{\overline{w}_1, \overline{w}_2,...,\overline{w}_{23}\}$\\
$\rho$                    & Pearson's correlation coefficient\\
$\omega_{TC}$             & Depth conditional probability of occurrence of TC events; see Eq. (\ref{eq:weight_scenario})\\
$t$         & time\\
\noalign{\vskip 0.04in}
\hline
\end{tabular}
\\
$^*$ Variables in the text with an overline bar indicate a spatial average value on a unit-area basis.
\end{minipage}
\end{table}


\section{Background on the Joint Probability Method}
The JPM models tropical cyclone occurrence at a coastal reference location (CRL) as a marked Poisson process: storms arrive randomly in time at rate $\lambda_{TC}$, and  each event is assigned a mark consisting of the key storm characteristics
\begin{equation}
    \mathbf{x}_{JPM} = \left\{ x_o, c_p, \theta, R_{\max}, v_f \right\},
\end{equation}
that include the landfall location, \( x_o \), relative to the CRL, the storm’s central pressure deficit, \( c_p \), the track angle relative to the coast, \( \theta \), the radius of maximum wind speed, \( R_{\max} \), and the forward speed, $v_f$.
These  JPM parameters define idealized tropical cyclone represented as a synthetic storm with a track defined by the landfall location and track angle. In turn, these parameters define the synthetic storm spatiotemporal wind and pressure fields that force storm surge, which is typically simulated using a coupled hydrodynamic model such as ADCIRC \cite{luettich1991solution}. Flood probabilities are then obtained by weighting each simulated storm-surge response according to the joint probability distribution of the storm characteristics,  \( p(\mathbf{x}_{JPM}) \) which is estimated from historical storm records. This yields the probability distribution of storm surge flood depths at locations around the CRL, represented either as a probability density function (PDF), \( p(\eta_{\max}) \), or a cumulative distribution function (CDF), \( P(\eta_{\max}) \). In this way, the JPM provides a probabilistic model relating the distribution of storm characteristics to the distribution of storm-surge flood depths. In the next section, we extend this probabilistic framework to compound flooding by incorporating non-tropical cyclones and extending the joint probability formulation with stochastic rainfall fields and stochastic antecedent hydrologic conditions. 

\section{Extended JPM Framework for Compound Flooding}
\label{sec:theory}
We extend the JPM to regional compound flooding---riverine, pluvial, and storm surge---from both tropical and non-tropical storms. The framework consists of defining the compound flood-depth probability distribution (Section \ref{sec:theory_depths}), which in turn provides the foundation for a probabilistic characterization of compound flooding (Section \ref{sec:theory_characterization}). The extension of the JPM to compound flood-depth frequency in Section \ref{sec:theory_depths} proceeds in three steps (see Fig. \ref{fig:1}): defining the annual flood depth distribution that combines tropical and non-tropical events (Eqs. \ref{eq:pA}-\ref{eq:peta}), establishing that the flood response models are integrated over the probability distributions of the JPM variables (Eqs. \ref{eq:pTC}-\ref{eq:pNT}), and finally extending the JPM storm variables to include rainfall and hydrologic drivers with an explicit probabilistic structure (Eqs. \ref{eq:TC_chararacteristics}-\ref{eq:p_swsq}). The resulting compound flood-depth probability distribution then forms the basis for the probabilistic characterization of compound flooding in Section \ref{sec:theory_characterization}, consisting of CFTZ delineation, flood-depth attribution, and response-based design storm selection.

\subsection{Compound Flood-Depth Probability Distribution}
\label{sec:theory_depths}

\subsubsection{Annual Flood Depth Probability Distribution}

The extended JPM is formulated in terms of the probability distribution of annual maximum flood depth. At each point, the annual maximum flood depth CDF, $P_A(\eta_{\max})$, where the subscript `\emph{A}' denotes annual, combines tropical and non-tropical storms, with storms arriving at the overall frequency $\lambda$. This CDF is obtained by summing (over all possible numbers of storms, $n$) the probability that each of the $n$ storms does not exceed a compound flood depth, $P(\eta_{\max})^n$, weighted by the Poisson probability of $n$ storms in an annual time interval $[0,T]$:
\begin{align}
P_A(\eta_{\max}) =\sum_{n=0}^{+\infty}\frac{\left [ \ \lambda T\right ]^n}{n!}e^{-\lambda T} \left(P(\eta_{\max})\right)^n= e^{-\lambda T (1-P(\eta_{\max}))},
\label{eq:pA}
\end{align}
where $T$ is equal to 365 days when the units of $\lambda$ are 1/day. The corresponding  quantile function  provides the maximum flood depth, $\eta_{\max}$, associated with an AEP, $p_{\text{AEP}}$, i.e.,
\begin{align}
\eta_{\max} = Q(p_{\text{AEP}}) &= P_A^{-1}(1-p_{AEP}),
\label{eq:Qp}
\end{align}
where $P_A^{-1}(\cdot)$ is the inverse of the compound flood depth annual non-exceedance probability function of Eq. (\ref{eq:pA}). The corresponding return period is $R_p = 1/p_{\text{AEP}}$. For a given AEP, evaluating Eq. (\ref{eq:Qp}) at each computational grid point yields the AEP flood-depth (or water-surface elevation) map.

The extended JPM expresses the CDF $P(\eta_{\max})$ and PDF $p(\eta_{\max})$  as the weighted mixture of a tropical cyclone component, $p_{TC}(\eta_{\max})$, and a non-tropical storm component $p_{NT}(\eta_{\max})$, i.e., 
\begin{align}
p(\eta_{\max})=\frac{\lambda_{TC}}{\lambda}p_{TC}(\eta_{\max})+\frac{\lambda_{NT}}{\lambda}p_{NT}(\eta_{\max}),
\label{eq:peta}
\end{align}
where  $\lambda= \lambda_{TC}+\lambda_{NT}$ is the total storm arrival rate, with $\lambda_{TC}$ and $\lambda_{NT}$ denoting the mean arrival rates of tropical and non-tropical storms, respectively. Both $\lambda_{TC}$ and $\lambda_{NT}$ represent time-averaged frequencies over the observation period, i.e., $\lambda_{TC} = \frac{1}{T_o}\int_0^{T_o}\lambda_{TC,t}(t),dt$ and $\lambda_{NT} = \frac{1}{T_o}\int_0^{T_o}\lambda_{NT,t}(t),dt$, where $T_o$ is the period of observation \cite<e.g.,>{bartlett2025stochasticE}. In this sense, these time-averaged frequencies accurately represent the event process, but they must be paired with probabilistic characteristics derived from time-varying probability distributions weighted toward periods of higher relative storm frequency, as discussed later in the hydrologic formulation. The tropical and non-tropical component PDFs, $p_{TC}(\eta_{\max})$ and $p_{NT}(\eta_{\max})$, are derived separately in the following sections.

\subsubsection{Compound Flood-Depth Distributions for TC and Non-TC Storms}

The probability distributions of maximum flood depth for tropical and non-tropical storms, $p_{TC}(\eta_{\max})$ and $p_{NT}(\eta_{\max})$, are obtained by integrating the respective conditional flood responses, $p_{TC}(\eta_{\max}|\mathbf{x}_{TC}(t))$ and $p_{NT}(\eta_{\max}|\mathbf{x}_{NT}(t))$ over the probability distributions of the corresponding forcing and boundary condition characteristics, $p(\mathbf{x}_{TC}(t))$ and $p(\mathbf{x}_{NT}(t))$:
\begin{align}
p_{TC}(\eta_{\max})&=\int...\int p_{TC}(\eta_{\max}|\mathbf{x}_{TC}(t)) p_{TC}(\mathbf{x}_{TC}(t)) d^n\mathbf{x}_{TC} \label{eq:pTC} \\
p_{NT}(\eta_{\max})&=\int...\int p_{NT}(\eta_{\max}|\mathbf{x}_{NT}(t))p_{NT}(\mathbf{x}_{NT}(t)) d^n\mathbf{x}_{NT}, \label{eq:pNT}
\end{align}
where  $\mathbf{x}_{TC}(t)$ and $\mathbf{x}_{NT}(t)$  denote the respective tropical cyclone and non-tropical storm characteristics, some of which may vary over time during the event (e.g., the rainfall field). The differentials \( d^n \mathbf{x}_{TC} \) and \( d^n \mathbf{x}_{NT} \) denote integration over the $n$-dimensional space of these characteristics. In the traditional JPM framework, the conditional flood responses  $p_{TC}(\eta_{\max}|\mathbf{x}_{TC}(t))$ and $p_{NT}(\eta_{\max}|\mathbf{x}_{NT}(t))$ are deterministic (e.g., as determined from ADCIRC simulations), so that so these conditional PDFs reduce to Dirac delta functions \cite<e.g.,>{resio2007white}. The present formulation generalizes this framework by allowing the conditional response to remain probabilistic, thereby accommodating uncertainty in the modeled flood response while recovering the deterministic JPM as a special case.

\subsubsection{Probabilistic Compound Flood Forcing}

When the conditional response in Eqs. (\ref{eq:pTC}) and (\ref{eq:pNT}) is deterministic---as in the traditional JPM---the flood-depth distribution is determined entirely by the probability distribution of the forcing characteristics, $p_{TC}(\mathbf{x}_{TC}(t))$ and $p_{NT}(\mathbf{x}_{NT}(t))$. The principal extension developed here is therefore not the response model itself, but the probabilistic description of the forcing and antecedent hydrologic conditions governing compound flooding. Specifically, the probabilistic description of the JPM storm characteristics is generalized to include rainfall and the hydrologic variables governing compound flooding while preserving the probabilistic structure of the original framework. 

\begin{figure}
    \centering
    \includegraphics[width=6in]{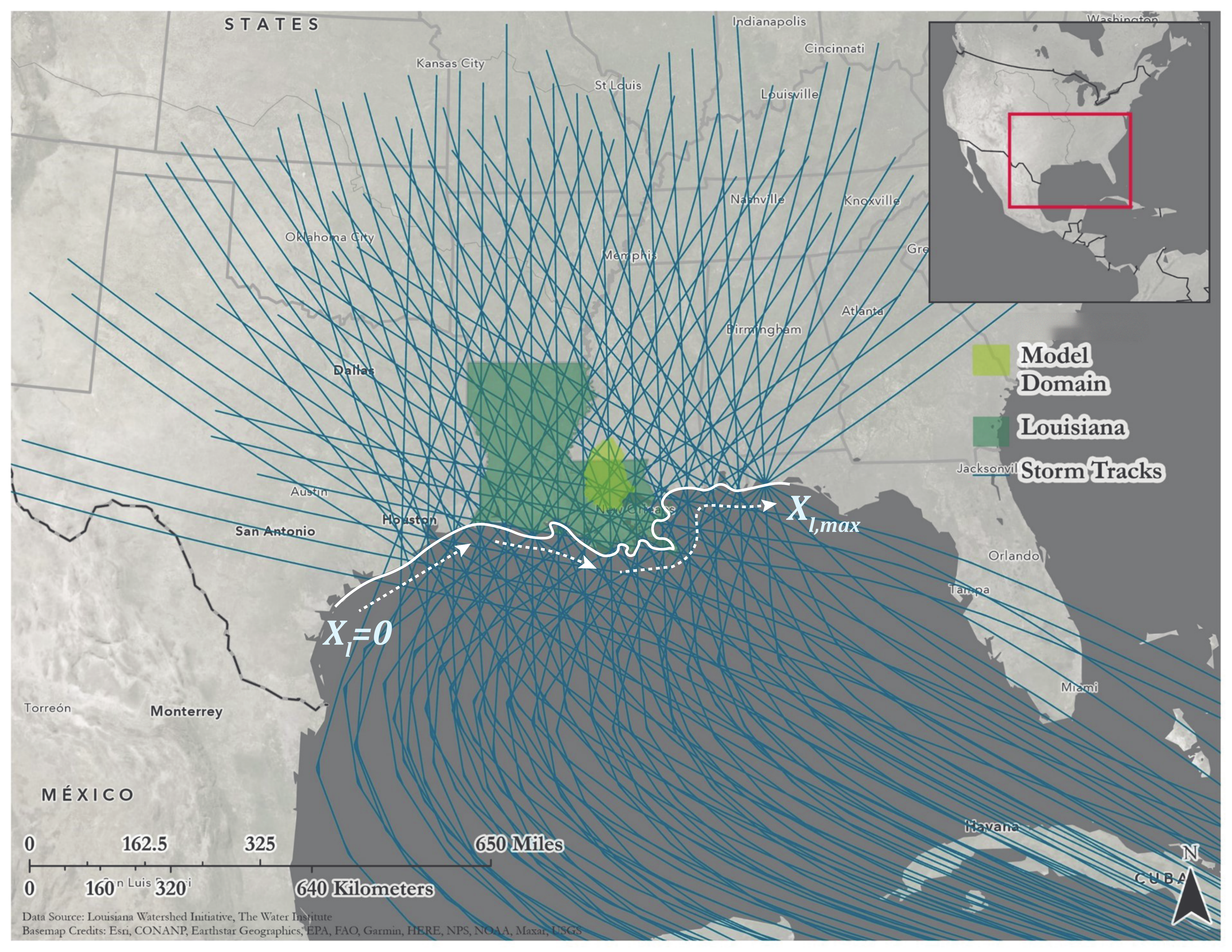}
    \caption{Synthetic TC tracks (blue lines), roughly representing historical TC track patterns, over the HEC-HMS and HEC-RAS model domain (yellow shading) for the Lake Maurepas case study (Section \ref{sec:case_study}) within Louisiana (green shading). Unlike the traditional JPM approach, which considers track frequency relative to a coastal reference location, here track frequency is relative to a coastal coordinate, $x_l$ (white line), extending from 0 to $x_{l,\max}$.}. 
    \label{fig:2}
\end{figure}

Within this extended JPM framework, TC and non-TC storms are each represented by a probability distribution of storm characteristics, here denoted generically by $p({\mathbf{x}_{Storm}})$. The specific storm characteristics comprising $\mathbf{x}_{Storm}$ depend on the physical processes modeled for each event type. While the traditional JPM is formulated with respect to a discrete coastal reference location (CRL), this discretization is not intrinsic to the theory. We therefore introduce a continuous coastline coordinate, $x_l$, representing landfall along the regional shoreline (Fig. \ref{fig:2}). For tropical cyclones, $p({\mathbf{x}_{Storm}})$ is therefore given by the traditional JPM PDF
\begin{align}
p({\mathbf{x}_{Storm}}) = p(c_p, \theta, R_{max}, v_f | x_l)p(x_l),
\end{align}
where $p(x_l)$ describes the distribution of the landfall location and $p(c_p, \theta, R_{max}, v_f | x_l)$ is the conventional JPM distribution of storm parameters conditioned on the landfall location \cite{resio2007white}. For non-TC storms, the storm characteristics, $\mathbf{x}_{Storm}$, represent the boundary conditions used to drive the hydrodynamic model. In the present framework, these consist of the wind field $\mathbf{u}(t)$ together with a generalized coastal stage hydrograph parameterized by a peaking factor,  $\kappa$, a lag time, $\tau_l$, between the non-tidal residual and peak river flow. Accordingly,
\begin{align}
p({\mathbf{x}_{Storm}})=p(\kappa, \tau_l, \mathbf{u}(t)),
\end{align}
though these characteristics should be viewed as interchangeable with other attributes when called for by the modeling specifics of a given geography. Thus, extending the JPM to non-tropical storms requires only specifying the storm characteristics appropriate for modeling the physical forcing while preserving the underlying probabilistic framework.


For both TC and non-TC storms, the storm characteristics are augmented to include additional variables governing pluvial and fluvial flooding. Pluvial flooding arises from runoff generated within the watershed, whereas fluvial flooding results from the aggregation of runoff through the drainage network. Runoff volume is governed primarily by rainfall, $\mathbf{r}(t)$, and the available soil water storage capacity (i.e., the initial deficit), which depends on antecedent soil moisture, $\mathbf{s}$, and the storage capacity depth, $\mathbf{w}$, where soil moisture is expressed as the vertically averaged volumetric water content normalized between 0 and 1 \cite{porporato2022ecohydrology}. These variables exhibit the dominant event-to-event variability controlling runoff generation. In contrast, the timing and routing of runoff are governed primarily by watershed characteristics such as drainage area, topography, land use, and channel geometry, which vary comparatively little from one event to another. The objective is not to prescribe a particular hydrologic model, but rather to define a parsimonious stochastic description of the dominant event-to-event variability required to initialize a broad class of rainfall-runoff and hydraulic models within the JPM framework. Accordingly, the storm characteristics, $\mathbf{x}_{Storm}$, are extended to include the additional variables governing inland flooding. These include:
\begin{itemize}
    \item Rainfall field over watershed points, \( \mathbf{r}(t) \), conditioned on its (spatial) average \( \overline{\mathbf{r}}(t) \).
    \item Soil moisture over watershed points, \( \mathbf{s} \), conditioned on the watershed basin (spatial) average \( \overline{\mathbf{s}} \).
    \item Water storage capacity over watershed points, \( \mathbf{w} \), conditioned on the watershed basin (spatial) average \( \overline{\mathbf{w}} \).
    \item Baseflow contributions from rivers and streams, \( \mathbf{\overline{q}}_b \), on a unit-area basis per watershed.
\end{itemize}
Soil moisture and baseflow represent antecedent watershed conditions immediately prior to storm initiation, whereas rainfall evolves throughout the event \cite{bartlett2015stochastic, bartlett2025stochasticE, kavetski2003semidistributed}. The antecedent conditions may be resolved according to the temporal resolution of the available observations, from hourly to daily values.

These variables form the state description used by many conceptual and physically based hydrologic models and methods, including the SCS-CN method, VIC, PDM, TOPMODEL, and soil-moisture accounting approaches \cite{beven1979physically,beven2012rainfall,liang1994simple,troy2008efficient,ponce1996runoff,kavetski2003semidistributed,burnash1973generalized}. At the event scale, the available storage deficit, $(1-\mathbf{s})\mathbf{w}$, governs saturation-excess runoff generation. Over many events, the total storage capacity variables, $\mathbf{w}$ and $\overline{\mathbf{w}}$ regulate the long-term partitioning of rainfall between infiltration and runoff, thereby implicitly accounting for variations in infiltration capacity without explicitly introducing Hortonian infiltration variables. Although such variables could be incorporated explicitly within the framework, this parsimonious representation captures the dominant event-scale controls on runoff generation when soil moisture and storage parameters are properly calibrated \cite{rigby2006simplified}. As demonstrated in \citeA{bartlett2025stochasticE}, these variables provide sufficient parametrization to capture rainfall-runoff dynamics across CFTZ watersheds.

With these additional hydrologic characteristics, the overall set of characteristics for each event type is generally given by
\begin{align}
\label{eq:TC_chararacteristics}
\mathbf{x}_{(\cdot)}(t) =& \{ \mathbf{x}_{Storm}, \mathbf{r}(t), \mathbf{\overline{r}}(t), \mathbf{s}, \mathbf{\overline{s}}, \mathbf{w},\mathbf{\overline{w}}, \mathbf{\overline{q}}_b \}
\end{align}
where $_{(\cdot)}$ is a placeholder for tropical cyclone (TC) and non-tropical (NT) storms, with  $\mathbf{x}_{Storm} = \mathbf{x}_{JPM}$ for tropical cyclones and $\mathbf{x}_{Storm} = \{\kappa, \tau_l, \mathbf{u}(t)\}$ for non-tropical storms (Table \ref{tab:vars_params}). Accordingly, for compound flooding, the PDFs of tropical and non-tropical event characteristics, $p(\mathbf{x}_{TC}(t))$ and $p(\mathbf{x}_{NT}(t))$, now generally consist of 1) storm parameters, e.g., JPM parameters, governing the coastal storm surge, 2) a random field of rainfall, and 3) antecedent hydrology governing the land surface conditions, i.e.,
\begin{align}
p_{(\cdot)}(\mathbf{x}_{(\cdot)}(t)) = & \overbrace{p(\mathbf{x}_{Storm})}^{\substack{\text{Storm} \\\text{Parameters} }}\overbrace{
p(\mathbf{\overline{r}}(t)|\mathbf{x}_{Storm})p(\mathbf{r}(t)|\mathbf{\overline{r}}(t))}^{\substack{\text{Rainfall} \\ \text{Random Field}}}\overbrace{p(\mathbf{s},\mathbf{w},\mathbf{\overline{s}},\mathbf{\overline{q}}_b)}^\text{Hydrology}.
\label{eq:pcdot}
\end{align}
where $_{(\cdot)}\in \{\text{TC}, \text{NT}\}$  for respective tropical cyclone and non-tropical descriptions. 
In Eq. (\ref{eq:pcdot}), the rainfall representation follows the standard hierarchical construction in which the spatially averaged rainfall is conditioned on storm characteristics, $p(\mathbf{\overline{r}}(t)|\mathbf{x}_{Storm})$, while the random field is conditioned on the spatial average,  $p(\mathbf{r}(t)|\mathbf{\overline{r}}(t))$ \cite<e.g.,>{villarini2022probabilistic,kleiber2023stochastic}.

The PDF of the soil moisture and baseflow attributes is independent of the other PDFs since it may be reasonably assumed that the occurrence and timing of storm events are independent of the soil moisture and baseflow conditions over an area \cite{bartlett2015stochastic,bartlett2025stochasticE}. This hydrologic PDF is given by
\begin{align}
p_{(\cdot)}(\mathbf{s},\mathbf{w},\mathbf{\overline{s}},\mathbf{\overline{q}}_b) =p(\mathbf{s}|\mathbf{\overline{s}})p(\mathbf{\overline{q}}_b|\mathbf{\overline{s}})p(\mathbf{\overline{s}})p(\mathbf{w}|\mathbf{\overline{w}}),
\label{eq:p_swsq}
\end{align}
where $_{(\cdot)}$ is a placeholder for TC and NT for respective tropical cyclone and non-tropical descriptions, $p(\mathbf{\overline{s}})$ is the PDF of the spatial average soil moisture for the watersheds,  $p(\mathbf{s}|\mathbf{\overline{s}})$ is the PDF describing the random field of watershed soil moisture conditional on the average value, $p(\mathbf{\overline{q}}_b|\mathbf{\overline{s}})$ is the PDF of the baseflow produced within the watershed conditional on the watershed spatial average soil moisture, and $p(\mathbf{w}|\mathbf{\overline{w}})$ is the PDF describing the distribution of storage capacity for each watershed with an average value of $\mathbf{\overline{w}}$. This conditional dependence structure follows from established approaches in spatially lumped hydrologic modeling that links large‐scale (unit-area) watershed behavior to local, point‐scale processes \cite{bartlett2015stochastic,bartlett2016mean,bartlett2015unified2,bartlett2017reply,bartlett2025stochasticE}.

The soil moisture PDF, $p_{(\cdot)}(\mathbf{\overline{s}})$, represents variations in soil moisture driven by the continuous hydrologic response across all storm events. It is obtained by a time integration of the instantaneous soil moisture PDF, $p(\mathbf{\overline{s}};t)$, weighted by the normalized time-varying frequency of each storm type, i.e., $p(\mathbf{\overline{s}})  = \int_0^{T_o} p(\mathbf{\overline{s}};t)\frac{\lambda_{TC,t}(t)}{\lambda_{TC}\:T_o
}dt$ for TC events and  $p(\mathbf{\overline{s}})  = \int_0^{T_o} p(\mathbf{\overline{s}};t)\frac{\lambda_{NT,t}(t)}{\lambda_{NT}\:T_o}dt$ for non-TC events, where $\lambda_{TC}$ and $\lambda_{NT}$ are the average frequencies over the period of observation, $T_o$ \cite<e.g.,>{bartlett2025stochasticE}. In practice, these time integrations need not be performed explicitly because the required empirical PDFs are estimated directly from the corresponding historical event populations, thereby implicitly accounting for the time-varying occurrence of each storm type.

\subsubsection{Numerical Discretization}

The flood-depth PDFs follow directly from Eqs. (\ref{eq:pTC}) and (\ref{eq:pNT}) by integrating the responses over the continuous driver distributions of Eq. (\ref{eq:pcdot}). However, when the response operator is a numerical hydrodynamic model, these integrals cannot be evaluated analytically. Instead, they are approximated by numerical quadrature, whereby the continuous driver distributions are discretized into a finite collection of weighted driver realizations that are propagated through the flood-response model to approximate the flood-depth PDFs.

The continuous joint driver distribution of Eq. (\ref{eq:pcdot}) then is represented by the discrete measure,
\begin{align}
p_{\mathbf{x}}(\mathbf{x}) \approx \sum_{i} \omega_{i} \delta(\mathbf{x}- \mathbf{x}_{i})
\label{eq:p_discrete}
\end{align}
where $\mathbf{x}_i$ denotes the $i$th realization of the joint driver vector and $\sum_i \omega_i=1$. Consequently, all response integrals reduce to weighted summations over hydrodynamic model realizations. This discrete representation is used throughout the remainder of the framework to evaluate flood-depth distributions, conditional attribution distributions, and response-based design storms. 


\subsection{Probabilistic Characterization of the Compound Flood Response}
\label{sec:theory_characterization}

The compound flood-depth distribution developed in Section \ref{sec:theory_depths} provides a probabilistic description of the flood response from which several additional probabilistic characterizations follow naturally. These include the statistical delineation of compound flood transition zones, probabilistic attribution of flood depths to hydrologic and coastal processes, and response-based design-storm selection. Together, these quantities provide information beyond flood-depth frequency alone and are developed in the following sections.

\subsubsection{Statistical CFTZ Definition}

The extended JPM defines the compound flood transition zone (CFTZ) directly from the probability distribution of flood depths rather than from individual compound flood events. For an AEP, the compound flood depth is obtained from the quantile function $Q(R_p)$ of Eq. (\ref{eq:Qp}). Likewise, the corresponding rainfall-, fluvial-, and storm-surge-only quantiles, $Q_R (R_p)$, $Q_F (R_p)$, and $Q_S (R_p)$, are obtained by restricting the flood-response operator to the individual driver. The CFTZ therefore emerges from the probability distribution of flood depths as the set of locations where compound flooding systematically increases AEP flood depths beyond those produced by any individual flood-generation mechanism in isolation, i.e.,
\begin{align}
\text{CFTZ}(R_p) &=
\begin{cases}
1, & Q(R_p) - \max\left[ Q_R(R_p),Q_S(R_p),Q_F(R_p)\right]  > \epsilon, \\[4pt]
\text{None}, & \text{otherwise},
\end{cases}
\label{eq:CFTA_stat}
\end{align}
where $\epsilon$ is a user-defined threshold representing the minimum increase in flood depth considered practically significant.

Unlike previous event-based definitions, which identify the CFTZ from the amplification associated with an individual storm \cite<e.g.>{bilskie2018defining, bilskie2021enhancing,gori2020tropical,shen2019flood,han2024compound}, the present formulation is derived from the long-term flood-depth distribution. Consequently, the CFTZ reflects the increased frequency of flood depths arising from multiple probabilistic pathways rather than the nonlinear amplification associated with any particular event. As shown by the results in Section \ref{sec:compare_event}, this distinction leads to materially different CFTZ delineations and establishes the probabilistic basis for response-based design.

\subsubsection{Flood Depth Attribution}
Conditioning the extended JPM on a specified compound flood depth (or equivalently, AEP) naturally yields the probability distribution of flood-depth attribution. Because the framework derives the joint probability distribution of flood depths and storm characteristics (Section 3.1), conditioning on a target flood depth yields the probability distribution of storm realizations capable of producing that depth,
\begin{align}
p_{(\cdot)}(\mathbf{x}_{(\cdot)}\mid\eta_{\max}) = \frac{p(\eta_{\max},\mathbf{x}_{(\cdot)})}{p(\eta_{\max})}
\label{eq:p(x|eta}
\end{align}
where the denominator is the corresponding flood-depth PDF, and the numerator is the joint flood-depth/storm distribution implied by the integrands of Eqs. (\ref{eq:pTC}) and (\ref{eq:pNT}), which is the respective response PDF function multiplied by the general PDF of Eq. (\ref{eq:pcdot}). As previously stated, $_{(\cdot)}$ is a placeholder for TC and NT, for two different event types.

Propagating the conditional storm distribution through the surge-only response operator yields the conditional probability distribution of surge-attributed flood depth,
\begin{align}
p_{S,(\cdot)}(\eta_{\max,S}|\eta_{\max})= \int...\int p_{(\cdot)}(\eta_{\max,S}|\mathbf{x}_{(\cdot)})p_{(\cdot)}(\mathbf{x}_{(\cdot)}|\eta_{\max})d^n\mathbf{x}_{(\cdot)},
\label{eq:pTC,S}
\end{align}
where $p_{(\cdot)}(\eta_{\max,S}|\mathbf{x}_{(\cdot)})$ is the surge-only (or coastal-water-only) response. The hydrologic attribution is defined as the complementary increase in flood depth relative to the surge-only response, $\eta_{\max,H}=\eta_{\max}-\eta_{\max,S}$. Formulating the attribution relative to the surge-only response, $\eta_{\max,S}$, reflects the inherent asymmetry between coastal and hydrologic processes. Storm surge dynamics are largely unaffected by the presence or absence of hydrologic processes, whereas hydrologic processes---through drainage blockage, backwater effects, and elevated tailwater conditions---are strongly dependent on coastal water levels \cite<e.g.,>{feng2022investigating}. This formulation also aligns with the typical sequence of coastal flooding, in which storm surge generally precedes peak hydrologic response under elevated coastal water levels. \cite{green2025comprehensive,tanim2021developing}. The resulting joint conditional distribution is
\begin{align}
p_{(\cdot)}(\eta_{\max,H},\eta_{\max,S}|\eta_{\max}) = \delta(\eta_{\max,H}+\eta_{\max,S}-\eta_{\max})p_{S, (\cdot)}(\eta_{\max,S}|\eta_{\max}),
\label{eq:contributions_H+S}
\end{align}
where the fully conditioned PDF $p(\eta_{\max,H}\mid\eta_{\max,S},\eta_{\max})$ is represented by the Dirac delta function
$\delta(\eta_{\max,H}+\eta_{\max,S}-\eta_{\max})$, which constrains the probability mass to the line $\eta_{\max,H}+\eta_{\max,S}=\eta_{\max}$.

The PDF of the hydrologic attributed percentage, $p_{H_{\%},(\cdot)}(H_{\%}|\eta_{\max})$, follows by multiplying Eq. (\ref{eq:contributions_H+S}) by the transformation $\delta\left(H_{\%}-\frac{\eta_{\max,H}}{\eta_{\max}}100\right)$ and integrating over $\eta_{\max,S}$ and $\eta_{\max,H}$, yielding, 
\begin{align}
p_{H_{\%},(\cdot)}(H_{\%}|\eta_{\max}) = \frac{\eta_{\max}}{100}p_{S, (\cdot)}\left(\eta_{\max}\left(1-\frac{H_{\%}}{100}\right)|\eta_{\max}\right),
\label{eq:pHpercent}
\end{align}
where $H_{\%}$ is conditional on the total compound depth $\eta_{\max}$ and reflects the variability of surge contributions as described by Eq. (\ref{eq:pTC,S}). The complementary storm surge attributed percentage, $S_{\%}$, has a PDF given by 
\begin{align}
p_{S_{\%},(\cdot)}(S_{\%}|\eta_{\max})=p_{H_{\%}, (\cdot)}(100-S_{\%}|\eta_{\max}).
\label{eq:pSpercent}
\end{align}
The corresponding attribution PDFs over all storm types are obtained by weighting the tropical cyclone and non-tropical results by their
depth-conditioned probabilities of occurrence, i.e.,
\begin{align}
p_{H_{\%}}(H_{\%}|\eta_{\max}) &=\omega_{TC}\,p_{H_{\%},TC}(H_{\%}|\eta_{\max})+(1-\omega_{TC}) p_{H_{\%},NT}(H_{\%}|\eta_{\max})\\
p_{S_{\%}}(S_{\%}|\eta_{\max}) &=\omega_{TC}\,p_{S_{\%},TC}(S_{\%}|\eta_{\max})+(1-\omega_{TC})p_{S_{\%},NT}(S_{\%}|\eta_{\max}),
\label{eq:pSall_Scenario}
\end{align}
where the weight is given by 
\begin{align}
\omega_{TC} = \frac{\lambda_{TC}\, p_{TC}(\eta_{\max})}{\lambda \, p(\eta_{\max})},
\label{eq:weight_scenario}
\end{align}
in which $p_{TC}(\eta_{\max})$ is the flood-depth PDF for tropical cyclones and $p(\eta_{\max})$ is the overall flood-depth PDF of Eq.~(\ref{eq:peta}).

For an individual storm type (TC or non-TC), Eqs. (\ref{eq:p(x|eta})--(\ref{eq:pSpercent}) define the continuous attribution framework and Steps 1--4 of Table~\ref{tab:attribution_steps} provide the corresponding step-by-step discretized implementation based on the weighted event catalog, with each step cross-referenced to its continuous counterpart. 

\begin{table}[ht!]
\begin{minipage}{\textwidth}
\linespread{.5}\selectfont
\caption{Procedure for computing driver attribution and selecting equiprobable design-storm sets at a point, for a target compound flood depth $\eta_{\max}$ (e.g., an AEP depth). Each step is the discrete counterpart of the continuous formulation given by the cited equation.\label{tab:attribution_steps}}
\centering
\renewcommand{\arraystretch}{.8}
\small
\begin{tabular}{@{} S D M @{}}
\toprule
\textbf{Step} & \textbf{Description} & \textbf{Discretized representation} \\
\midrule
1 & \textbf{Filter by tolerance band [Eq. (\ref{eq:p(x|eta})].} Select the storms (indexed by $i$ with weight $\omega_i$) whose compound flood depth $\eta_{\max,i}$ falls within a tolerance $\Delta\eta_{\max}$ of the target depth $\eta_{\max}$, and renormalize their weights over the selected set $ST_{\Delta}$. &
$\displaystyle ST_{\Delta} = \bigl\{\, i \;\bigm|\; |\eta_{\max,i}-\eta_{\max}| \le \Delta\eta_{\max} \,\bigr\}$
\par\smallskip
$\displaystyle \bar\omega_i = \omega_i\Big/\!\!\sum_{j\in ST_{\Delta}}\!\!\omega_j\,,\qquad i\in ST_{\Delta}$ \\[0.6em]
\addlinespace
2 & \textbf{Build the conditional surge PDF [Eq. (\ref{eq:pTC,S})].} For each $i$-th storm, the model transforms the storm into a surge-only flood depth $\eta_{\max,S,i}$; the renormalized weights are aggregated into the PDF of surge-only depth conditional on the compound depth. &
$\displaystyle p_{S,(
\cdot
)}(\eta_{\max,S}\mid\eta_{\max}) = \!\!\sum_{i\in ST_{\Delta}}\!\! \bar\omega_i\,\delta\!\left(\eta_{\max,S}-\eta_{\max,S,i}\right)$ \\[0.6em]
\addlinespace
3 & \textbf{Convert to hydrologic attribution \% [Eq. (\ref{eq:pHpercent})].} Each surge-only depth is transformed into  a percent hydrologic attribution,  $H_\% = 100\,\left(1-\frac{\eta_{\max,S,i}}{\eta_{\max}}\right)$ . &
$\displaystyle p_{H_\%,(\cdot)}(H_\%\mid\eta_{\max}) = \!\!\sum_{i\in ST_{\Delta}}\!\! \bar\omega_i\,\delta\!\left(H_\% - 100\bigl(1-\tfrac{\eta_{\max,S,i}}{\eta_{\max}}\bigr)\right)$ \\[0.6em]
\addlinespace
4 & \textbf{Convert to surge attribution \% [Eq. (\ref{eq:pSpercent})].} Each surge-only depth is likewise transformed into the complementary percent surge attribution, $S_\% = 100 - H_\%$. &
$\displaystyle p_{S_\%,(\cdot)}(S_\%\mid\eta_{\max}) = \!\!\sum_{i\in ST_{\Delta}}\!\! \bar\omega_i\,\delta\!\left(S_\% - 100\,\tfrac{\eta_{\max,S,i}}{\eta_{\max}}\right)$ \\[0.6em]
\addlinespace
5 & \textbf{Partition into $K$ equiprobable sets [Eqs. (\ref{eq:pScenario}) and (\ref{eq:p_xstorm})].} Sort the filtered storms by ascending surge attribution, accumulate the renormalized weights, and partition at equiprobable increments (e.g., $K=5$ sets of probability $\approx 0.2$). &
$\displaystyle ST_k = \Bigl\{\, i \;\bigm|\; \tfrac{k-1}{K} < \!\!\sum_{j\le i}\!\bar\omega_j \le \tfrac{k}{K} \Bigr\},\; k=1,\dots,K$
\stnote{Storms sorted by $S_{\%,i}$.} \\[0.6em]
\addlinespace
6 & \textbf{Select a representative event [Eqs. (\ref{eq:design_storm}) and (\ref{eq:q_med})].} From each set $ST_k$, select one design storm (e.g., median or modal attribution), uniquely identified by its JPM storm, rainfall field, and antecedent condition. &
$\displaystyle i_k^\ast = \operatorname*{arg\,median}_{i\in\,ST_k} S_{\%,i}\,,\qquad$
\stnote{Representative event: $\mathbf{x}_{i_k^\ast}$}\tabularnewline
\bottomrule
\multicolumn{3}{@{}p{\linewidth}@{}}{\footnotesize $^*$The placeholder $_{(\cdot)}$ is for tropical cyclone (TC) and non tropical (NT) storms.} \\
\end{tabular}
\end{minipage}
\end{table}

\subsubsection{Response-based design storm selection}

A flood depth response generally results from many distinct storm realizations, each occurring with different probability and exhibiting different balances between coastal and hydrologic forcing. These realizations naturally define a set of probabilistic design scenarios, thereby extending the traditional deterministic design storm concept to a probabilistic description of physically plausible storm scenarios.

The probabilistic design scenarios are constructed by partitioning the surge-attribution PDF of Eq. (\ref{eq:pSpercent}) into $K$ equiprobable attribution scenarios,
\begin{align}
\int_{l_{k-1}}^{l_k}
p_{S_\%, (\cdot)}(S_\%|\eta_{\max})\,dS_{\%,(\cdot)}
=
\frac{1}{K},
\qquad
l_k=P^{-1}_{S_\%, (\cdot)}\!\left(\frac{k}{K}\Big|\eta_{\max}\right),
\label{eq:pScenario}
\end{align}
where $k=1,\ldots,K$, $P^{-1}_{S_\%, (\cdot)}\!\left(\cdot\right)$ is the inverse CDF, and the integration limits $l_{k-1}$ and $l_k$ define the  bounds of each attribution scenario (in terms of surge attribution percentage). Each interval therefore represents an equally likely class of storm attribution conditioned on the specified compound flood depth. Associated with each attribution scenario is a probability distribution of storm characteristics, 
\begin{align}
p_{k,(\cdot)}(\mathbf{x})
=
K
\int_{l_{k-1}}^{l_k}
p_{(\cdot)}(\mathbf{x}_{(\cdot)}|\eta_{\max},S_{\%,(\cdot)})
\,p_{S_\%,(\cdot)}(S_{\%,(\cdot)}|\eta_{\max})
\,dS_{\%,(\cdot)}.
\label{eq:p_xstorm}
\end{align}
where $p_{(\cdot)}(\mathbf{x}_{(\cdot)}|\eta_{\max},S_{\%})$ is the conditional PDF of storm characteristics given both the target flood depth and surge attribution,
\begin{align}
p_{(\cdot)}(\mathbf{x}_{(\cdot)} \mid \eta_{\max}, S_{\%}) = \frac{\frac{\eta_{\max}}{100}\,p_{(\cdot)}\!\left(\eta_{\max}\tfrac{S_\%}{100}\mid\mathbf{x}_{(\cdot)}\right)p_{(\cdot)}(\mathbf{x}_{(\cdot)}|\eta_{\max})}{p_{S_{\%},(\cdot)}(S_{\%} \mid \eta_{\max})}.
\end{align}
where $\frac{\eta_{\max}}{100}\,p_{(\cdot)}\!\left(\eta_{\max}\tfrac{S_\%}{100}\mid\mathbf{x}_{(\cdot)}\right)$ is the response function within Eqs. (\ref{eq:pTC}) and (\ref{eq:pNT}) transformed by a change of variables based on $S_{\%}=\frac{\eta_{\max,S}}{\eta_{\max}}100$. 
For the discretized implementation of each scenario, the previous PDF of storm characteristics of Eq. (\ref{eq:p_xstorm}), is represented by a weighted set of design storms (Step 5 of Table~\ref{tab:attribution_steps}).

For each scenario $k$, a representative design storm, $\mathbf{x}_{k}^{*}$, is selected as the mode (or alternatively the mean) of the corresponding design-storm scenario PDF, $p_{k,(\cdot)}(\mathbf{x}_{(\cdot)}|\eta_{\max},S_\%)$,  evaluated at the median attribution of the interval,
\begin{align}
\mathbf{x}_{k,(\cdot)}^{*}
=
\operatorname*{arg\,max}_{\mathbf{x}_{(\cdot)}}
p_{k,(\cdot)}(\mathbf{x}_{(\cdot)}\mid\eta_{\max},l_k^{\mathrm{med}}),
\label{eq:design_storm}
\end{align}
where the median attribution satisfies
\begin{align}
l_k^{\mathrm{med}}=K
\int_{l_{k-1}}^{l_k^{\mathrm{med}}}
p_{S_{\%},(\cdot)}(S_\%|\eta_{\max})\,dS_{\%,(\cdot)}
=
\frac{1}{2}.
\label{eq:q_med}
\end{align}
In the case of the discrete set of design storms, the representative design storm is the median (or modal) storm of the set (see Step 6 of Table \ref{tab:attribution_steps}). The resulting collection of representative storms spans the range of physically distinct yet equiprobable mechanisms capable of producing the specified AEP flood depth.  The continuous forms referenced in Table \ref{tab:attribution_steps} provide the basis for distribution fitting (e.g., at Step 1) that may be used to overcome situations with too few suitable candidate storms being selected.

While Eqs. (\ref{eq:pScenario})--(\ref{eq:q_med}) define response-based design storms for an individual storm type (e.g., non-TC and TC), the framework extends naturally to the combined storm climatology by defining the attribution scenarios from the overall surge-attribution PDF of Eq. (\ref{eq:pSall_Scenario}) rather than from the storm-type-specific attribution PDFs. Specifically, Eq. (\ref{eq:pSall_Scenario}) replaces $p_{S_\%}(S_\%|\eta_{\max})$ in Eq. (\ref{eq:pScenario}) to determine the attribution-scenario boundaries, $l_k$. These common boundaries are then used to separately construct the storm-characteristic PDFs of Eq. (\ref{eq:p_xstorm}) for tropical cyclone and non-tropical storms. Consequently, while the attribution scenarios remain equiprobable by construction, the relative probability of tropical cyclone and non-tropical design storms within each scenario is governed by their depth-conditioned likelihoods of occurrence, $\omega_{TC}$ and $(1-\omega_{TC})$, respectively, where $\omega_{TC}$ is defined by Eq.~(\ref{eq:weight_scenario}).



\begin{figure}
    \centering
    \includegraphics[width=1.0\linewidth]{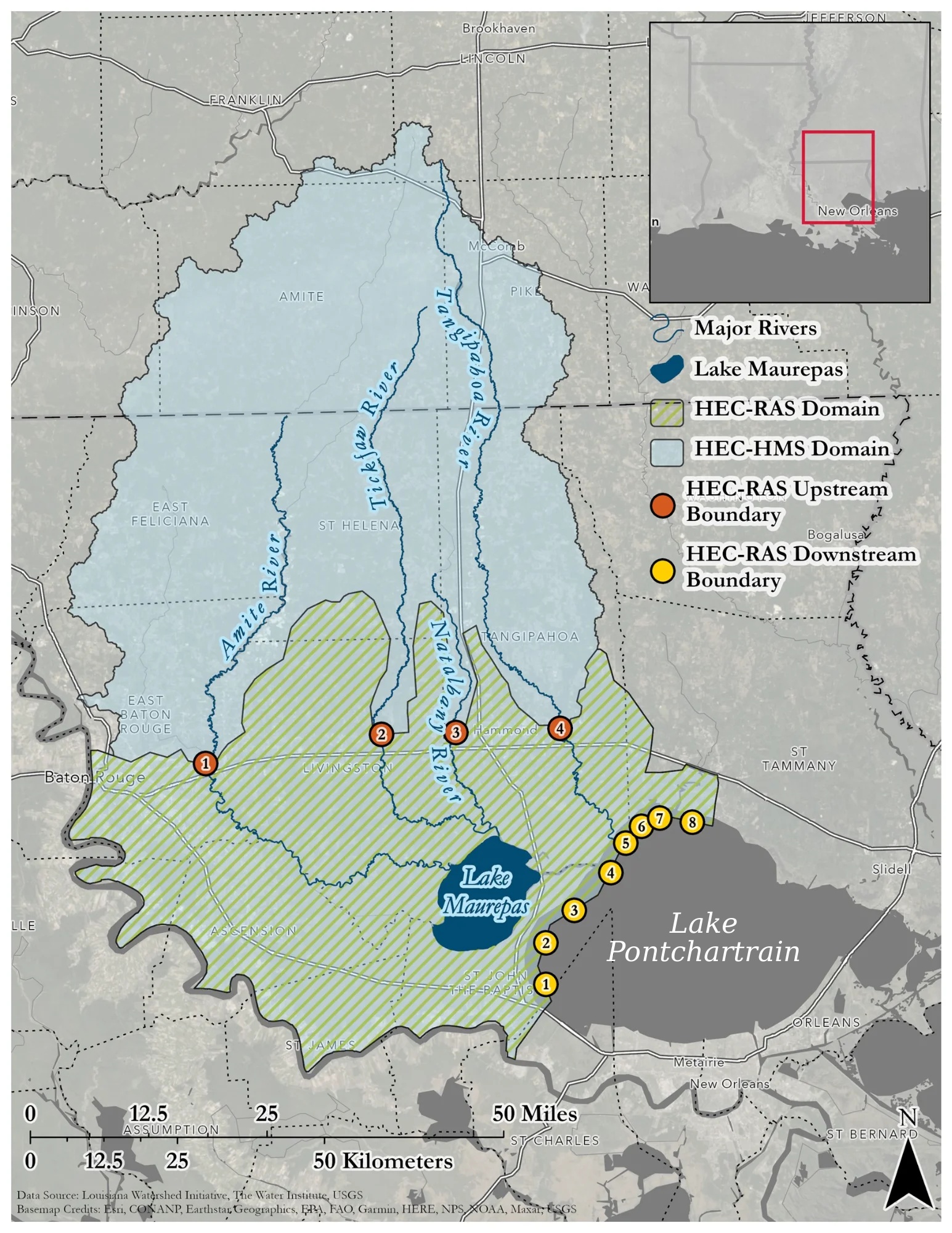}
    \caption{The HEC-RAS 2D model domain with inflows from the 1D HEC-HMS at four major river inflows shown in red and coastal storm surge boundary conditions from ADCIRC at line segments with centers shown in yellow.}
    \label{fig:domain}
\end{figure}
\section{Case Study}
\label{sec:case_study}

The probabilistic framework developed in the preceding Section \ref{sec:theory} is intentionally formulated independently of any particular hydrologic or hydraulic model. Its objective is to define the minimum stochastic descriptions of storm forcing, rainfall fields, and antecedent hydrologic state required to initialize compound flood analyses within an extended JPM framework. The case study presented here represents one implementation of that framework using the probabilistic assumptions summarized in Table \ref{tab:assumptions}. These assumptions are adopted primarily for parsimony and computational efficiency, providing a tractable demonstration of the framework rather than prescribing a preferred model formulation.

 \begin{landscape}

\begin{table}[p]
\begin{minipage}{23 cm}
\linespread{.5}\selectfont
\centering
\caption{Probabilistic assumptions adopted for the case study; none is a requirement of the framework of Section 3. Alternative assumptions would change the flood-depth frequency estimates while preserving CFTZ delineation, flood-driver attribution, and response-based design storm selection.}
\label{tab:assumptions}

\renewcommand{\arraystretch}{1.25}

\begin{threeparttable}
\begin{tabularx}{\linewidth}
{>{\raggedright\arraybackslash}p{6.2cm}
 >{\raggedright\arraybackslash}p{6.8cm}
 >{\raggedright\arraybackslash}X}
\toprule

\textbf{Assumption} &
\textbf{Rationale} &
\textbf{Expected Direction of Bias} \\

\midrule

Storm occurrence is independent of antecedent hydrologic conditions &
Consistent with established stochastic hydrology, where storm occurrence and watershed state are treated as independent processes. &
May underestimate compound flood frequency if storms preferentially occur during wetter antecedent conditions. \\

\addlinespace
\addlinespace
Historical climate is treated as stationary &
Consistent with conventional JPM applications and appropriate for demonstrating the framework. &
No systematic bias over the historical period, but may underestimate future flood frequencies under nonstationary climate conditions. \\

\addlinespace
\addlinespace
Non-tropical coastal forcing is represented by a generalized stage hydrograph parameterized by $\kappa$ and $\tau_l$ &
Provides a parsimonious representation of regional non-tropical coastal forcing. &
No expected systematic bias; primarily affects event-specific flood depths. \\

\addlinespace
\addlinespace
Antecedent soil moisture is represented by a single soil layer &
Captures the first-order controls on runoff generation while maintaining parsimony. &
Likely underestimates flood depths where deeper storage or prolonged wetness contributes significantly to runoff. \\

\addlinespace
\addlinespace
Spatially averaged antecedent soil moisture ($\mathbf{s}=\overline{\mathbf{s}}$) is rank-correlated across subwatersheds ($\rho=1$) &
Approximately consistent with available data \cite{koster2019length} and adopted for parsimony. &
May overestimate flood depths by increasing synchronization of runoff contributions. \\

\addlinespace

Fully saturated soil ($\mathbf{\overline{s}} = 1$) assumed within the HEC-RAS domain &
Near-saturated, low-storage wetland soils; infiltration fully represented in the upstream HEC-HMS subwatersheds. &
Overestimates flood depth from local rainfall within the HEC-RAS domain.\\

\addlinespace
\addlinespace
Baseflow is an average taken across multiple events &

Minimizes complexity, as baseflow has a second-order effect on flood peaks relative to storm-runoff \cite<e.g.,>{berghuijs2019relative}. &
Reduces antecedent streamflow variability and therefore likely underestimates the upper tail of flood depths, reducing predicted flood depth frequencies.\\

\addlinespace
\addlinespace
Infiltration is represented implicitly through storage capacity &
Storage capacity captures the dominant event-scale control on runoff generation required for model initialization. &
Little systematic bias following calibration; explicit infiltration models would slightly redistribute runoff in time \cite{rigby2006simplified}. \\

\addlinespace
\addlinespace
Hydrologic and hydraulic model implementation &
The framework is model-independent; the selected models provide one parsimonious implementation. &
Model dependent; affects quantitative flood-depth estimates but no expected systematic bias in CFTZ delineation or attribution. \\

\bottomrule

\end{tabularx}
\end{threeparttable}
\end{minipage}
\end{table}

\end{landscape}

\subsection{Study Area}

The study area is an archetypal CFTZ consisting of the watersheds draining into Lake Maurepas, including the Amite, Natalbany, Tangipahoa, and Tickfaw Rivers. This region exemplifies the multi-driver flooding dynamics addressed by the proposed framework (Fig. \ref{fig:domain}) and has experienced severe flooding from pluvial, fluvial, and coastal processes. Over a 48-hour period, the August 2016 non-tropical event produced more than 20 inches of rainfall over the Amite Basin and caused an estimated \$10 billion in damages \cite{watson2017characterization}.  Hurricane Isaac (2012) also generated widespread flooding through the combined effects of storm surge, heavy rainfall, and riverine inflows  \cite{berg2013tropical,rahman2021investigation}. Compound flooding was evaluated within the downstream HEC-RAS model domain (Fig.~\ref{fig:domain}), which encompasses the portion of the study area previously identified as susceptible to compound flooding \cite{bilskie2018defining,santiago2021examination}. The upstream boundary of the HEC-RAS domain was defined by four river inflows routed from 22 upstream HEC-HMS subwatersheds (Appendix~\ref{sec:HEC-HMS}). These inflow locations were selected to coincide with USGS streamflow gages located upstream of the compound flooding region. The downstream boundary was represented by eight water-level boundary conditions along Lake Pontchartrain, which hydraulically connects to Lake Maurepas through Pass Manchac and North Pass.

\subsection{Storm Frequencies}
\label{sec:freq}

The event frequencies were estimated separately for tropical and non-tropical storms. The frequency of tropical cyclones, $\lambda_{TC}$, was estimated from the HURDAT2 dataset by counting historical tropical cyclone crossings of the reference transect shown in Fig.~\ref{fig:2}, yielding $\lambda_{TC} = 1.184$~storms/yr. over the 72 years of record (1950-2022). Non-tropical storms were identified using a peak-over-threshold approach applied to total storm precipitation, after excluding tropical cyclone events identified from the HURDAT2 dataset \cite{landsea2013atlantic}. Applying a total-storm rainfall threshold of 80 mm identified 44 storms over 18 years, resulting in $\lambda_{NT} = 2.444$~storms/yr, which was sufficient to capture extremes greater than a 1-year event. These 44 storms are listed in Table \ref{tab:historic_storms} of \ref{sec:Integrated_modeling}.  The total storm frequency was therefore $\lambda = \lambda_{TC} + \lambda_{NT} = 3.628$~storms/yr.

\subsection{Implementation of Compound Flood Depth Response}
\label{sec:flood_response}

The response of the TC and non-TC events was represented by models as further described in \ref{sec:Integrated_modeling}.
For the deterministic models used in this case study, the tropical cyclone flood-depth response of Eq.~(\ref{eq:pTC}) was represented by

\begin{align}
p_{TC}(\eta_{\max}|\mathbf{x}_{TC}(t)) =& \int\!\dots\!\int
\delta\Bigg(\eta_{\max} - \max_{t \in [0,T_d]}  
\overbrace{f(\mathbf{x}_S(t),\mathbf{r}(t), \mathbf{s},\mathbf{\overline{q}}_b,\mathbf{q}(t))}^\text{HEC-RAS} \Bigg) 
\delta\Big(\mathbf{x}_S(t) - 
\longoverbrace{f(\mathbf{x}_{JPM})}{\text{ADCIRC+SWAN w/ PBL}} \Big)  \notag \\
&\quad\times \delta\Big(\mathbf{q}(t) - 
\underbrace{f(\mathbf{r}(t), \mathbf{s},\mathbf{\overline{q}}_b)}_{\text{HEC-HMS}} \Big)   
d^n\mathbf{x}_S\, d^n\mathbf{q},
\label{eq:petaTC_PS}
\end{align}
where $\eta_{\max}$ denotes the maximum flood depth  over the storm duration $T_d$, evaluated on the DEM grid. The integration was performed over all upstream inflows $\mathbf{q}(t)$ and downstream surge conditions $\mathbf{x}_S(t)$ provided as inputs to the HEC-RAS model. Winds were applied throughout the HEC-RAS domain. The Dirac delta functions $\delta(\cdot)$ enforce deterministic constraints, ensuring consistency with outputs from HEC-RAS, HEC-HMS, and ADCIRC+SWAN \cite{bartles2022hydrologic,brunner2021hecras,dietrich2011modeling,dietrich2012performance,westerink1994adcirc,booij1999third}. River inflow hydrographs $\mathbf{q}(t)$ were generated by HEC-HMS over the 22 upstream watersheds, while storm surge and wind fields $\mathbf{x}_S(t) = \{\boldsymbol{\eta}_s(t), \mathbf{u}(t)\}$ were obtained from ADCIRC+SWAN simulations configured for the 2023 Louisiana Coastal Master Plan \cite{cobell2023coastal}. Synthetic storm wind and pressure fields were provided by OceanWeather Inc. from their PBL model \cite{cialone2015north} and incorporated into HEC-RAS in a Lagrangian reference frame accounting for wind magnitude and direction, with wind stress parameterized using the Garrett drag formulation. The ADCIRC+SWAN water levels were applied at eight  boundary locations $\boldsymbol{\eta}_s = \{\eta_{s,1},\dots,\eta_{s,8}\}$, while river inflows from HEC-HMS were applied at four points $\mathbf{q} = \{q_1,\dots,q_4\}$ (Fig. \ref{fig:domain}). For all model realizations, HEC-RAS v6.1 was used because of its Linux implementation, which was required for the large simulation ensemble. Since this version did not support spatially varying downstream boundary conditions, the continuous coastal boundary was discretized into the eight stage-hydrograph offshore segments. For simulations driven solely by coastal forcing, this approximation produced negligible differences in simulated water levels within the HEC-RAS domain relative to those obtained using the original ADCIRC model and boundary conditions.

For non-tropical events, an analogous formulation was used, except that ADCIRC+SWAN storm-surge boundary conditions were replaced by a generalized stage hydrograph derived from historical observations. The hydrograph consisted of a generic stage pattern made event-specific through a peaking factor $\kappa$ and lag time $\tau_l$, which respectively controlled the stage magnitude and timing relative to peak river flow in the Amite River. The non-tropical flood depth response of Eq.~(\ref{eq:pNT}) was

\begin{align}
p_{NT}(\eta_{\max}|\mathbf{x}_{NT}(t)) =& \int\!\dots\!\int 
\delta\Bigg(\eta_{\max}-\max_{t \in [0,T_d]} \overbrace{f(\boldsymbol{\eta}_s(t), \mathbf{u}(t), \mathbf{r}(t), \mathbf{s},\mathbf{\overline{q}}_b,\mathbf{q}(t))}^\text{HEC-RAS} \Bigg) \notag \\
&\quad\times \delta\Big(\boldsymbol{\eta}_s-\longoverbrace{f(\tau_l,\kappa; t)}{\text{General Stage Hydrograph}}\Big) 
\,\delta\big(\mathbf{q}(t)-\underbrace{f(\mathbf{r}(t), \mathbf{s},\mathbf{\overline{q}}_b)}_{\text{HEC-HMS}}\big)\,
d^n\boldsymbol{\eta}_s\, d^n\mathbf{q},
\label{eq:petaNT_PS}
\end{align}
where, at the DEM resolution, $\eta_{\max}$ is again the maximum flood depth over $T_d$, and the Dirac delta functions enforce deterministic equivalence to the governing models. Each event included a wind field, $\mathbf{u}(t)$, from the “best reanalysis” wind fields (from OWI, Inc.). Here, the same eight coastal boundary locations were instead prescribed by the generalized stage hydrograph (see Fig.~\ref{fig:domain}). The generalized stage hydrograph was used in place of event-specific ADCIRC+SWAN water levels for the reasons given in Section \ref{sec:StormFreqProb} (Eq. (31)). The corresponding OWI wind field was nevertheless applied for each historical non-tropical event.

Respective details on the HEC-HMS and HEC-RAS model setups are in Appendices \ref{sec:HEC-HMS} and \ref{sec:HEC-RAS}.  For both tropical and non-tropical responses, infiltration within the HEC-RAS model and the 22 HEC-HMS watershed models was represented using the USACE constant-loss and initial-deficit method, with an initial deficit equivalent to $(1-\mathbf{s})\,\mathbf{w}$. These 23  models were initialized using the spatially averaged antecedent conditions:  
$\mathbf{\overline{s}} = \{\overline{s}_1,\dots,\overline{s}_{23}\}$,  
$\mathbf{\overline{q}}_b = \{q_{b,1},\dots,q_{b,23}\}$, and  
$\mathbf{\overline{w}} = \{\overline{w}_1,\dots,\overline{w}_{23}\}$, with respective PDFs as discussed in the next section.

\subsection{Implementation of the Extended Driver Probability Distributions}
\label{sec:StormFreqProb}

For both tropical and non-tropical events, we constructed a probabilistic characterization of the factors driving flood depths consisting of a joint PDF of the factors per storm event. This includes storm characteristics, stochastic rainfall, and hydrologic conditions conditioned on the spatially averaged soil moisture, baseflow, and storage depth of the 23 model units (1 HEC-RAS and 22 HEC-HMS). Although all storm characteristics evolve continuously in time, several quantities are represented by characteristic values for implementation. For example, antecedent soil moisture is specified immediately prior to storm onset, JPM parameters are evaluated at landfall, whereas rainfall fields evolve continuously throughout each event.

Over all 23 watershed models of the study, point values of soil moisture and storage depth were taken equal to their spatial averages, i.e., $\mathbf{s}=\mathbf{\overline{s}}$ and $\mathbf{w}=\mathbf{\overline{w}}$, and baseflow was represented by the average across the historical events listed in Tables   \ref{tab:calibration_storms} and \ref{tab:historic_storms} (see Table \ref{tab:assumptions} for assumption details).  Accordingly, PDFs for point-wise soil moisture and storage depth, as well as the baseflow PDF were represented by probability point masses $\delta(\mathbf{s}-\mathbf{\overline{s}})$, $\delta(\mathbf{w}-\mathbf{\overline{w}})$, and $\delta(\mathbf{\overline{q}}_b-\mathbf{\overline{q}}_{b,avg})$,  with Dirac delta functions indicating that the point value equals the spatial average with probability 1.  The hydrology PDF of Eq. (\ref{eq:p_swsq}) then took the overall form of $\delta(\mathbf{s}-\mathbf{\overline{s}})p(\mathbf{\overline{s}})\,\delta(\mathbf{w}-\mathbf{\overline{w}})\delta(\mathbf{\overline{q}}_b-\mathbf{\overline{q}}_{b,avg})$.

The PDF of tropical cyclone storm and land surface characteristics, $p_{TC}(\mathbf{x}_{TC}(t))$, included the PDF of storm characteristics represented by the JPM tropical cyclone characteristics, $p(\mathbf{x}_{JPM})$, augmented with PDFs for the random field of rainfall, $\delta(\mathbf{\overline{r}}(t)|\mathbf{x}_{JPM})p(\mathbf{r}(t)|\mathbf{\overline{r}}(t))$, and a PDF of hydrologic attributes, $p_{TC}(\mathbf{s},\mathbf{\overline{s}},\mathbf{\overline{q}}_b,\mathbf{\overline{w}})$, i.e.,

\begin{align}
p_{TC}(\mathbf{x}_{TC}(t)) = & \overbrace{p(\mathbf{x}_{JPM})}^{\substack{\text{JPM} \\ \text{Storms}}}\overbrace{
\delta(\mathbf{\overline{r}}(t)|\mathbf{x}_{JPM})}^{\substack{\text{IPET} \\ \text{Rainfall Model}}}
\overbrace{p(\mathbf{r}(t)|\mathbf{\overline{r}}(t))}^{\substack{\text{Rain-} \\ \text{fall} \\ \text{Field} }}\overbrace{\delta(\mathbf{s}-\mathbf{\overline{s}})p(\mathbf{\overline{s}})\delta(\mathbf{w}-\mathbf{\overline{w}})\delta(\mathbf{\overline{q}}_b-\mathbf{\overline{q}}_{b,avg})}^\text{Hydrology}.
\label{eq:pTC_PS}
\end{align}
where $p(\mathbf{x}_{JPM})$ was represented by the 645 JPM storms from the USACE coastal hazard study of Louisiana \cite{nadal2022coastal}.

\begin{figure}
    \centering
    \includegraphics[width=6in]{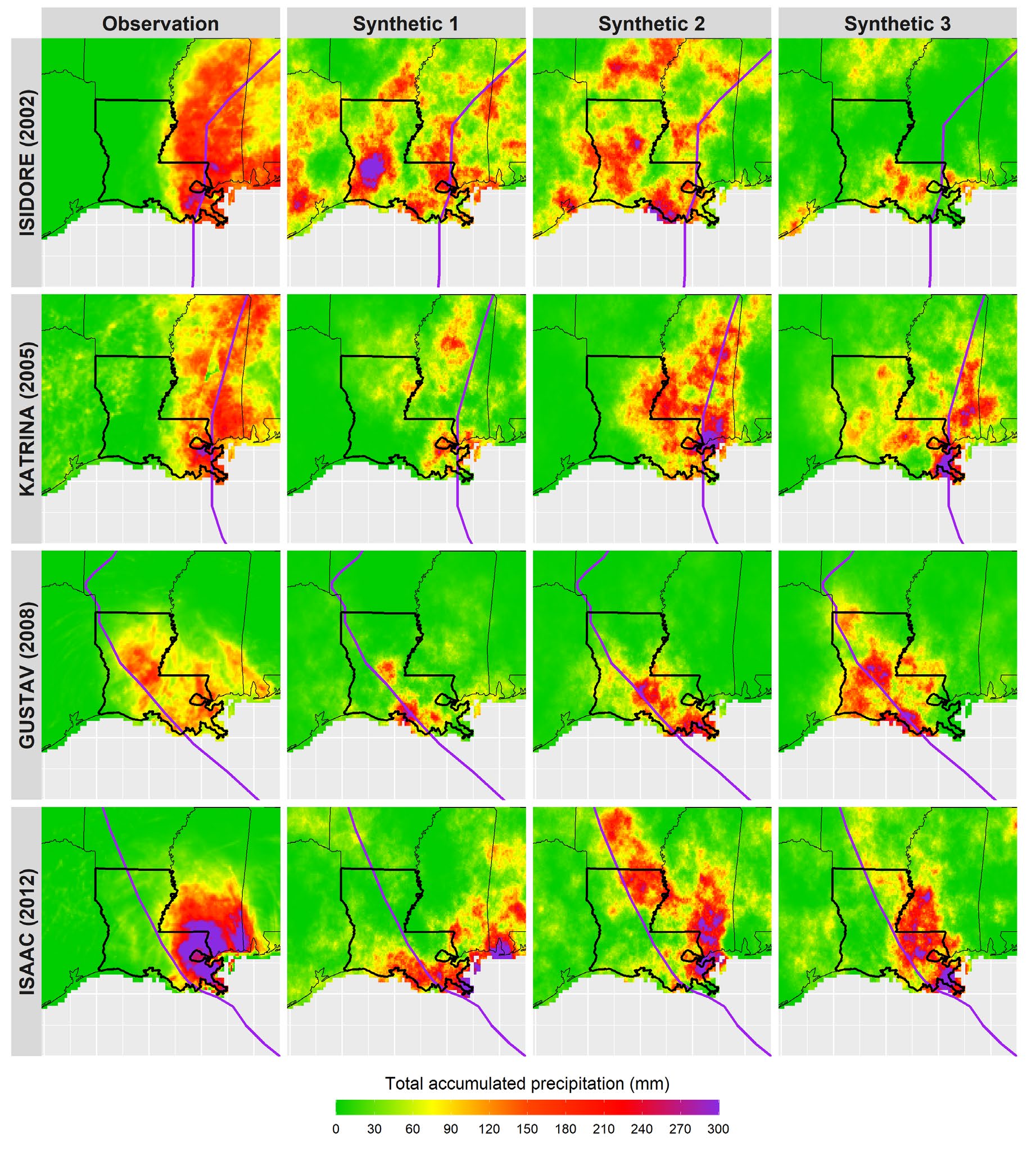}
    \caption{Total accumulated rainfall from Analysis of Record for Calibration (AORC) observations (first column) and example synthetic ensemble rainfall fields (next three columns) generated from the parametric generator for Hurricane Isidore, Hurricane Katrina, Hurricane Gustav and Hurricane Isaac \cite{villarini2022probabilistic}. TC tracks are shown as the magenta line. }
    \label{fig:3}
\end{figure}

Following \citeA{villarini2022probabilistic}, the rainfall field was represented as a bias-corrected IPET model, $\delta(\mathbf{\overline{r}}(t)|\mathbf{x}_{JPM})$, modulated by a spatially correlated multiplicative noise, $p(\mathbf{r}(t)|\mathbf{\overline{r}}(t))$ that is equal to $\delta(\mathbf{r}(t)-h[\mathbf{\overline{r}}(t)]\epsilon)p(\epsilon)$. Here, 
the function $h(\cdot)$ was a deterministic bias correction term while $\epsilon$ was a random component with a PDF $p(\epsilon)$ represented by a mixture of Gaussian PDFs \cite{villarini2022probabilistic}. When the total rainfall resulting from the random field, i.e., $\int_0^{T_d}\delta(\mathbf{r}(t)-h[\mathbf{\overline{r}}(t)]\epsilon)p(\epsilon)dt$, was plotted in comparison to observed storms, the rainfall totals exhibited broad spatial patterns similar to those observed in historical tropical cyclones (Fig. \ref{fig:3}). 


The soil moisture PDF, $p(\mathbf{\overline{s}})$, was fitted to the soil moisture values that allowed the HEC-HMS models to best replicate the observed discharge for 20 of the tropical storm events in Table~\ref{tab:calibration_storms} of Appendix \ref{sec:calibration}.  
For each historical event, baseflow was calculated with the Eckhardt separation method, yielding $\mathbf{\overline{q}}_b$ \cite{eckhardt2005construct,eckhardt2008comparison}.  The resulting average across all events, $\mathbf{\overline{q}}_{b,\mathrm{avg}}$ was used in the probabilistic description in the component,  $\delta(\mathbf{\overline{q}}_b-\mathbf{\overline{q}}_{b,\mathrm{avg}})$.  
The JPM storm attribute PDF $p(\mathbf{x}_{JPM})$ consisted of an empirical landfall-location distribution $p(x_l)$  and continuous multivariate PDF for the other JPM attributes that are conditional on the landfall location. These continuous PDFs (conditional on the landfall location) were fit to the historical data of HURDAT2 (see \ref{sec:JPM_PDFs}).

For the HEC-RAS model domain, a fully saturated soil ($\mathbf{\overline{s}}=1$)  was assumed to establish a conservative flood depth estimate, consistent with near-saturated, low storage soils of this coastal zone. Infiltration is fully represented in the upstream HEC-HMS subwatersheds generating the riverine inflows, so the fluvial driver is loss-corrected and the assumption affects only local rainfall within the HEC-RAS domain. This infiltration assumption (over just the HEC-RAS domain) was applied identically across all simulations, and consequently, it only introduces a common-mode bias that largely cancels in the relative comparisons defining the transition zone  (Eq. \ref{eq:CFTA_stat}) and driver attribution (Eqs. \ref{eq:p(x|eta} - \ref{eq:pSpercent}).

Based on a  comparison with the soil moisture values inferred from observed data (via HEC-HMS best fits to USGS gage data), the probability distribution of spatially averaged soil moisture, $p(\mathbf{\overline{s}})$, was best represented by a mixture distribution consisting of two parts: 1) $ \delta(\mathbf{\overline{s}}-1)P(\mathbf{\overline{s}}=1)$, representing the discrete probability of soil saturation, and 2) a truncated normal PDF, representing (unit-area) soil moisture values within the range $(0,1]$, $p(\mathbf{x}; \boldsymbol{\mu}, \Sigma)$, weighted by the probability the soil is not saturated, i.e.,  $(1-P(\mathbf{\overline{s}}=1))$. The overall soil moisture PDF thus took the form
\begin{align}
p(\mathbf{\overline{s}}) = \delta(\mathbf{\overline{s}}-1)P(\mathbf{\overline{s}}=1)+(1-P(\mathbf{\overline{s}}=1))p(\overline{s}; \boldsymbol{\mu}, \Sigma).
\end{align}
where $\boldsymbol{\mu} = \{\mu_1, \mu_2,...,\mu_{22}\}$ denotes the vector of ensemble mean soil moisture values, and the covariance matrix, $\Sigma$, is parameterized by the standard deviations $\boldsymbol{\sigma} = \{\sigma_1, \sigma_2,...,\sigma_{22}\}$ and a common rank correlation coefficient, $\rho$, set to $\rho=1$ (see Table \ref{tab:assumptions}). This implies that each subwatershed follows a common latent wetness percentile, while allowing the absolute soil moisture values to differ according to their respective marginal distributions. The probability of saturation, $P(\mathbf{\overline{s}}=1)$, the basin average values, $\boldsymbol{\mu}$, and the standard deviations, $\boldsymbol{\sigma}$, were found by fitting the 22 marginal distributions of $p(\mathbf{\overline{s}})$ to the respective empirically derived (historical) soil moisture values, $\mathbf{\overline{s}}$, of each of the 22 subwatersheds (see Table \ref{tab:calibration_storms}). The 23rd watershed (the HEC-RAS domain) was assigned the corresponding value of $\mathbf{\overline{s}}=1$ as described previously.

The PDF of non-tropical storm event characteristics, $p_{NT}(\mathbf{x}_{NT};t)$ included the PDFs for the non-tropical storm surge characteristics, $p(\kappa)p(\tau_l)$, the PDF of the wind field conditional on the storm rainfall, $p(\mathbf{u}(t)|\mathbf{\overline{r}}(t))$, PDFs for the random field of rainfall, $p(\mathbf{\overline{r}}; t)p(\mathbf{r}|\mathbf{\overline{r}}; t)$, and a PDF of soil moisture and baseflow hydrologic attributes, $p_{NT}(\mathbf{s},\mathbf{\overline{s}},\mathbf{\overline{q}}_b,\mathbf{\overline{w}})$ i.e.,

\begin{align}
p_{NT}(\mathbf{x}_{NT}(t)) = & \overbrace{p(\kappa)p(\tau_l)}^{\substack{\text{Non-tidal} \\ \text{Residual}}}\overbrace{
p(\mathbf{u}(t)|\mathbf{\overline{r}}(t))}^{\substack{\text{Historic} \\ \text{Wind Fields}}}\overbrace{
p(\mathbf{\overline{r}}(t))p(\mathbf{r}(t)|\mathbf{\overline{r}}(t))}^{\substack{\text{Historic} \\ \text{Rainfall Fields}}}\overbrace{\delta(\mathbf{s}-\mathbf{\overline{s}})p(\mathbf{\overline{s}})\delta(\mathbf{w}-\mathbf{\overline{w}})p(\mathbf{\overline{q}}_b|\mathbf{\overline{s}})}^{\substack{\text{Historic} \\ \text{Hydrology}}},
\label{eq:pNT_PS}
\end{align}
where the wind, rainfall, and antecedent hydrologic components were represented empirically using the catalog of historic non-TC events (Table \ref{tab:historic_storms}).  The PDF of the spatially averaged rainfall, $p(\overline{\mathbf{r}}(t))$, was assumed independent of the non-tidal residual, and the peaking factor $\kappa$ and lag time $\tau_l$ were treated as mutually independent, because in each case the historical data exhibited negligible dependence (Kendall's $\tau_K\approx0$). These weak dependencies motivated representing the downstream boundary using randomized generalized stage hydrographs rather than event-specific ADCIRC+SWAN water-level time series.

The empirical representations of the wind field, rainfall field, and antecedent hydrologic conditions were obtained from the 44 largest non-tropical storm events identified in Table \ref{tab:historic_storms}. For each event, rainfall was represented using the Analysis of Record for Calibration (AORC) dataset (5 km, hourly; \citeA{fall2023office}), which provides a continuous record beginning in 1979 and exhibits the strongest agreement with observed precipitation in Louisiana (correlation coefficient $r>0.75$) among available gridded products \cite{kim2022evaluation}. Event-specific wind forcing was obtained from the corresponding OWI atmospheric reanalysis, while the antecedent baseflow and watershed moisture deficit were inferred using the calibrated HEC-HMS model driven by AORC rainfall and constrained by USGS streamflow observations.

\subsection{Numerical Evaluation of the Flood-Depth Distributions}

The flood-depth PDFs $p_{TC}(\eta_{\max})$ and $p_{NT}(\eta_{\max})$ of Eqs. (\ref{eq:pTC}) and (\ref{eq:pNT}) were evaluated by propagating discretized realizations of the driver PDFs of   \( p_{TC}(\mathbf{x}_{TC}(t)) \) and \( p_{NT}(\mathbf{x}_{NT}(t)) \) through the response models of Eqs. (\ref{eq:petaTC_PS}) and (\ref{eq:petaNT_PS}), following the discrete representation of Eq. (\ref{eq:p_discrete}).  For the tropical events, the PDF $p_{TC}(\mathbf{x}_{TC}(t))$ of Eq. (\ref{eq:pTC_PS}) was quantized into 322,500 events---the 645 JPM storms of the USACE Louisiana coastal hazard study \cite{nadal2022coastal} $\times$ 100 equiprobable rainfall fields $\times$ 5 antecedent soil moisture conditions---with each event assigned the product weight $\omega_i \cdot (1/100) \cdot \omega_m$, where $\omega_i$ is the probability weight of the $i$-th JPM storm, adopted directly from the JPM quadrature developed for the 2023 Louisiana Coastal Master Plan (Johnson et al., 2023), and $\omega_m$ is the weight of the $m$-th soil-moisture condition.  The non-tropical PDF $p_{NT}(\mathbf{x}_{NT}(t))$ was quantized into $1,100$ events -- $44$ historical storms $\times$ $5$ non-tidal-residual peaking factors $\times$ $5$ lag times -- with each event assigned the equal weight $(1/44) \cdot (1/5) \cdot (1/5) = 1/1100$. The individual discretized PDFs are constructed in Appendix~C (Eqs.~(\ref{eq:JPMset})--(\ref{eq:NT_dis}); Table~\ref{tab:vars_params_weights}).

The number of realizations in each dimension was established through preliminary convergence testing with random subsamples drawn from an ensemble of 1,500 HEC-HMS realizations. For tropical cyclones, 100 rainfall fields and 5 soil-moisture conditions per JPM storm (500 rainfall–soil-moisture realizations per storm) were sufficient to capture 90--98\% of the variance in simulated peak discharge across the three historical events tested (Tropical Storm Matthew and Hurricanes Rita and Isaac). For non-tropical events, the 5 non-tidal-residual peaking factors and 5 lag times were likewise sufficient, although the available historical record limited the number of storm events to 44.

The $322,500$ tropical and $1,100$ non-tropical events were used to initialize the respective response models, from which the maximum flood depth was obtained using the HEC-RAS model. For both event types, the pairs of flood depths and their corresponding probability weights were sorted by depth. The cumulative sum of the probability weights provided the CDFs, $P_{TC}(\eta_{\max})$ and $P_{NT}(\eta_{\max})$, from which the derivatives yielded the PDFs, $p_{TC}(\eta_{\max})$ and $p_{NT}(\eta_{\text{max}})$. These distributions, together with the storm occurrence frequencies, were then used to construct the annual maximum flood-depth CDF, $P_A(\eta_{\text{max}})$, following Eqs. (\ref{eq:pA}) and (\ref{eq:peta}).

\begin{figure}[b!]
  \centering
  \includegraphics[width=1.0\linewidth]{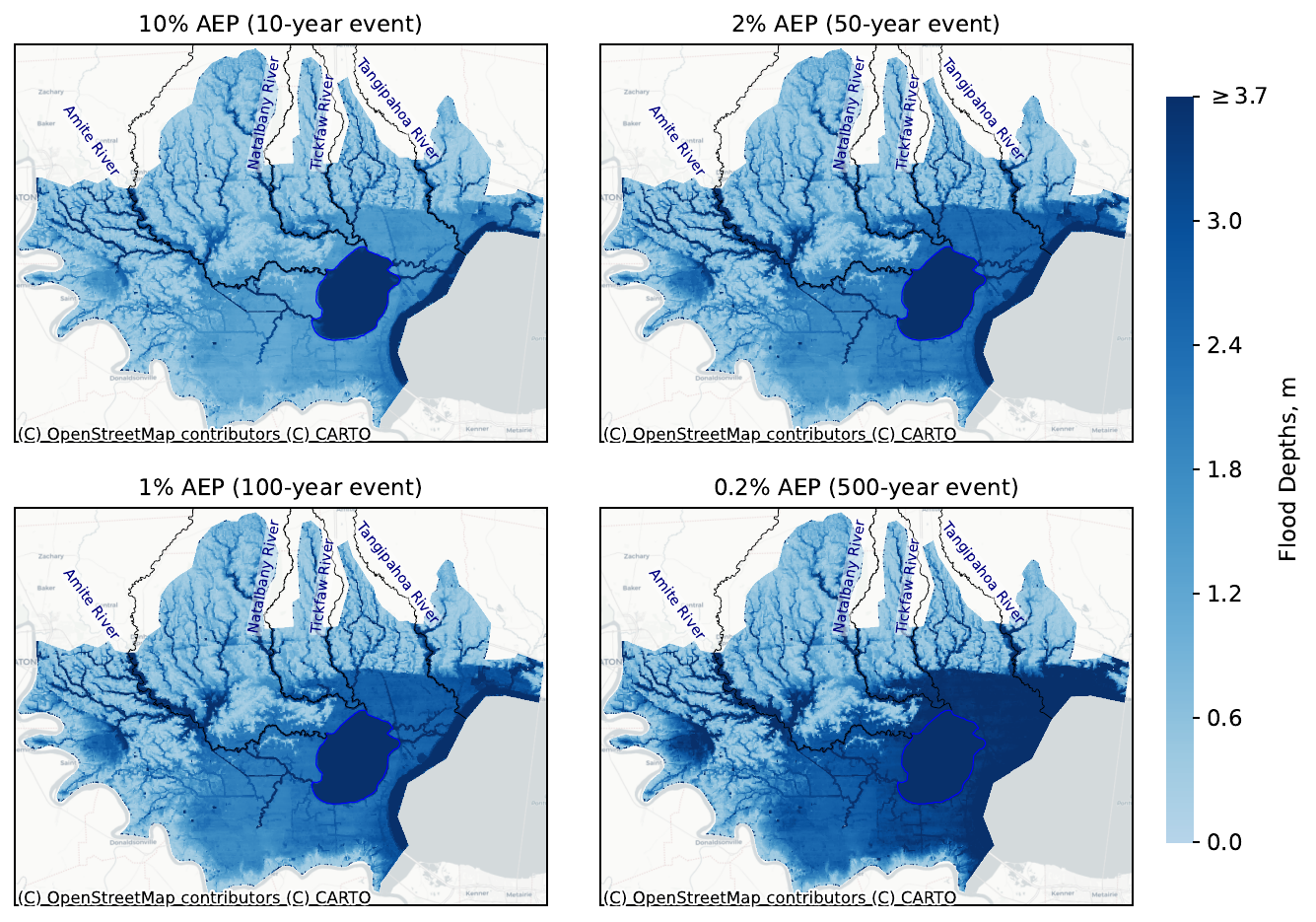}
  \caption{Compound flood depths for different AEPs as derived from the PDF of Eq. (\ref{eq:pA}), which accounts for both tropical and non-tropical storms based on the storm responses of Eqs. (\ref{eq:petaTC_PS}) and (\ref{eq:petaNT_PS}) and the likelihood of the storm characteristics as described by the PDFs $p_{TC}(\mathbf{x}_{TC}(t))$ and $p_{NT}(\mathbf{x}_{NT}(t))$ of  Eqs. (\ref{eq:pTC_PS}) and (\ref{eq:pNT_PS}), as discretized in \ref{sec:PDF_discretized}.} 
  \label{fig:compound_flooding}
\end{figure}

\section{Results}
\label{sec:results}

As discussed in the previous section, we conducted model simulations for both tropical and non-tropical storm events. To quantify the contributions of individual flood drivers, three additional sets of tropical cyclone simulations were performed with pluvial, fluvial, and storm surge processes modeled independently.  Together, these simulations produced five sets of flood responses: compound flooding from tropical cyclones, compound flooding from non-tropical storms, and three driver-isolated tropical cyclone responses.  Annualized flood-depth probability density functions (PDFs) were then computed at each model grid point following Eqs. (\ref{eq:pA}) and (\ref{eq:peta}).

As expected, compound flood depths increase with decreasing AEP, with the 10\%, 2\%, 1\%, and 0.2\% AEP events exhibiting progressively greater flood extents and depths (Fig. \ref{fig:compound_flooding}). The study area spans the transition between inland riverine and coastal flooding, where upland flows discharge onto a broad wetland plain connected to the shallow, brackish estuarine system surrounding Lake Maurepas and influenced by coastal flooding propagating inland from Lake Pontchartrain. Within this transition region, AEP flood depths are relatively uniform, increasing from approximately 2.1 m for the 10\% AEP event to 3.7 m for the 0.2\% AEP event (Fig. \ref{fig:compound_flooding}).

\begin{figure}[h!]
  \centering
  \includegraphics[width=1.0\linewidth]{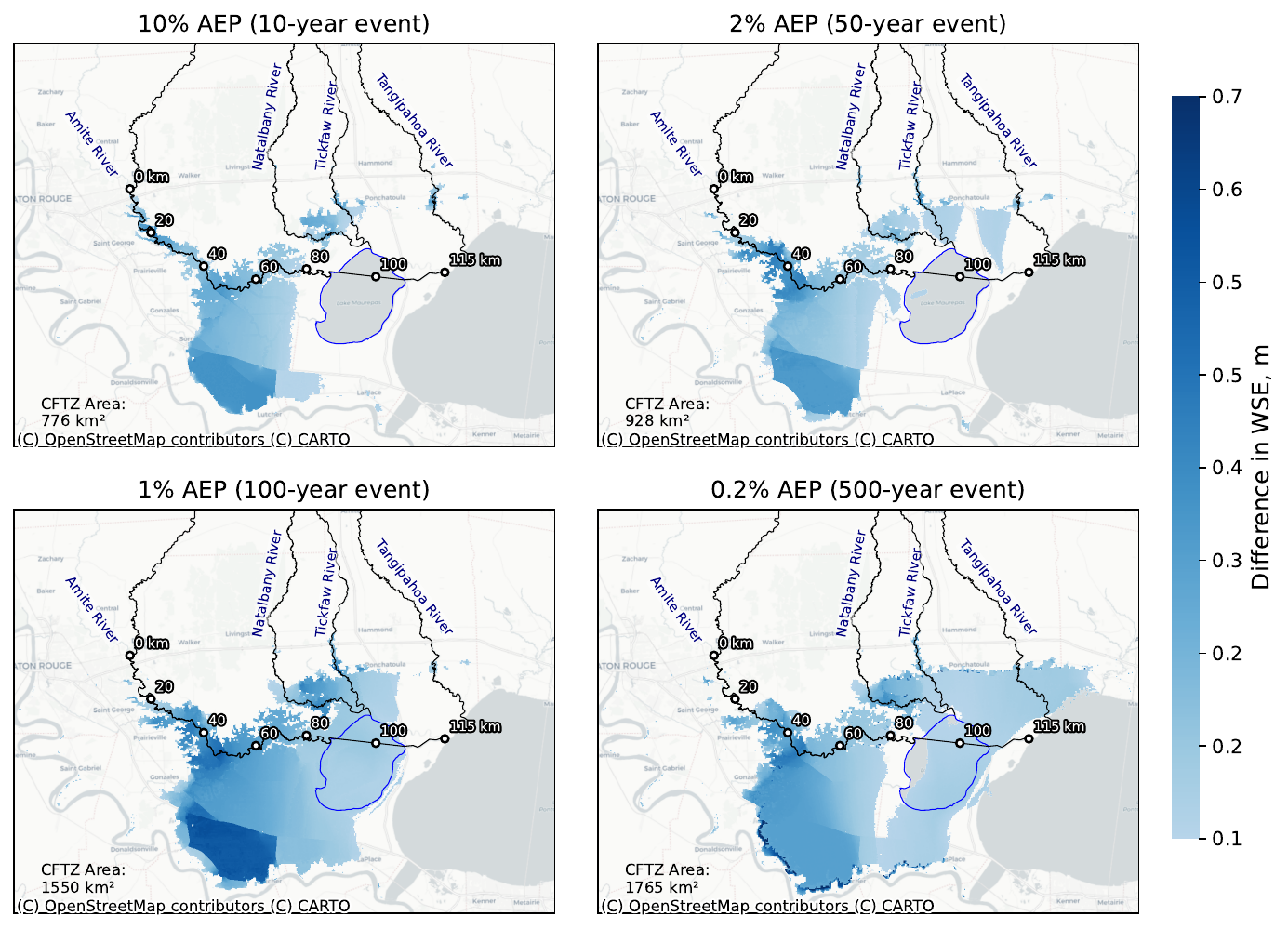}
  \caption{For different tropical cyclone AEP events, the CFTZ defined using the statistical criterion of Eq. (\ref{eq:CFTA_stat}) with a threshold, $\epsilon$, equal to 0.1 m. The transect path along the Amite River is shown in the figure, while the corresponding flood elevations and flood depth differences along this 115 km transect are presented in Fig. \ref{fig:amite_transect} }
  \label{fig:stat_cftz}
\end{figure}

While the AEP flood maps provide one characterization of the underlying flood-depth probability distributions, the same probabilistic flood-response framework also enables statistical delineation of the CFTZ, quantification of flood-driver attribution, and identification of response-based design storms. These results are presented in the following sections.

\begin{figure}
  \centering
  \includegraphics[width=1.0\linewidth]{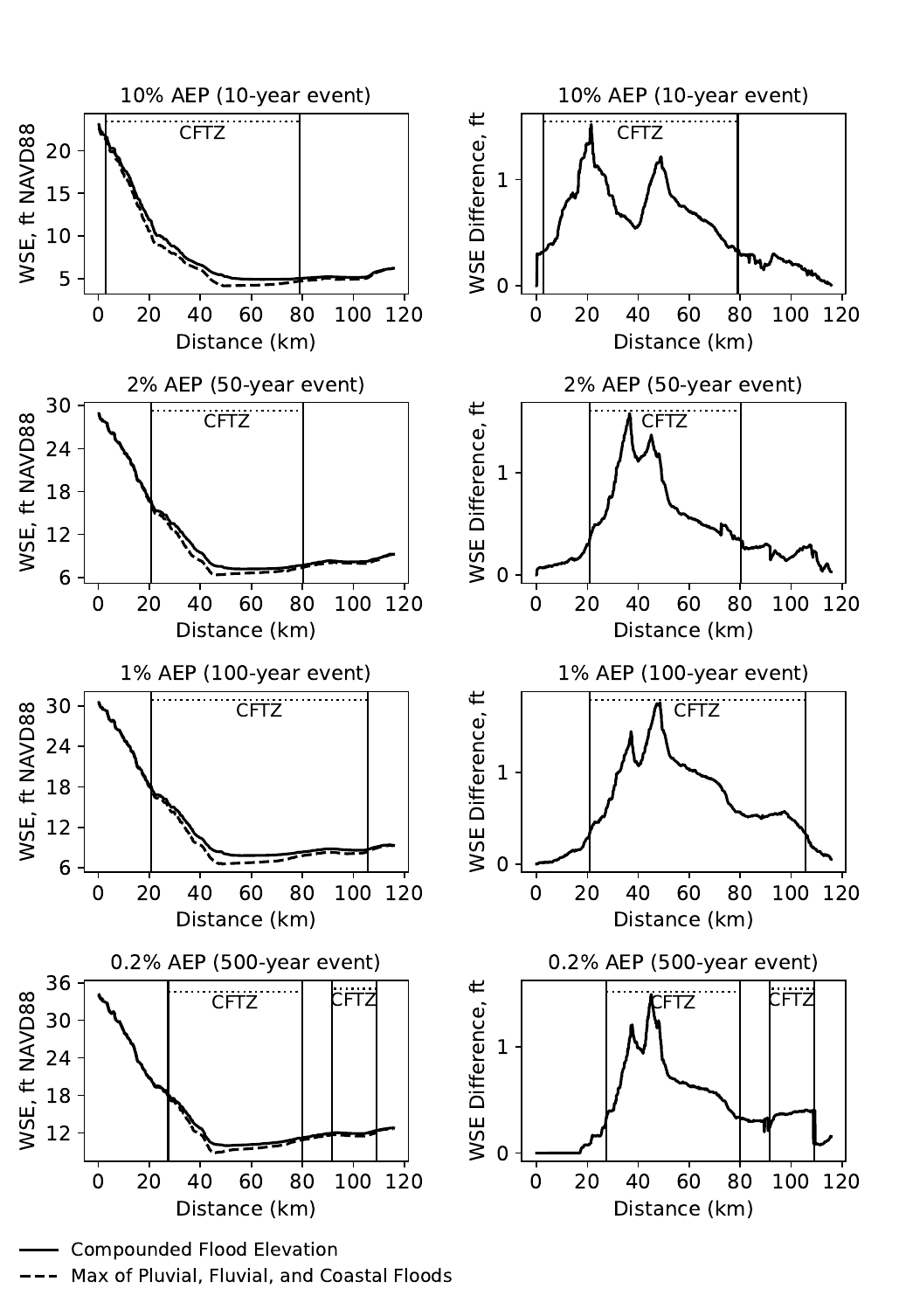}
  \caption{Along the Amite River transect of Figure  \ref{fig:stat_cftz}, a comparison of compound flooding WSEs with the envelope (pointwise maximum; dashed line) of the independently simulated pluvial-only, fluvial-only, and coastal-only WSEs (left column). The difference in WSEs, as defined in Eq. (\ref{eq:CFTA_stat}), identifies the CFTZ when it exceeds the threshold $\epsilon$, here set to 0.1 m.}
  \label{fig:amite_transect}
\end{figure}

\subsection{Defining the Compound Flood Transition Zone (CFTZ)}

\begin{figure}
  \centering
  \includegraphics[width=1.0\linewidth]{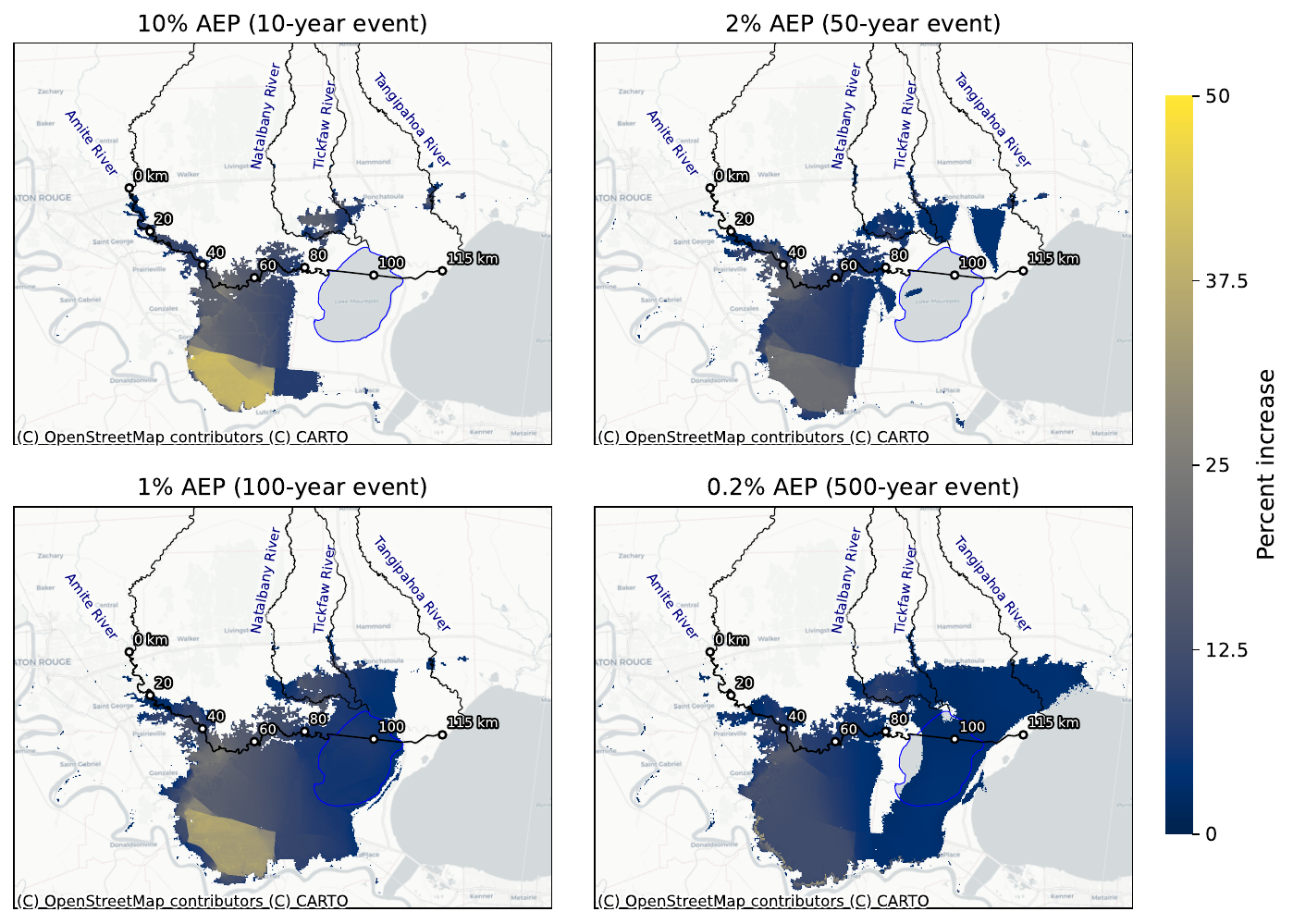}
  \caption{For different tropical cyclone AEPs and the CFTZ as defined in Fig. \ref{fig:stat_cftz}, the percent increase in the WSE resulting from the compound flood surface  in comparison to the WSE from the maximum of the pluvial-only, fluvial-only, and coastal-only surfaces. The transect path of Fig. \ref{fig:amite_transect} is shown for reference.}
  \label{fig:cftz_percent_increase}
\end{figure}

Applying the statistical CFTZ definition of Eq. (\ref{eq:CFTA_stat}) with $\epsilon=0.1$ m, CFTZs were delineated for the 10\%, 2\%, 1\%, and 0.2\% AEP flood events (Fig. \ref{fig:stat_cftz}). Water surface elevations (WSEs) and the corresponding compound-flood depth increases are shown along the Amite River transect (Fig. \ref{fig:amite_transect}). Across all AEPs, a pronounced discontinuity in flood depth occurs near the I-10 freeway and Highway 641 because of flow blockage, with consistently greater flood depths south and west of these transportation corridors. In the upper Amite River, compound flooding is primarily controlled by hydrologic processes, whereas around Lake Maurepas the combined influence of hydrologic and coastal processes produces substantially greater flood depths. With decreasing AEP, the CFTZ expands into the Lake Maurepas wetlands while contracting downstream along the Amite River (Figs. \ref{fig:stat_cftz} and \ref{fig:amite_transect}). The Amite River transect further illustrates these spatial patterns. The largest increases in WSE occur where flooding is already frequent, with the most pronounced changes occurring near the 50-km mark of Amite River transect (Fig. \ref{fig:amite_transect}). 

\begin{figure}
  \centering
  \includegraphics[width=1.0\linewidth]{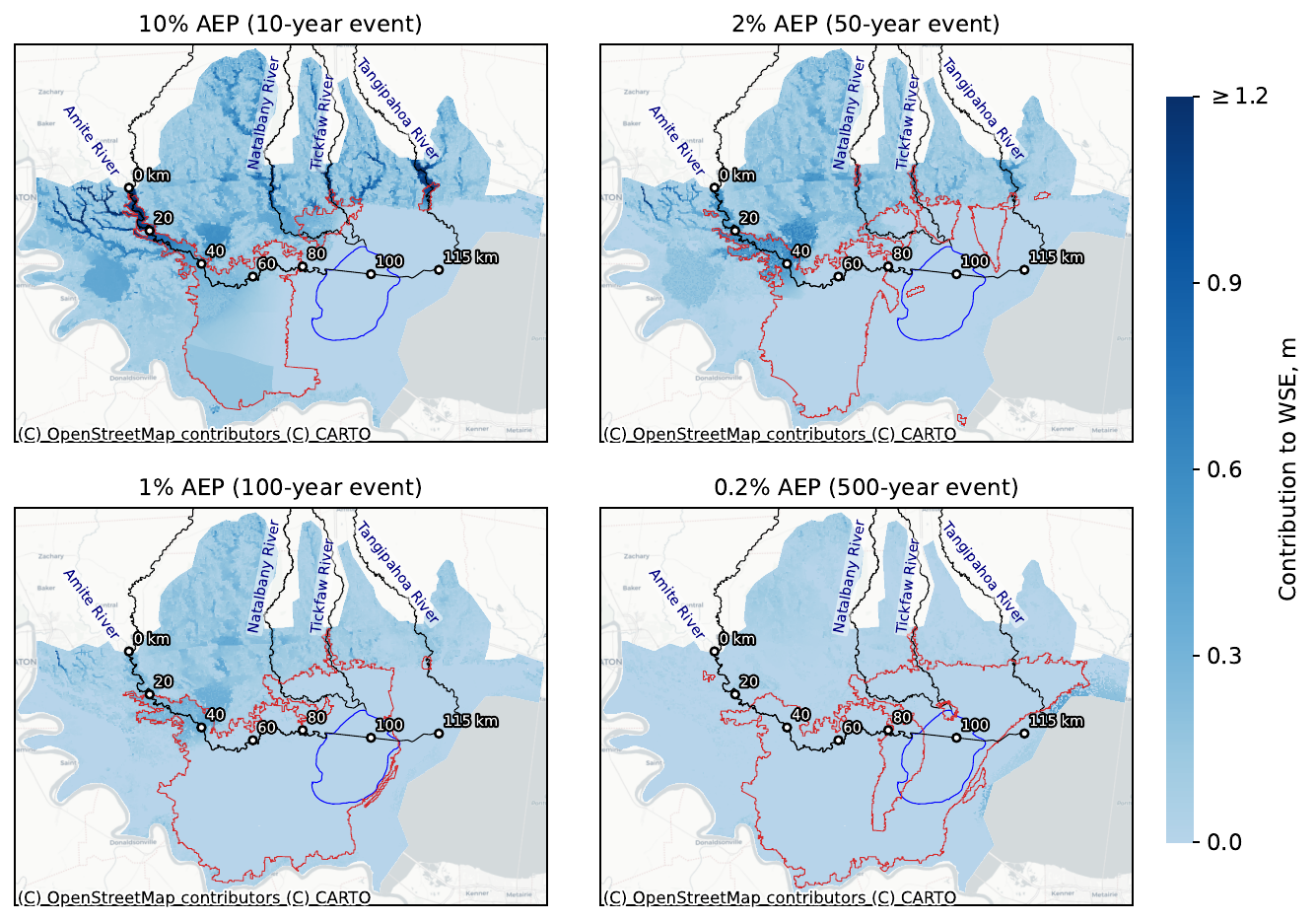}
  \caption{The increase in the compound flood depth from the inclusion of non-tropical events, which is the difference of the overall compound flood depths minus the compound flood depths calculated from the tropical cyclones alone. For comparison, the CFTZ (based on the tropical cyclones) is plotted (red boundaries).}
  \label{fig:cftz_nontrop_contrib}
\end{figure}

Across the study area, compound flooding increases AEP flood depths by approximately 5--50\%, with some areas south of I-10 exhibiting increases approaching 50\% (Fig. \ref{fig:cftz_percent_increase}). similarly, the areal extent of the CFTZ likewise increases with AEP, expanding from 776 km$^2$ for the 10\% AEP to 928 km$^2$, 1550 km$^2$, and 1765 km$^2$ for the 2\%, 1\%, and 0.2\% AEP events, respectively (Fig. \ref{fig:stat_cftz}). Non-tropical storms have only a modest influence on the CFTZ outside the primary river corridors but contribute appreciably to compound flood depths along the Amite River (Fig. \ref{fig:cftz_nontrop_contrib}). Along the Amite River transect, the non-tropical contribution increases from approximately 0 m to 1.2 m for the 10\% AEP event, and remains approximately 0.6 m and 0.3 m for the 2\% and 1\% AEP events, respectively.

\subsection{Comparison with Event-Based CFTZ Delineation}
\label{sec:compare_event}

\begin{figure}[h!]
  \centering
  \includegraphics[width=1.0\linewidth]{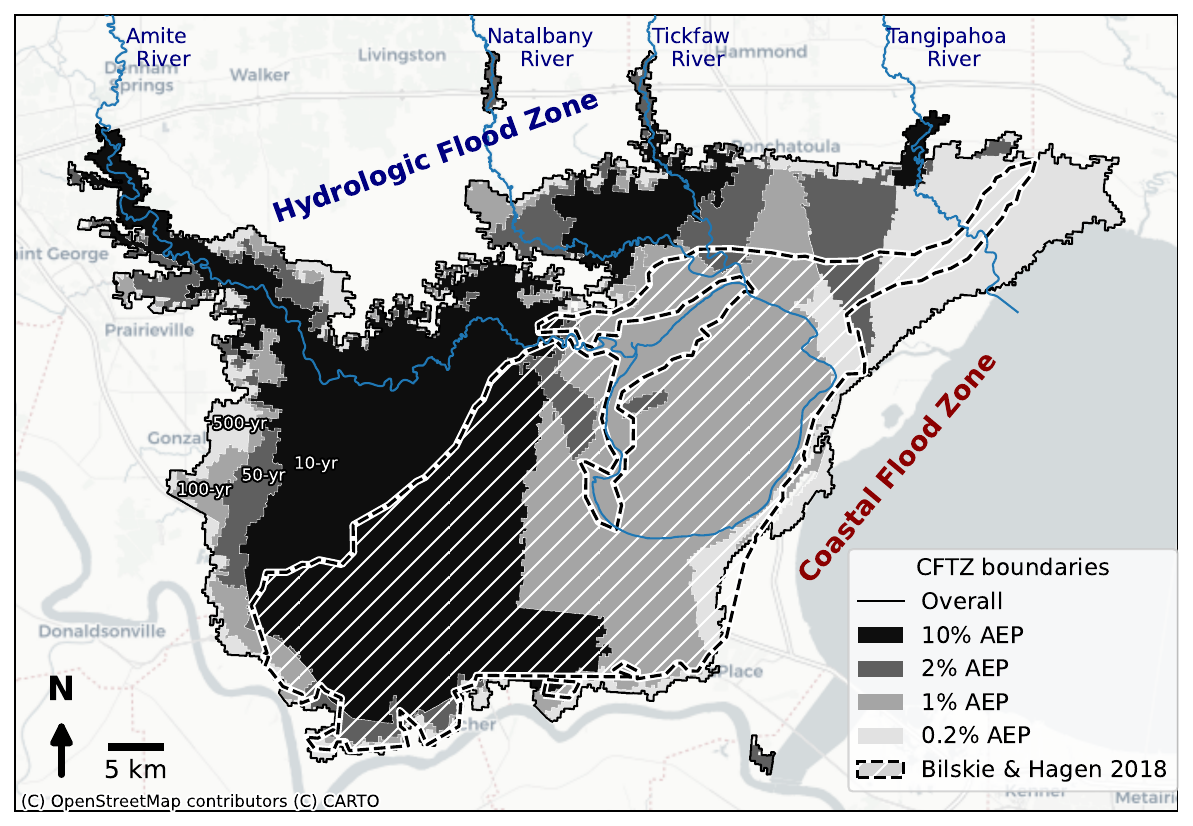}
  \caption{Comparison of the CFTZ from this study, shown by AEP, with high-frequency events (e.g., 10\% AEP) progressively stacked on top of low-frequency events (e.g., 1\% AEP), against the CFTZ definition of \citeA{bilskie2018defining}. The CFTZ covers the area between a coastal flood zone and a hydrologic flood zone. The event-based CFTZ of \citeA{bilskie2018defining} is approximately 938 km$^2$, while the overall statistical-based CFTZ of this study is 2038 km$^2$}
  \label{fig:cftz_overall}
\end{figure}

Figure \ref{fig:cftz_overall} compares the statistical CFTZ proposed here with the event-based CFTZ of \citeA{bilskie2018defining}. The event-based approach identifies locations where a particular compound event produces greater flooding than the corresponding single-driver events, whereas the statistical formulation of Eq. (\ref{eq:CFTA_stat}) identifies locations where the frequency of large flood depths increases because multiple hydrologic and coastal flood pathways contribute to the same flood response across many possible events.

Under the statistical definition developed here, the CFTZ is dynamic, extending farther downstream and broadening coastward as the AEP decreases (i.e., as event magnitude increases). In contrast, the CFTZ of \citeA{bilskie2018defining} was delineated from a single synthetic event consisting of the August 2016 rainfall combined with a Category 4 Hurricane Gustav storm surge. That event produced approximately 0.2\% AEP flows in the Amite, Natalbany, and Tangipahoa Rivers and a 1\% AEP flow in the Tchefuncte River. Consequently, its CFTZ overlaps primarily with the 1\% and 0.2\% AEP statistical CFTZs identified here, but encompasses only 938 km$^2$ compared with the overall statistical CFTZ area of 2038 km$^2$.

\begin{figure}[h!]
  \centering
  \includegraphics[width=1.0\linewidth]{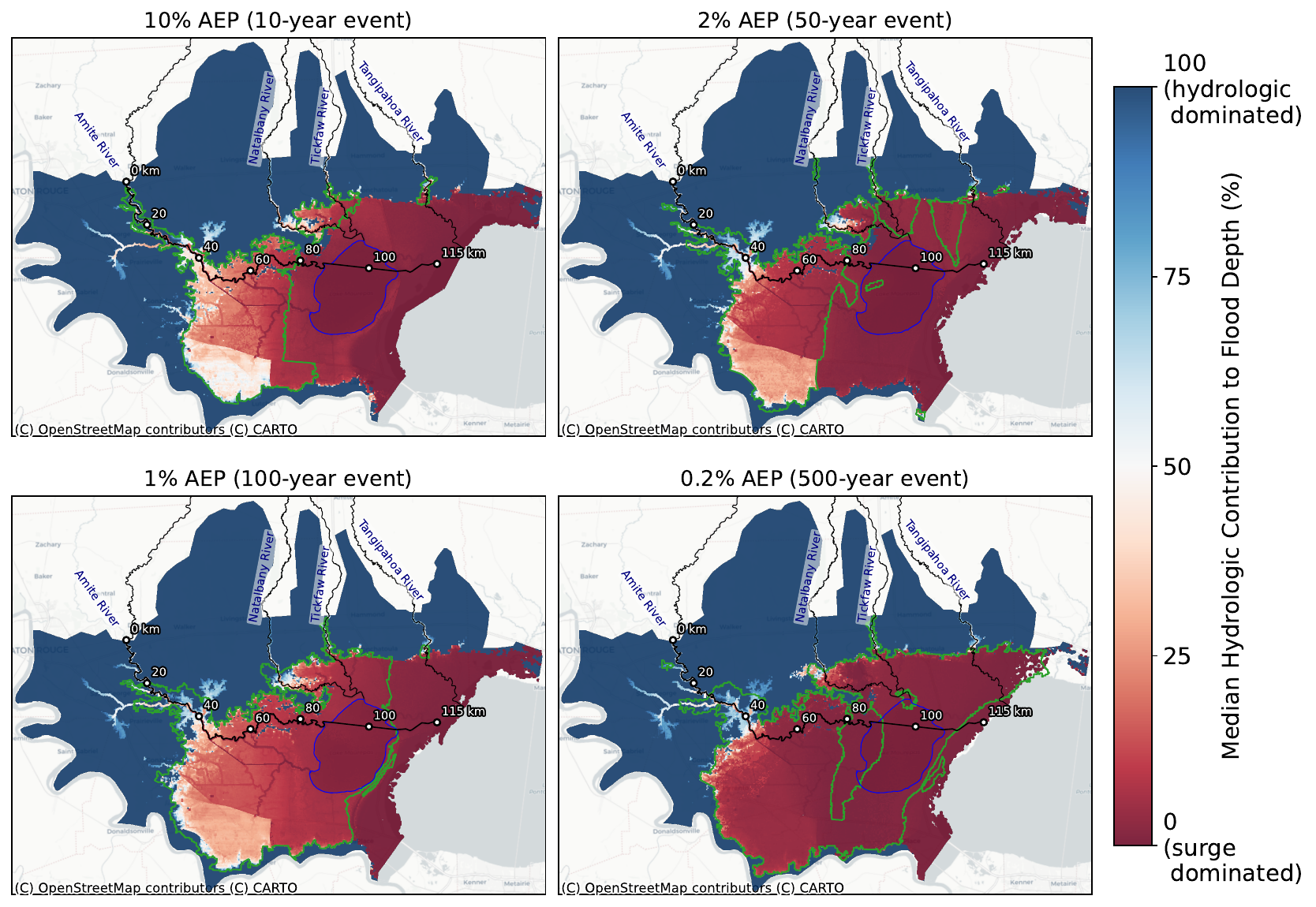}
  \caption{The median hydrologic contribution to flood depth for the 10\%, 2\%, 1\% , and 0.2\% AEP flood events. The green contour denotes the CFTZ, illustrating how the transition between coastal and hydrologic dominance evolves with AEP.}
  \label{fig:H_contrib}
\end{figure}

\begin{figure}[h!]
  \centering
  \includegraphics[width=1.0\linewidth]{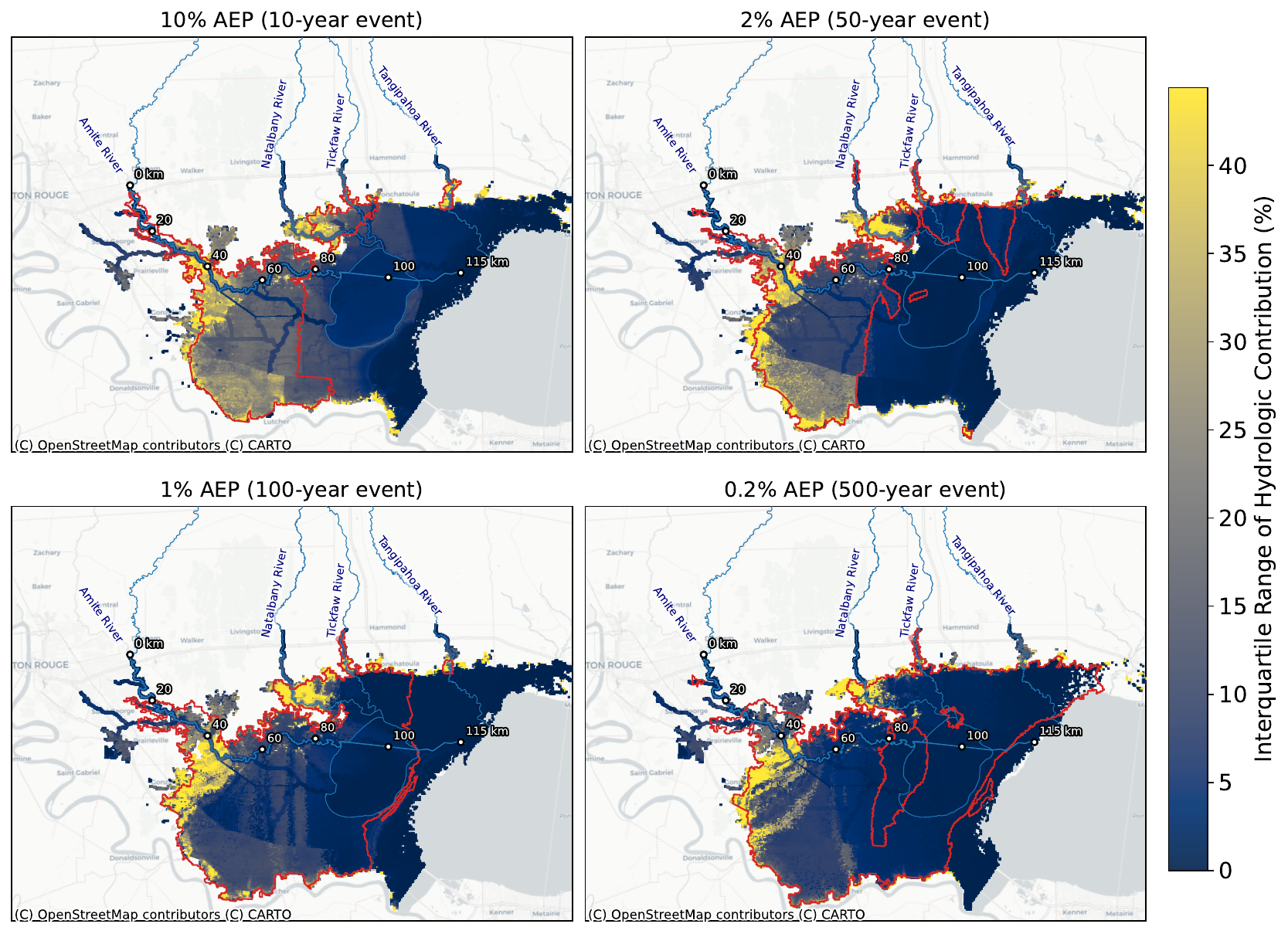}
  \caption{The interquartile range of the probabilistic hydrologic contribution computed from all storm realizations associated with the 10\%, 2\%, 1\% , and 0.2\% AEP flood events. Larger values indicate greater variability in the estimated hydrologic contribution across storm realizations producing similar flood depths. The red contour denotes the CFTZ.}
  \label{fig:H_IQR}
\end{figure}

\subsection{Attribution of Flood Drivers}

To characterize the attribution of flood drivers, we applied the probabilistic framework of Table \ref{tab:attribution_steps}, using a flood-depth tolerance of $\Delta \eta_{\max}=\pm1.25$ cm to identify storm realizations associated with each target compound flood depth. Figure \ref{fig:H_contrib} presents the resulting median hydrologic contribution for the 10\%, 2\%, 1\%, and 0.2\% AEP flood events. Hydrologic dominance decreases toward the coast, while storm-surge dominance increases, producing a continuous transition between the two flood-generation mechanisms. The CFTZ consistently occupies this transition region, where both processes contribute appreciably to the resulting flood depth. As event magnitude increases (i.e., from the 10\% to the 0.2\% AEP event), the median hydrologic contribution within the CFTZ decreases, reflecting the increasing influence of storm surge.

Figure \ref{fig:H_IQR} presents the corresponding interquartile range of the hydrologic contribution. Outside the CFTZ, attribution is relatively deterministic because flooding is overwhelmingly controlled by either hydrologic processes or storm surge. Within the CFTZ, however, the interquartile range increases substantially, indicating that similar flood depths can arise from many different combinations of hydrologic and coastal forcing. Interestingly, the largest increases in AEP flood depth (Figs. \ref{fig:stat_cftz} and \ref{fig:cftz_percent_increase}) generally coincide with regions exhibiting both large median hydrologic contributions and large interquartile ranges. This indicates that the greatest amplification of compound flooding occurs where hydrologic processes remain an important contributor and multiple attribution pathways exist.

\begin{figure}[h!]
  \centering
  \includegraphics[width=1.0\linewidth]{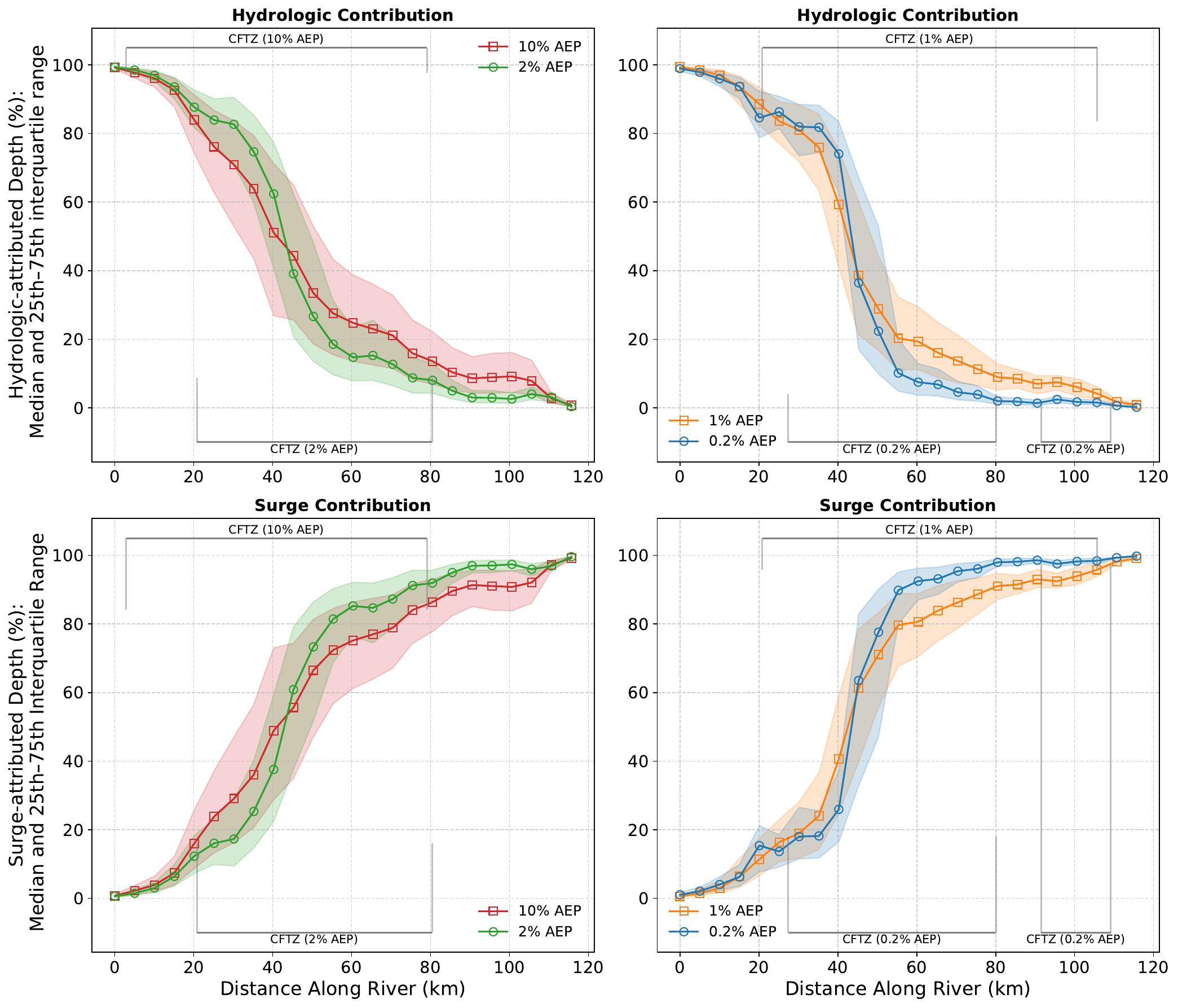} 
  \caption{Along the Amite River transect (see Figs. \ref{fig:H_contrib} and \ref{fig:H_IQR}), the percentage of compound flood depth attributed to hydrologic processes (rainfall over antecedent moisture conditions), as well as the complimentary surge contribution across different AEPs. For reference, the CFTZ is identified for each AEP.}
  \label{fig::hydro-contrib}
\end{figure}

These spatial patterns are further illustrated by comparing the increase in the compound flood depths along the Amite River transect (Fig. \ref{fig:amite_transect}) with the attribution characterization shown on Figs. \ref{fig:H_contrib} and \ref{fig:H_IQR}. Along the transect, the hydrologic contribution decreases from nearly 100\% in the upstream watershed to nearly 0\% near the coast, with a complementary increase in storm-surge attribution (Fig. \ref{fig::hydro-contrib}). For the 10\% AEP event, even a modest reduction in hydrologic contribution is sufficient to initiate the CFTZ, whereas for the larger events the upstream CFTZ boundary typically occurs when the median hydrologic contribution remains around 80--90\%. As event magnitude increases, the downstream CFTZ boundary extends toward progressively smaller hydrologic contributions, reflecting the increasing inland influence of storm surge. A notable exception occurs for the 0.2\% AEP event, where a break in the CFTZ coincides with increased hydrologic contribution near Lake Maurepas (approximately 90 km along the transect).

\begin{figure}[h!]
  \centering
  \includegraphics[width=1.0\linewidth]{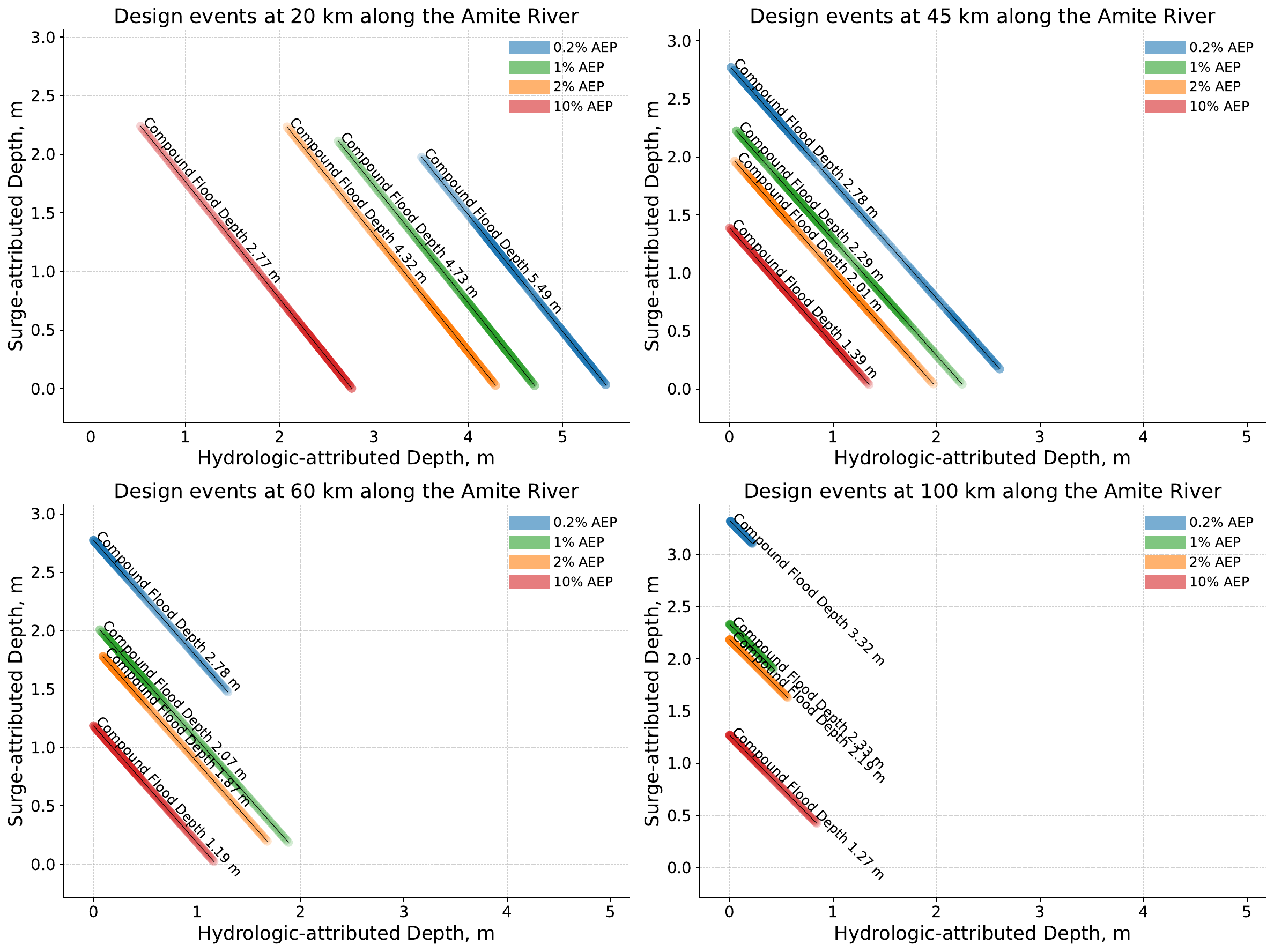}
  \caption{Along different distances of the Amite River transect (Figure \ref{fig:amite_transect}) and for various AEPs, the range of flood depth attributions from modeled storm events with more likely attributions shown by a darker shading of the line. A compound flood depth (±1.25 cm) can result from different event types with varying contributions from storm surge and hydrologic processes. For a given AEP, the relative likelihood of these event types is shown in Figure \ref{fig::hydro-contrib}, with hydrologically driven floods more common upstream and surge-driven floods more common downstream along the Amite River. All lines have a 1:1 slope, and the flood depth is the increase in the WSE beyond the 1-year event.}
  \label{fig:design_storms}
\end{figure}

The probabilistic attribution is examined in greater detail at selected locations along the Amite River transect in Fig. \ref{fig:design_storms}. Each line represents the range of hydrologic and storm-surge contributions capable of producing a given AEP flood depth, with longer lines indicating a broader range of attribution pathways and shorter lines indicating a more constrained set of flood-generating mechanisms. The shading denotes the relative likelihood of each attribution pathway, with darker segments indicating more probable combinations. Near the center of the CFTZ (approximately 60 km along the transect), the 2\% and 1\% AEP events exhibit a relatively broad range of attribution pathways, whereas at approximately 100 km near Lake Maurepas the 0.2\% AEP event is associated with a much narrower range of hydrologic contributions. These attribution distributions demonstrate that a given AEP flood depth may arise from multiple combinations of hydrologic and coastal forcing, providing the probabilistic basis for identifying representative response-based design storms discussed in the following section.

\subsection{Response-based Design Storms}

Continuing the probabilistic framework of Table \ref{tab:attribution_steps}, the same flood-depth tolerance of $\Delta \eta_{\max}=\pm1.25$ cm is used to identify storm realizations associated with a target AEP flood depth. Whereas the previous section used these events to characterize the probabilistic attribution of flood drivers (Steps 1--4), the final steps of the framework (Steps 5 and 6) partitions the identified storm realizations into equiprobable sets of representative response-based design storms.

\begin{figure}[h!]
  \centering
  \includegraphics[width=1.0\linewidth]{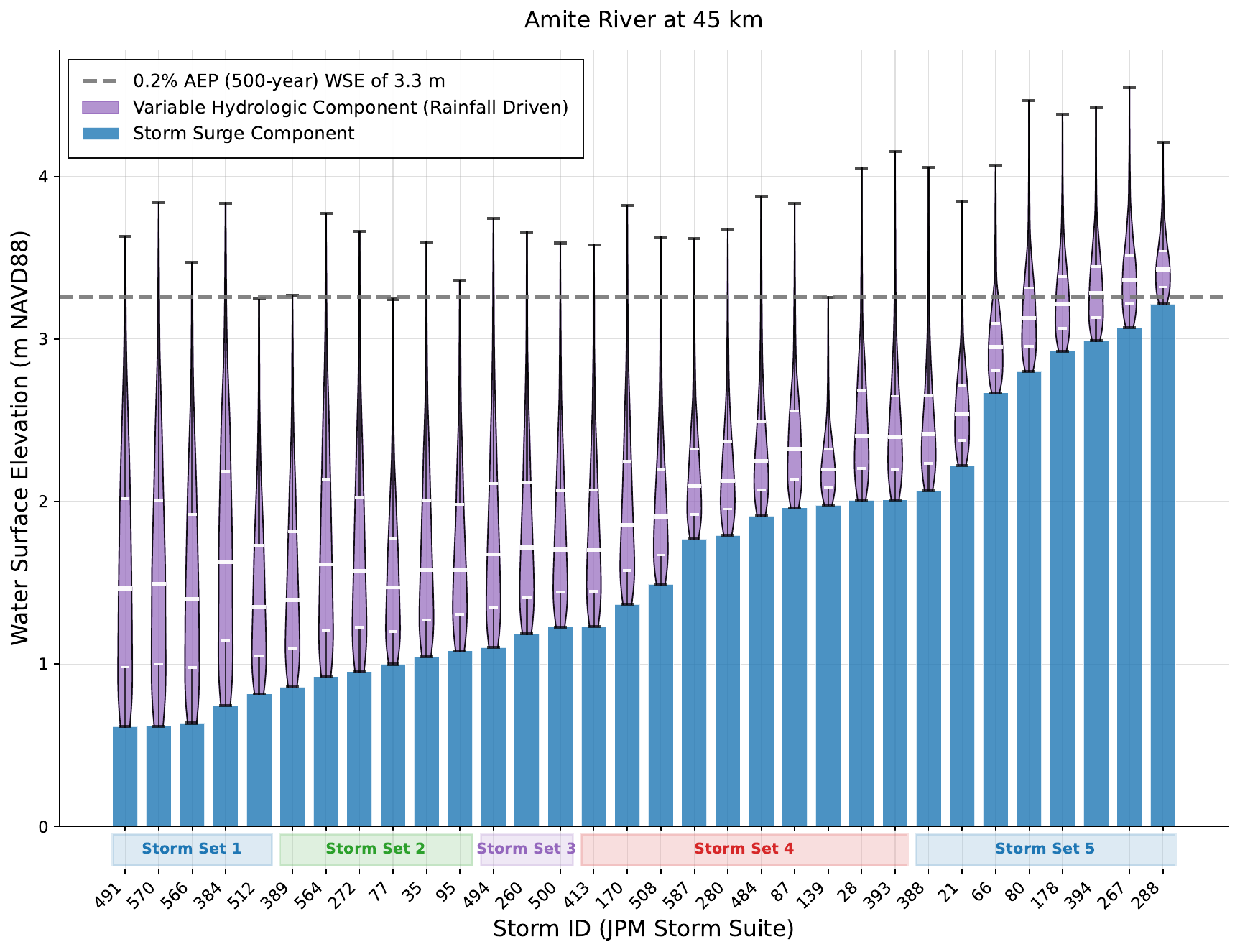}
  \caption{At 45 km along the Amite River transect (Fig. \ref{fig:amite_transect}), a subset of the 645 modeled JPM storms with a water surface elevation (WSE) equal to or exceeding the 0.2\% AEP (500-year) event WSE within a tolerance of $\Delta\eta_{\max}=\pm$1.25 cm. Each storm consists of a constant surge-driven WSE (when rainfall is absent) plus a variable rainfall-driven WSE. The rainfall component is represented by violin plots summarizing 500 realizations of rainfall fields under different antecedent moisture conditions. White bars within the violins indicate the median and interquartile range (25th–75th percentiles). These storms are grouped into equiprobable sets of design events, from which an engineer may select a representative event (five are shown here) tailored to the design case.}
  \label{fig:design_storm_500_yr}
\end{figure}

An example, we identify the response-based design storms for the the 0.2\% AEP flood event at the 45 km mark of the Amite River transect (Fig. \ref{fig:design_storm_500_yr}). At this location, 32 discrete JPM storms produce water surface elevations within the specified tolerance of the target flood depth. These storms are partitioned into five equiprobable design storm sets, each representing a 20\% probability of occurrence. The first set is characterized by predominantly hydrologic forcing with minimal storm surge, whereas the fourth set reflects approximately equal hydrologic and coastal contributions, and the fifth set is primarily surge dominated.

Although the overall attribution varies across the five design storm sets, each individual event remains fully defined by a unique JPM storm identifier (0--645), rainfall realization (1--100), and antecedent hydrologic condition (1--5). Rather than selecting a single deterministic design event, the proposed framework therefore identifies multiple representative storms that produce the same design flood with known probabilities. These response-based design storms provide practitioners with probabilistically justified alternatives for evaluating infrastructure performance, allowing mitigation measures to be assessed against the range of physically plausible flood-generating mechanisms associated with a specified design flood.

\subsection{CFTZ and Design Storm Sensitivity}

The probabilistic characterization of compound flooding depends on two user-defined parameters: the threshold, $\epsilon$, used to delineate the CFTZ, and the flood-depth tolerance, $\Delta \eta_{\max}$, used to identify storm realizations associated with a target flood depth. The threshold $\epsilon$ filters locations where compound flooding produces only negligible differences relative to traditional single-driver flood analyses, whereas the tolerance $\Delta \eta_{\max}$ governs the number of storm realizations available for probabilistic flood-depth attribution and response-based design storm selection.

\begin{figure}[h!]
  \centering
  \includegraphics[width=1.0\linewidth]{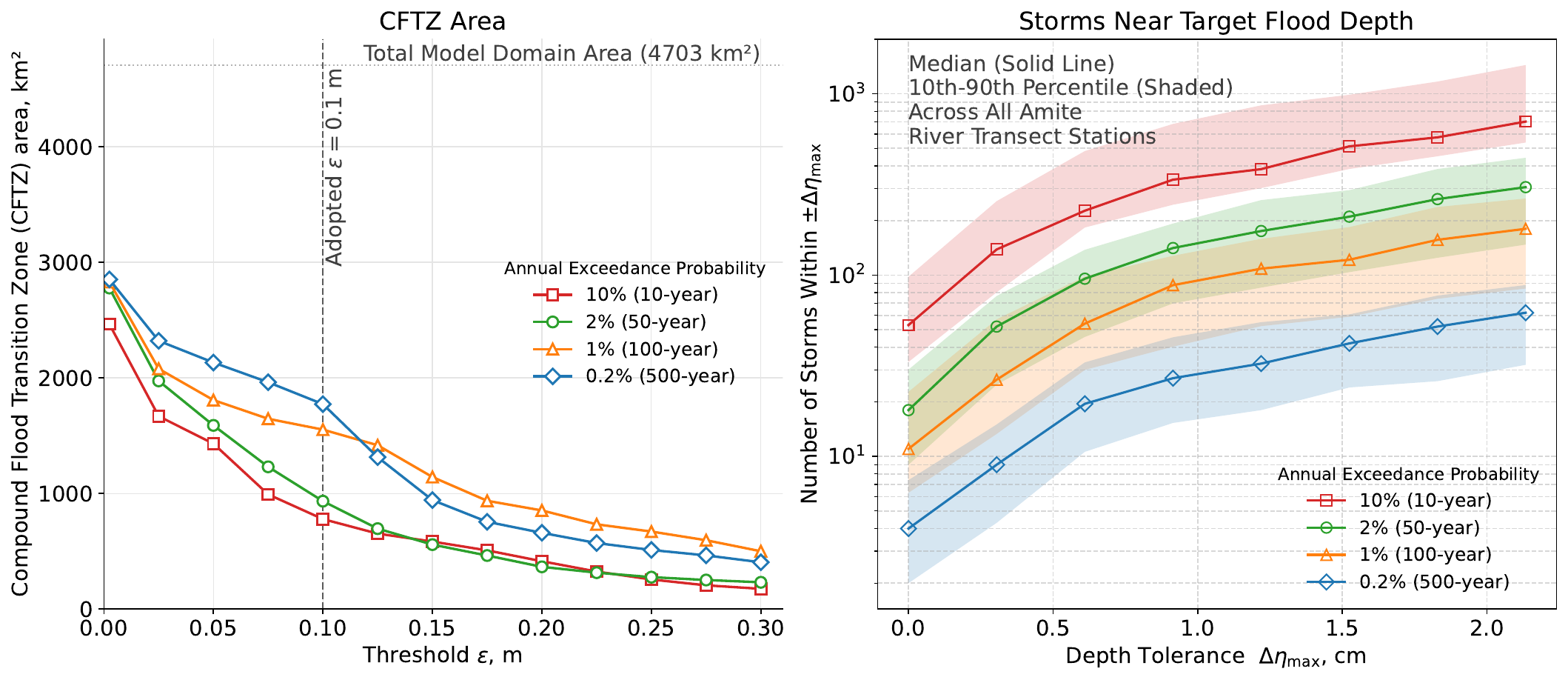}
  \caption{Sensitivity of the probabilistic flood-response characterization to the selected thresholds. Left panel: Area classified as the CFTZ as a function of the flood-depth threshold, $\epsilon$, for 10\%, 2\%, 1\% , and 0.2\% AEP flood events. Right panel: Number of storm events with a simulated flood depth within a tolerance of $\pm \Delta\eta_{\max}$ about the target flood depth used for probabilistic flood-depth attribution. }
  \label{fig:thresholds}
\end{figure}

Figure \ref{fig:thresholds} illustrates the sensitivity of the framework to these parameters. As expected, the CFTZ area is highly sensitive to the selected threshold $\epsilon$. When $\epsilon \rightarrow 0$, nearly any positive increase in flood depth due to compounding is classified as part of the CFTZ, resulting in approximately 63\% of the HEC-RAS domain (roughly 3000 km$^2$) being identified as CFTZ for the 0.2\% AEP event. Conversely, increasing the threshold progressively restricts the CFTZ to locations where compound flooding produces larger flood-depth increases. The adopted value of $\epsilon=0.1$ m represents a practical compromise: it excludes negligible increases while preserving the principal transition zone. Larger thresholds begin to eliminate meaningful portions of the CFTZ. For example, when $\epsilon>0.3$ m, the 0.2\% AEP event produces a smaller CFTZ area than the more frequent 1\% AEP event (Fig. \ref{fig:thresholds}), indicating that the threshold has become sufficiently restrictive to mask the underlying probabilistic structure.


The flood-depth tolerance $\Delta \eta_{\max}$ exhibits a different behavior. Increasing $\Delta \eta_{\max}$ increases the number of storm realizations available to estimate the conditional attribution distributions and identify response-based design storms. Excessively small tolerances provide too few realizations for robust statistical estimation, whereas excessively large tolerances admit storms associated with increasingly different flood depths. A tolerance of $\Delta \eta_{\max}=\pm1.25$ cm provided sufficient storm realizations across the Amite River transect while remaining small relative to the simulated flood-depth range. Moreover, this value lies near the point of diminishing returns, beyond which further increases in $\Delta \eta_{\max}$ yield relatively modest gains in the number of storm realizations (Fig. \ref{fig:thresholds}). Overall, both sensitivity analyses exhibit smooth, monotonic behavior without abrupt transitions, indicating that the resulting CFTZ delineation, probabilistic flood-depth attribution, and response-based design storm selection are robust to reasonable choices of the framework parameters.

\section{Discussion }
\label{sec:discussion}

The central contribution of this study is not the addition of rainfall and hydrology to the Joint Probability Method, but the development of a probabilistic framework that propagates the stochastic variability of rainfall fields, antecedent hydrologic conditions, and storm characteristics through the flood response. This probabilistic representation provides the common foundation for statistical delineation of the CFTZ, probabilistic flood-depth attribution, and response-based design storm selection. Without explicitly representing the stochastic variability of rainfall and antecedent hydrology, each JPM storm would be associated with a single deterministic hydrologic realization, fundamentally altering the frequency distribution of compound flood depths, collapsing the range of possible flood attributions, and reducing response-based design-storm scenarios to a less representative set of events with predetermined characteristics. 

\subsection{Flood Depth Frequency Amplification}

Previous studies of compound flooding have primarily emphasized the nonlinear amplification of individual flood events resulting from interactions among storm surge, runoff, and riverine flooding. From this perspective, compound flooding is viewed principally as a process that increases flood depths during a given event. The probabilistic framework developed here provides a complementary interpretation: when viewed across many possible storm realizations, compound flood hazard increases not only the magnitude of individual flood events but also the frequency with which extreme flood depths occur. This interpretation is supported by the attribution results, where the largest increases in compound flood depth coincide with regions exhibiting both large median hydrologic contributions and large interquartile ranges in attribution. The median contribution identifies where hydrologic processes remain important, whereas the interquartile range reflects the diversity of hydrologic-surge attribution pathways capable of producing similar flood depths. Together, these results suggest that compound flood hazard amplification is fundamentally probabilistic: extreme flood depths become more frequent because multiple hydrologic and surge attribution pathways can produce the same flood response.

\subsection{Statistical Versus Event-Based CFTZ Delineation}

Previous studies have generally identified CFTZs on an event-by-event basis by comparing flood responses from compound and isolated flood-driver simulations \cite<e.g.,>{bilskie2021enhancing,gori2020tropical,shen2019flood,han2024compound}. Such approaches identify where nonlinear interactions amplify or attenuate flooding during a particular event. However, understanding long-term compound flood hazard requires identifying where compound processes systematically modify the frequency distribution of flood-depth exceedance.

The statistical CFTZ introduced here addresses this limitation by identifying locations where compound flooding alters the probability distribution of flood depths relative to traditional single-driver analyses. Unlike an event-based delineation, which represents the response to a single realization, the statistical CFTZ captures regions where multiple hydrologic and coastal forcing combinations contribute to similar flood responses. This broader representation provides complementary information for flood-risk assessment: event-based approaches identify where flood depths are amplified during specific scenarios, whereas the statistical formulation identifies where the likelihood of experiencing those flood depths is increased because multiple flood-generation pathways converge.

\subsection{Response-based Design Storms}

Traditional design storm practices implicitly assumes a correspondence between the AEP of the flood drivers and the AEP of the resulting flood response. This assumption generally holds when flooding is controlled by a single dominant mechanism, such as river discharge or coastal surge. Within CFTZs, however, multiple flood drivers amplify the frequency of the flood depths and this correspondence no longer holds. Identical AEP flood depths may arise from substantially different combinations of rainfall, antecedent hydrologic conditions, runoff, and storm surge.

Most multivariate approaches address this problem within the driver space by defining joint AEPs for rainfall, discharge, storm surge, or other forcing variables through dependence models such as copulas. These methods have substantially advanced compound flood frequency analysis by identifying statistically consistent combinations of flood drivers \cite{salvadori2004frequency,salvadori2011return,salvadori2013multivariate}. Nevertheless, the selected design events remain conditioned on the probability of the forcing variables rather than the probability of the flood response itself. Consequently, the AEP (or return period) of the selected drivers does not necessarily correspond to the AEP (or return period) of the resulting flood depth \cite<e.g.,>{graler2013multivariate,bender2016multivariate, pena2022}.

The extended JPM developed here addresses this limitation by conditioning directly on the flood response. The Dirac delta formulation embeds the deterministic flood-response model within the joint probability distribution of the forcing variables, allowing both the probability distribution of flood depth and the conditional distribution of the flood drivers to be obtained from the same probabilistic framework. Rather than beginning with driver AEPs and predicting an uncertain flood response, the analysis begins with the flood depth response of engineering (or planning) interest and identifies the range of storm realizations capable of producing that response.

The resulting design philosophy differs fundamentally from traditional approaches. Instead of identifying a single deterministic design storm, the framework identifies multiple response-based design storms together with their probabilities of occurrence. Engineers may therefore evaluate infrastructure performance against representative storm realizations spanning hydrologically dominated, surge-dominated, and mixed flood-generation mechanisms while retaining a direct probabilistic interpretation. More broadly, this formulation suggests that design storms should be viewed not as unique realizations but as probability distributions over physically plausible storm events conditioned on the target flood response.

\subsection{Connections with Stochastic Ecohydrology}

The framework is extended with the generic probabilistic description of hydrology given by Eq. (\ref{eq:p_swsq}), which is intentionally independent of any particular hydrologic model but reflects the state descriptions commonly used in stochastic ecohydrology. Stochastic ecohydrology is concerned with directly deriving the probability distributions of hydrologic processes. Although the present implementation estimated the long-term soil-moisture and baseflow distributions from explicit simulation of a finite ensemble of historical storm events, equivalent probability distributions arise naturally from stochastic ecohydrologic formulations that evolve the governing hydrologic statistics through time.

Recent developments in stochastic ecohydrology directly provide the probabilistic description of Eq. (\ref{eq:p_swsq}) \cite{bartlett2025stochasticE}. Rather than inferring antecedent conditions from an ensemble of simulations, these formulations derive analytical probability distributions by evolving the governing hydrologic statistics over the full historical sequence of climatic forcing. The resulting theory provides analytical expressions for soil moisture, $p(\mathbf{s}|\overline{\mathbf{s}})$, and storage, $p(\mathbf{w}|\overline{\mathbf{w}})$, linking basin-scale watershed states to point-scale hydrologic variability. These analytical distributions could replace the empirically derived PDFs adopted in the present case study, while also relaxing simplifying assumptions such as prescribing a fixed average baseflow. 

The probabilistic structure of Eq. (\ref{eq:p_swsq}) also naturally accommodates richer descriptions of watershed processes. For example, the single-layer soil-moisture distribution, $p_{(\cdot)}(\mathbf{\overline{s}})$, may be generalized to a hierarchical two-layer formulation, $p_{(\cdot)}(\mathbf{\overline{s}_1}|\mathbf{\overline{s}_0})$, allowing the upper layer to represent surface-water storage in wetlands or marshes while the lower layer represents subsurface soil moisture \cite<e.g.,>{bartlett2025stochasticE}. Likewise, the analytical state descriptions may be mapped to distributed wetness-index fields for direct initialization of distributed hydraulic models \cite<e.g.,>{bartlett2025physically,beven1979physically,beven2012rainfall}. More broadly, the framework developed here provides a probabilistic bridge between the long-term statistical descriptions of stochastic ecohydrology and the event-based probabilistic descriptions of the Joint Probability Method.

\subsection{Extensions of the Framework}

Extensions of the framework may include more computationally efficient flood-response models, richer probabilistic descriptions of the hydrologic and rainfall forcing, and improved methods for defining response-based design storm scenarios when lower-frequency AEP events result in a sparse set of realizations. 
Building on the stochastic ecohydrologic extensions discussed in the previous section \cite{bartlett2025stochasticE}, richer probabilistic representations of rainfall and other forcing components may likewise be incorporated without altering the underlying formulation. In the present implementation, the principal limitation of the tropical cyclone rainfall representation lies not in the expected rainfall model, $\delta(\mathbf{\overline{r}}(t)|\mathbf{x}_{JPM})$, for which \citeA{villarini2022probabilistic} identified the bias-corrected IPET formulation as the best-performing parametric rainfall model evaluated, but in the stochastic rainfall representation, $p(\mathbf{r}(t)|\mathbf{\overline{r}}(t))$. The temporal evolution of rainfall remained governed by the expected IPET rainfall profile, $h[\mathbf{\overline{r}}(t)]$, so the stochastic rainfall model does not introduce additional subdaily variability beyond the smoothed expected profile. Consequently, short-duration rainfall extremes, and the resulting streamflow response, are likely muted. This limitation applies only to the tropical cyclone rainfall implementation; the non-tropical simulations employed observed rainfall directly and therefore retained the observed temporal variability. Future stochastic rainfall generators that preserve realistic subdaily variability could therefore increase the simulated tropical contribution to compound flooding while remaining fully consistent with the probabilistic framework developed here.

The principal computational expense of the present implementation lies in evaluating the flood response through large ensembles of numerical simulations. While previous work has focused primarily on reducing this cost through more efficient sampling of the joint driver distribution \cite<e.g.,>{toro2010efficient}, the present formulation highlights a complementary direction. The Dirac delta function explicitly defines the probabilistic mapping from storm characteristics to flood depth \cite{au1999transforming}. When the flood response is available only through numerical simulation, this probabilistic mapping is approximated by propagating a discretized representation of the JPM distribution through the numerical flood-response model. However, this mapping could instead be represented through physically based scaling relationships or reduced-order response models \cite<e.g.,>{bartlett2025physically}. The resulting joint probability distribution of the drivers and flood response may then be evaluated directly through numerical integration and, for sufficiently simple response models, even analytically. This would greatly accelerate CFTZ delineation, flood-depth attribution, and response-based design storm selection.

The same analytical formulation also suggests improvements to response-based design storm selection. In the present implementation, design storms are identified from a finite ensemble of simulated events. Future implementations could instead estimate continuous probability distributions for the conditional storm characteristics associated with a target flood depth, thereby removing the dependence on a finite ensemble of simulated storms while retaining the probabilistic interpretation of response-based design events.  More generally, because the framework explicitly separates the probabilistic forcing and boundary conditions from the flood-response model, advances in either component can be incorporated without altering the underlying probabilistic structure, providing a flexible and extensible foundation for future developments in compound flood hazard analysis.

\section{Concluding Remarks}

This study presents a probabilistic response-based framework that extends the Joint Probability Method (JPM) to compound flooding by explicitly representing the stochastic variability of rainfall fields, antecedent hydrologic conditions, and coastal storm characteristics within the JPM probabilistic formulation. Rather than augmenting the traditional JPM with additional flood drivers, the framework propagates the combined uncertainty in hydrologic and coastal forcing through the flood response, enabling a probabilistic characterization of compound flooding that includes (1) statistical delineation of the CFTZ, (2) probabilistic attribution of flood depths to hydrologic and coastal processes, and (3) response-based design storm selection for specified annual exceedance probabilities. Together, these capabilities shift compound flood analysis from the evaluation of individual events toward characterization of the long-term probabilistic structure governing flood hazards.

Application to the Lake Maurepas basin demonstrates that the greatest increases in AEP flood depths occur where substantial hydrologic contributions coincide with substantial variability in the possible attribution pathways. This finding suggests that, over long time scales, compound flood amplification is driven not solely by nonlinear interactions during individual events, but by the increased frequency with which multiple hydrologic and coastal pathways converge to produce the same flood depth. Consequently, the statistical CFTZ evolves with annual exceedance probability and encompasses more than twice the area identified by a representative event-based delineation.

More broadly, the framework preserves the probabilistic structure of the original JPM while naturally incorporating additional stochastic processes, including the long-term hydrologic variability described by stochastic ecohydrology. Rather than representing a final formulation, it provides a parsimonious probabilistic foundation upon which progressively more sophisticated descriptions of compound flooding can be systematically developed. This foundation also provides a natural pathway for improving the probabilistic characterization of compound flood transition zones, flood-depth attribution, and response-based design storm selection as more sophisticated process representations become available.




\acknowledgments
We acknowledge the support and guidance of Louisiana state agency personnel from the Office of Community Development (OCD), Louisiana Coastal Protection and Restoration Authority (CPRA), the Governor’s Office of Homeland Security and Emergency Preparedness (GOHSEP), and the Department of Transportation and Development (DOTD). This work was funded through a cooperative endeavor agreement between The Water Institute and Coastal Protection and Restoration Authority as part of the Louisiana Watershed Initiative (LWI) funded via the U.S. Department of Housing and Urban Development (HUD) Community Development Block Grant (CDBG) mitigation funds (Catalog of Federal Domestic Assistance (CFDA) 14.228 Grant B-18-DP-22-0001).   

\emph{Conflict of Interest Statement.} The authors declare no conflicts of interest.

\emph{Data Availability Statement.} The data used in this work is available on the Louisiana Watershed Initiative (LWI) Model data Amazon Web Services (AWS) s3 bucket \\
(https://registry.opendata.aws/lwi-model-data/). Specifically, the HEC-HMS and HEC-RAS models and associated forcing data are accessed at \\
https://lwi-model-data.s3.amazonaws.com/index.html\#endmc-managed/model\_application under the respective folders of 63cec0e23b68dc1a2aea077a and 672e862c47055c143d632a62, while the resulting datasets used in the analysis are accessed at\\ https://lwi-model-data.s3.amazonaws.com/index.html\#endmc-managed/dataset under folder 6932095f4c76280a3136c3fe \cite{lwi_endmc_2025}. All entries also are easily found by creating a free account at https://lwi.endmc.org/ and searching for 'Compound Flooding`, which retrieves all the model and data files. The Python scripts used to analyze the data and create the figures are available on Zenodo (zenodo.org)  \cite{Bartlett2025extended-jpm}.

\appendix 
\section{Integrated Modeling Framework}
\label{sec:Integrated_modeling}


\begin{figure}
    \centering
    \includegraphics[width=6in]{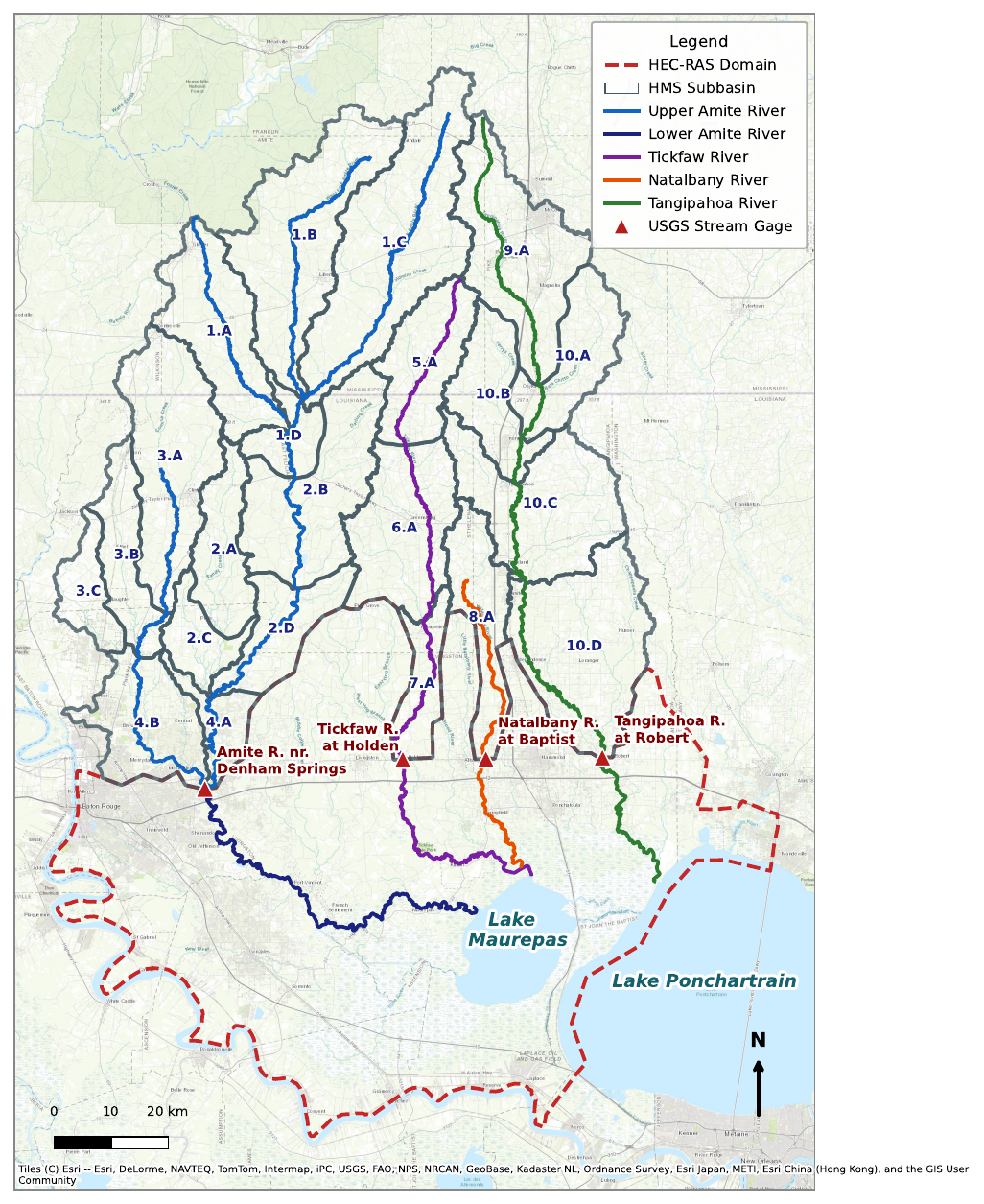}
    \caption{Locations of 22 HEC-HMS subwatersheds and the USGS gages used for calibration of the HEC-HMS models.}
    \label{fig:gagelocations}
\end{figure}

\subsection{HEC-HMS Setup}
\label{sec:HEC-HMS}

HEC-HMS \cite{bartles2022hydrologic} was used to simulate rainfall-runoff processes within the Amite, Tickfaw, Natalbany, and Tangipahoa River basins. A lumped representation consisting of 22 subwatersheds (Fig. \ref{fig:gagelocations}; Table \ref{tab:subbasins}) was adopted to efficiently simulate the large number of historical and synthetic storms required for the extended JPM analysis. Subwatershed boundaries were delineated based on topographic characteristics upstream of the USGS stream gages used for calibration of the model (Fig. \ref{fig:gagelocations}).

Runoff generation was represented using the USACE Deficit and Constant Loss method, consistent with guidance from the USACE Modeling, Mapping, and Consequence Center for Corps Water Management System model development \cite{bartles2022hydrologic}. Excess rainfall was transformed to runoff using the Clark Unit Hydrograph \cite{clark1945storage}, and channel routing was simulated with the Muskingum-Cunge method \cite{ponce1978muskingum,ponce1986diffusion}. Model parameters were initially estimated from watershed characteristics, including hydrologic soil groups for infiltration \cite{bartles2022hydrologic,NRCS2009HSG}, and subsequently refined through calibration against observed streamflow (Section \ref{sec:calibration}). The same constant event baseflow was prescribed for each simulation, as was discussed in Section \ref{sec:StormFreqProb}; Table \ref{tab:assumptions}.

\begin{table}[h!]
\begin{minipage}{\textwidth}
\linespread{.5}\selectfont
\caption{HEC-HMS subbasins of Figure~\ref{fig:gagelocations}, with contributing drainage basin,
drainage area, and percent impervious cover. \label{tab:subbasins} }
\noindent
\setlength{\tabcolsep}{2.5pt}
\begin{tabular}{c|l|l|c|c}
\hline
\noalign{\vskip 0.04in}
Subbasin ID & Subbasin Name & Drainage Basin & Drainage Area (mi$^2$) & Impervious (\%) \\
\hline
\noalign{\vskip 0.05in}
1.A & Beaver Creek & Amite River & 124.5 & 0.5 \\
1.B & West Fork & Amite River & 189.7 & 0.5 \\
1.C & East Fork & Amite River & 236.8 & 0.4 \\
1.D & Upper Amite & Amite River & 39.7 & 0.3 \\
2.A & Mills Creek & Amite River & 72.2 & 0.3 \\
2.B & Darlington Creek & Amite River & 154.3 & 0.4 \\
2.C & Sandy Creek & Amite River & 43.4 & 0.5 \\
2.D & Amite Baywood & Amite River & 73.9 & 1.0 \\
3.A & Upper Comite & Comite/Amite River & 161.0 & 0.5 \\
3.B & Redwood Creek & Comite/Amite River & 70.8 & 0.5 \\
3.C & White Bayou & Comite/Amite River & 44.0 & 1.0 \\
4.A & Amite Denham Springs & Amite River & 38.5 & 8.3 \\
4.B & Comite Central & Comite/Amite River & 103.6 & 10.5 \\
5.A & Upper Tickfaw & Tickfaw River & 93.9 & 0.5 \\
6.A & Twelve Mile Creek & Tickfaw River & 132.9 & 0.6 \\
7.A & Tickfaw Holden & Tickfaw River & 50.0 & 0.6 \\
8.A & Upper Natalbany & Natalbany River & 83.0 & 1.8 \\
9.A & Little Tangipahoa & Tangipahoa River & 158.5 & 2.8 \\
10.A & Bala Chitto Creek & Tangipahoa River & 82.0 & 0.9 \\
10.B & Terrys Creek & Tangipahoa River & 78.5 & 1.1 \\
10.C & Big Creek & Tangipahoa River & 154.5 & 0.7 \\\
10.D & Chappepeela & Tangipahoa River & 183.6 & 0.8 \\
\noalign{\vskip 0.04in}
\hline
\end{tabular}
\\
\end{minipage}
\end{table}

\begin{table}[h!]
\begin{minipage}{\textwidth}
\linespread{.5}\selectfont
\centering
\caption{Land use / land cover (LULC) classes from the 2016 National Land Cover Database (NLCD) and the corresponding Manning's roughness coefficients ($n$) assigned in the HEC-RAS 2D model prior to calibration. \label{tab:manning}}
\begin{tabular}{@{}lc@{}}
\toprule
\textbf{Land cover class (NLCD 2016)} & \textbf{Manning's $n$} \\
\midrule
Open Water                   & 0.025\\
Perennial Snow/Ice           & 0.022 \\
Developed, Open Space        & 0.040 \\
Developed, Low Intensity     & 0.100 \\
Developed, Medium Intensity  & 0.080 \\
Developed, High Intensity    & 0.150 \\
Barren Land                  & 0.028 \\
Deciduous Forest             & 0.160 \\
Evergreen Forest             & 0.180 \\
Mixed Forest                 & 0.170 \\
Shrub/Scrub                  & 0.100 \\
Herbaceous                   & 0.035 \\
Hay/Pasture                  & 0.033 \\
Cultivated Crops             & 0.038 \\
Woody Wetlands               & 0.120 \\
Emergent Herbaceous Wetlands & 0.070 \\
\bottomrule
\end{tabular}
\end{minipage}
\end{table}

\subsection{HEC-RAS Setup}
\label{sec:HEC-RAS}

Two-dimensional unsteady hydraulic simulations were performed in HEC-RAS v6.1 \cite{brunner2021hecras} over the coastal portion of the study domain. The model terrain was developed from the 3-m USGS Coastal National Elevation Database \cite{danielson2018coastal}, which merges nearshore topography and bathymetry. Channel bathymetry for the Amite River and Diversion Canal was reconstructed from the one-dimensional cross sections of the Amite River Basin Numeric Model \cite{dewberry2019amite}), while bed elevations in unsurveyed reaches were estimated from hydraulic geometry relationships \cite{leopold1953hydraulic,wilkerson2011physical}. The 3-m USGS terrain was further edited to remove LiDAR artifacts at bridge crossings and to incorporate levees and other hydraulic controls from the Coastal Master Plan ADCIRC mesh before being rendered at a 10 ft. × 10 ft. resolution \cite{cobell2023coastal}. Spatially distributed Manning's roughness coefficients were assigned from the 2016 National Land Cover Database following Chow et al. (1988), with values ranging from 0.025 for open water to 0.18 for forested land (Table \ref{tab:manning}).

The hydraulic model employed an unstructured two-dimensional mesh consisting of 80,413 computational cells, with a nominal cell size of 800 ft refined to 500 ft along major channels. Flow was simulated using the full Shallow Water Equations with the Eulerian–Lagrangian Method (SWE-ELM), which was selected because it provides stable and computationally efficient solutions for tidally influenced compound flooding while producing negligible differences in maximum water-surface elevations relative to the Eulerian formulation \cite{brunner2021hecras}. Simulations used a constant computational time step of 2 minutes. The resulting Courant numbers satisfied the recommended criterion of $C\leq1$ (maximum allowable $C=3$). The hydraulic model was forced by HEC-HMS inflow hydrographs and ADCIRC+SWAN storm-surge water levels at the boundaries described in Section \ref{sec:flood_response}. Spatially distributed rainfall was applied directly over the computational mesh, while wind forcing was specified using a Lagrangian reference frame.

\begin{table}[h!]
\begin{minipage}{\textwidth}
\linespread{.5}\selectfont
\caption{USGS gages used for calibration and validation of the integrated HEC-HMS/HEC-RAS model. HMS denotes streamgages used for HEC-HMS calibration, whereas HWM denotes streamgages used to validate peak water-surface elevations in the HEC-RAS model. \emph{Period} is the daily-value discharge record where available; otherwise, the daily-value stage record. \label{tab:gages} }
\noindent
\setlength{\tabcolsep}{4.5pt}
\begin{tabular}{c|p{2.75in}|c|c|c|c}
\hline
\noalign{\vskip 0.04in}
Site No. & Station Name & Lat. & Lon. & Use & Period \\
\hline
\noalign{\vskip 0.05in}
07375500 & Tangipahoa River at Robert & 30.5066 & $-90.3618$ & HMS & 1938--2026 \\
07376000 & Tickfaw River at Holden & 30.5038 & $-90.6773$ & HMS & 1940--2026 \\
07376500 & Natalbany River at Baptist & 30.5044 & $-90.5459$ & HMS & 1943--2026 \\
07378500 & Amite River near Denham Springs & 30.4641 & $-90.9904$ & HMS & 1938--2026 \\
07375230 & Tchefuncte River at Madisonville & 30.4041 & $-90.1548$ & HWM & 1958--2026 \\
07375650 & Tangipahoa River near Ponchatoula & 30.4437 & $-90.3346$ & HWM & 1962--2026 \\
07376300 & Tickfaw River near Springfield & 30.3766 & $-90.5511$ & HWM & 1948--2026 \\
07378650 & Jones Cr. at Old Hammond Hwy near Baton Rouge & 30.4407 & $-91.0445$ & HWM & 1962--2026 \\
07378722 & Claycut Bayou at Antioch Rd near Baton Rouge & 30.3869 & $-91.0073$ & HWM & 1995--2026 \\
07378745 & Alligator Bayou near Kleinpeter & 30.3213 & $-91.0207$ & HWM & 1955--2026 \\
07378746 & Bayou Manchac at Alligator B. near Kleinpeter & 30.3214 & $-91.0209$ & HWM & 1997--2026 \\
07378748 & Bluff Swamp near Kleinpeter & 30.3235 & $-91.0182$ & HWM & 1997--2026 \\
07378810 & Bayou Fountain at Bluebonnet Blvd near B.R. & 30.3585 & $-91.1082$ & HWM & 1995--2026 \\
07379000 & Ward Creek at Government St at Baton Rouge & 30.4445 & $-91.1432$ & HWM & 1954--2026 \\
07379050 & Ward Creek at Essen Lane near Baton Rouge & 30.4049 & $-91.1034$ & HWM & 1962--2026 \\
07379075 & Old Ward Crk Div. at Highland Rd near B.R. & 30.3552 & $-91.0151$ & HWM & 1999--2026 \\
07379100 & North Branch Ward Creek at Baton Rouge & 30.4180 & $-91.0915$ & HWM & 1962--2026 \\
07379960 & Dawson Cr. at Bluebonnet Blvd near Baton Rouge & 30.3824 & $-91.0943$ & HWM & 1995--2026 \\
07380101 & Bayou Manchac near Little Prairie & 30.3405 & $-90.9173$ & HWM & 1989--2026 \\
07380102 & Welsh Gully at J. Broussard Rd near Prairieville & 30.3369 & $-90.9690$ & HWM & 1998--2026 \\
07380120 & Amite River at Port Vincent & 30.3327 & $-90.8520$ & HWM & 1946--2026 \\
07380126 & Henderson Bayou near Port Vincent & 30.2974 & $-90.8837$ & HWM & 1980--2026 \\
07380127 & Henderson Bayou Pump Station near Port Vincent & 30.3183 & $-90.8599$ & HWM & 2015--2026 \\
07380200 & Amite River near French Settlement & 30.2755 & $-90.7793$ & HWM & 1950--2026 \\
07380212 & Airline Canal at Blind River near Gramercy & 30.1016 & $-90.7365$ & HWM & 2020--2026 \\
07380215 & Amite River at Hwy 22 near Maurepas & 30.3077 & $-90.6087$ & HWM & 1967--2026 \\
073802220 & Panama Canal at Hwy 44 near Gonzales & 30.1708 & $-90.9187$ & HWM & 1997--2026 \\
073802225 & Bayou Conway near Sorrento & 30.1733 & $-90.8612$ & HWM & 1998--2026 \\
0738022295 & Grand Goudine at Babin Rd near Duplessis & 30.2622 & $-90.9634$ & HWM & 1998--2026 \\
0738022395 & Black Bayou at Hwy 621 near Prairieville & 30.2694 & $-90.9169$ & HWM & 1997--2026 \\
073802245 & Black Bayou E of Gonzales & 30.2374 & $-90.8701$ & HWM & 1997--2026 \\
073802273 & Bayou Francois at Hwy 61 near Gonzales & 30.2274 & $-90.8995$ & HWM & 1997--2026 \\
073802280 & New River north of Sorrento & 30.2152 & $-90.8199$ & HWM & 2018--2026 \\
073802282 & New River Canal near Sorrento & 30.1894 & $-90.7862$ & HWM & 1997--2026 \\
073802284 & New River Canal east of Sorrento & 30.1893 & $-90.7813$ & HWM & 2017--2026 \\
\noalign{\vskip 0.04in}
\hline
\end{tabular}
\\
\end{minipage}
\end{table}

\subsection{Calibration and Validation with Historical Storms}
\label{sec:calibration}

The HEC-HMS and HEC-RAS models were calibrated and validated using historical TC and non-TC storm observations. Calibration relied on a network of USGS streamgages (Table \ref{tab:gages}) and high-water marks distributed throughout the study domain (Fig. \ref{fig:location_rascomp}), providing observations of streamflow, stage, and peak water-surface elevation for calibration and validation of the hydrologic and hydraulic models.

\begin{table}[h!]
\begin{minipage}{\textwidth}
\linespread{.5}\selectfont
\caption{Tropical cyclones used for the HEC-HMS model calibration and the development of antecedent conditions \label{tab:calibration_storms} }
\noindent
\begin{tabular}{c|c|c l }
\hline
\noalign{\vskip 0.04in}
TCs & Start Date & End Date \\
\hline
\noalign{\vskip 0.05in}
Tropical Storm Bertha & 08/03/2002 & 08/24/2002 \\
Hurricanes Isidore and Lili & 09/14/2002 & 10/26/2002 \\
Tropical Storm Bill & 06/27/2003 & 07/18/2003 \\
Tropical Storm Matthew & 10/05/2004 & 10/24/2004 \\
Hurricane Cindy & 07/03/2005 & 07/12/2005 \\ 
Hurricane Katrina & 08/20/2005 & 09/11/2005 \\
Hurricane Rita & 09/18/2005 & 10/10/2005 \\
Hurricanes Gustav and Ike & 08/15/2008 & 09/16/2008 \\
Tropical Storm Bonnie & 07/16/2010 & 07/26/2010 \\
Tropical Storm Lee & 09/01/2011 & 09/16/2011 \\
Hurricane Isaac & 08/15/2012 & 09/15/2012 \\
Tropical Storm Cindy & 06/20/2017 & 07/08/2017 \\
Hurricane Harvey & 08/15/2017 & 09/13/2017 \\
Hurricane Nate & 10/03/2017 & 10/18/2017 \\
Hurricane Barry & 07/11/2019 & 07/19/2019 \\
Tropical Storm Imelda & 09/15/2019 & 10/05/2019 \\
Tropical Storm Olga & 10/23/2019 & 10/31/2019 \\
Tropical Storm Cristobal & 06/01/2020 & 06/15/2020 \\
Hurricane Laura & 08/18/2020 & 09/08/2020 \\
Hurricane Sally and Tropical Storm Beta & 09/11/2020 & 09/30/2020 \\
Hurricane Delta & 10/02/2020 & 10/22/2020 \\
Hurricane Zeta & 10/23/2020 & 11/09/2020 \\
Hurricane Ida & 08/25/2021 & 09/08/2021 \\
\noalign{\vskip 0.04in}
\hline
\end{tabular}
\\
\end{minipage}
\end{table}

\begin{table}
\begin{minipage}{\textwidth}
\linespread{.5}\selectfont
\caption{The 44 observed non-tropical events used to characterize the PDF of the non-tropical storm event attributes. The events between 2002 and 2019 are ranked by Amite River discharge, measured at the USGS gauge at Denham Springs. \label{tab:historic_storms}}
\noindent
\begin{tabular}{c|c|c||c|c|c}
\hline
\noalign{\vskip 0.04in}
Rank & Date & Discharge (m$^3$/s) & Rank & Date & Discharge (m$^3$/s) \\
\hline
\noalign{\vskip 0.05in}
1  & 08/14/2016 & 5,720 & 23 & 01/22/2017 & 682 \\
2  & 03/13/2016 & 1,770 & 24 & 01/28/2012 & 671 \\
3  & 02/24/2003 & 1,420 & 25 & 04/16/2013 & 663 \\
4  & 03/30/2009 & 1,190 & 26 & 04/09/2003 & 660 \\
5  & 10/28/2015 & 1,100 & 27 & 04/05/2017 & 654 \\
6  & 05/12/2019 & 1,100 & 28 & 04/10/2002 & 637 \\
7  & 05/17/2004 & 1,090 & 29 & 01/01/2007 & 623 \\
8  & 01/11/2013 & 1,060 & 30 & 10/24/2017 & 623 \\
9  & 10/29/2006 &   969 & 31 & 06/27/2004 & 617 \\
10 & 02/20/2012 &   954 & 32 & 02/06/2010 & 603 \\
11 & 02/14/2004 &   932 & 33 & 03/23/2012 & 586 \\
12 & 02/14/2013 &   864 & 34 & 01/03/2017 & 575 \\
13 & 03/10/2011 &   830 & 35 & 02/26/2004 & 535 \\
14 & 01/29/2018 &   827 & 36 & 01/06/2019 & 530 \\
15 & 04/20/2019 &   827 & 37 & 02/05/2016 & 518 \\
16 & 02/23/2014 &   799 & 38 & 10/30/2002 & 510 \\
17 & 12/29/2018 &   799 & 39 & 05/16/2008 & 484 \\
18 & 11/20/2015 &   784 & 40 & 03/31/2014 & 436 \\
19 & 12/19/2009 &   759 & 41 & 02/06/2004 & 430 \\
20 & 03/04/2015 &   739 & 42 & 12/31/2012 & 428 \\
21 & 02/02/2005 &   697 & 43 & 01/05/2015 & 428 \\
22 & 11/07/2002 &   682 & 44 & 05/05/2007 & 391 \\
\noalign{\vskip 0.04in}
\hline
\end{tabular}
\end{minipage}
\end{table}

\begin{table}[ht!]
\begin{minipage}{\textwidth}
\linespread{.5}\selectfont
\centering
\caption{Goodness-of-fit statistics for simulated versus observed streamflow at each USGS gage at the HEC-RAS model boundary. Metrics are the Nash--Sutcliffe efficiency (NSE), coefficient of determination ($R^2$), percent bias (PBIAS), and the RMSE--observations standard deviation ratio (RSR).}
\label{tab:gof_per_gauge}
\small
\begin{tabular}{l l c c c c}
\hline
Computation point & Storm & NSE & $R^2$ & PBIAS (\%) & RSR \\
\hline
\multirow{5}{*}{Amite River} & Isaac & 0.92 & 0.93 & -14.68 & 0.29 \\
 & Katrina & 0.75 & 0.80 & -12.59 & 0.50 \\
 & Bill & 0.74 & 0.79 & -4.13 & 0.51 \\
 & Gustav/Ike & 0.63 & 0.81 & -39.96 & 0.61 \\
\cline{2-6}
\noalign{\vskip 0.04in}
 & \textbf{Average} & \textbf{0.76} & \textbf{0.83} & \textbf{-17.84} & \textbf{0.48} \\
\hline
\multirow{5}{*}{Tangipahoa River} & Isaac & 0.91 & 0.92 & -15.46 & 0.30 \\
 & Katrina & 0.79 & 0.80 & -4.04 & 0.46 \\
 & Bill & 0.88 & 0.88 & 1.81 & 0.35 \\
 & Gustav/Ike & 0.77 & 0.86 & 26.16 & 0.48 \\
\cline{2-6}
\noalign{\vskip 0.04in}
 & \textbf{Average} & \textbf{0.84} & \textbf{0.86} & \textbf{2.12} & \textbf{0.40} \\
\hline
\multirow{5}{*}{Tickfaw River} & Isaac & 0.77 & 0.81 & -29.33 & 0.47 \\
 & Katrina & 0.78 & 0.86 & -15.73 & 0.47 \\
 & Bill & 0.76 & 0.88 & -26.19 & 0.49 \\
 & Gustav/Ike & 0.60 & 0.62 & -19.69 & 0.63 \\
\cline{2-6}
\noalign{\vskip 0.04in}
 & \textbf{Average} & \textbf{0.73} & \textbf{0.79} & \textbf{-22.73} & \textbf{0.52} \\
\hline
\multirow{5}{*}{Natalbany River} & Isaac & 0.92 & 0.93 & -23.77 & 0.29 \\
 & Katrina & 0.46 & 0.57 & 21.75 & 0.73 \\
 & Bill & 0.82 & 0.84 & -22.49 & 0.43 \\
 & Gustav/Ike & 0.76 & 0.80 & -29.52 & 0.49 \\
\cline{2-6}
\noalign{\vskip 0.04in}
 & \textbf{Average} & \textbf{0.74} & \textbf{0.78} & \textbf{-13.51} & \textbf{0.48} \\
\hline
\end{tabular}
\end{minipage}
\end{table}

\begin{figure}
    \centering
    \includegraphics[width=6in]{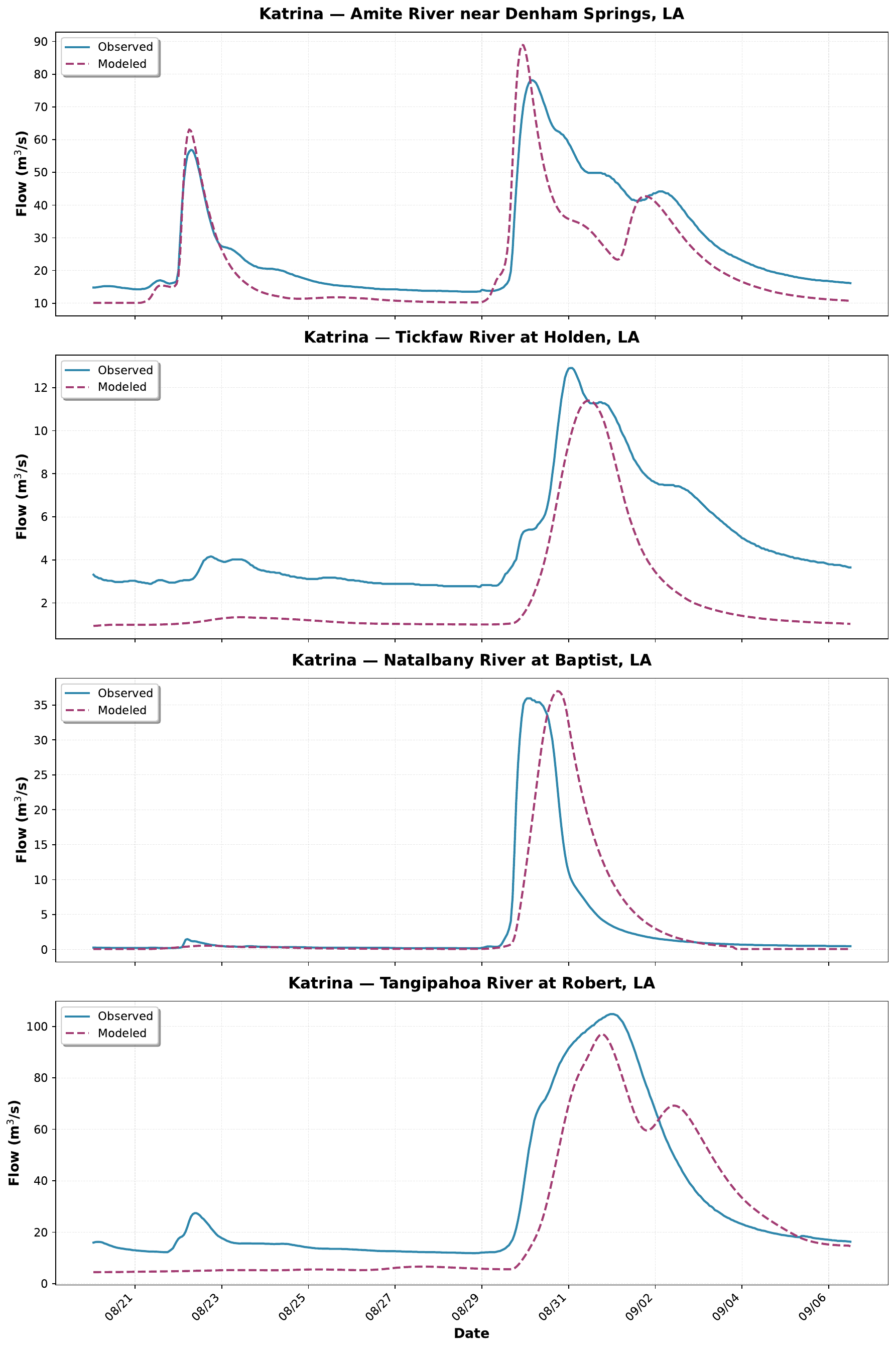}
    \caption{Comparison of HEC-HMS modeled and observed discharges for the 4 USGS gages at HEC-RAS domain upstream boundary. See Figure \ref{fig:gagelocations} for the location of the gages used for this comparison.}
    \label{fig:hmscalib}
\end{figure}

The HEC-HMS model was calibrated using four historical tropical cyclones: Tropical Storm Bill (2003), Hurricane Katrina (2005), Hurricanes Gustav and Ike (2008), and Hurricane Isaac (2012). These storms were selected from the 26 historical tropical cyclones (23 events, including three storm pairs) occurring between 2002 and 2021, while the remaining events were reserved for developing the antecedent-condition distributions (baseflow and watershed deficit) used in the extended JPM analysis (Table \ref{tab:calibration_storms}). Calibration against observed streamflow at the four HEC-HMS outlet gages (Table \ref{tab:gages}; Fig. \ref{fig:gagelocations}) yielded average Nash--Sutcliffe efficiencies of 0.76, 0.84, 0.73, and 0.74 for the Amite, Tangipahoa, Tickfaw, and Natalbany Rivers, respectively (Table \ref{tab:gof_per_gauge}). Figure \ref{fig:hmscalib} displays  representative comparisons between simulated and observed river discharges at the 4 USGS gage locations on the HEC-RAS domain boundary.

\begin{figure}
    \centering
    \includegraphics[width=6in]{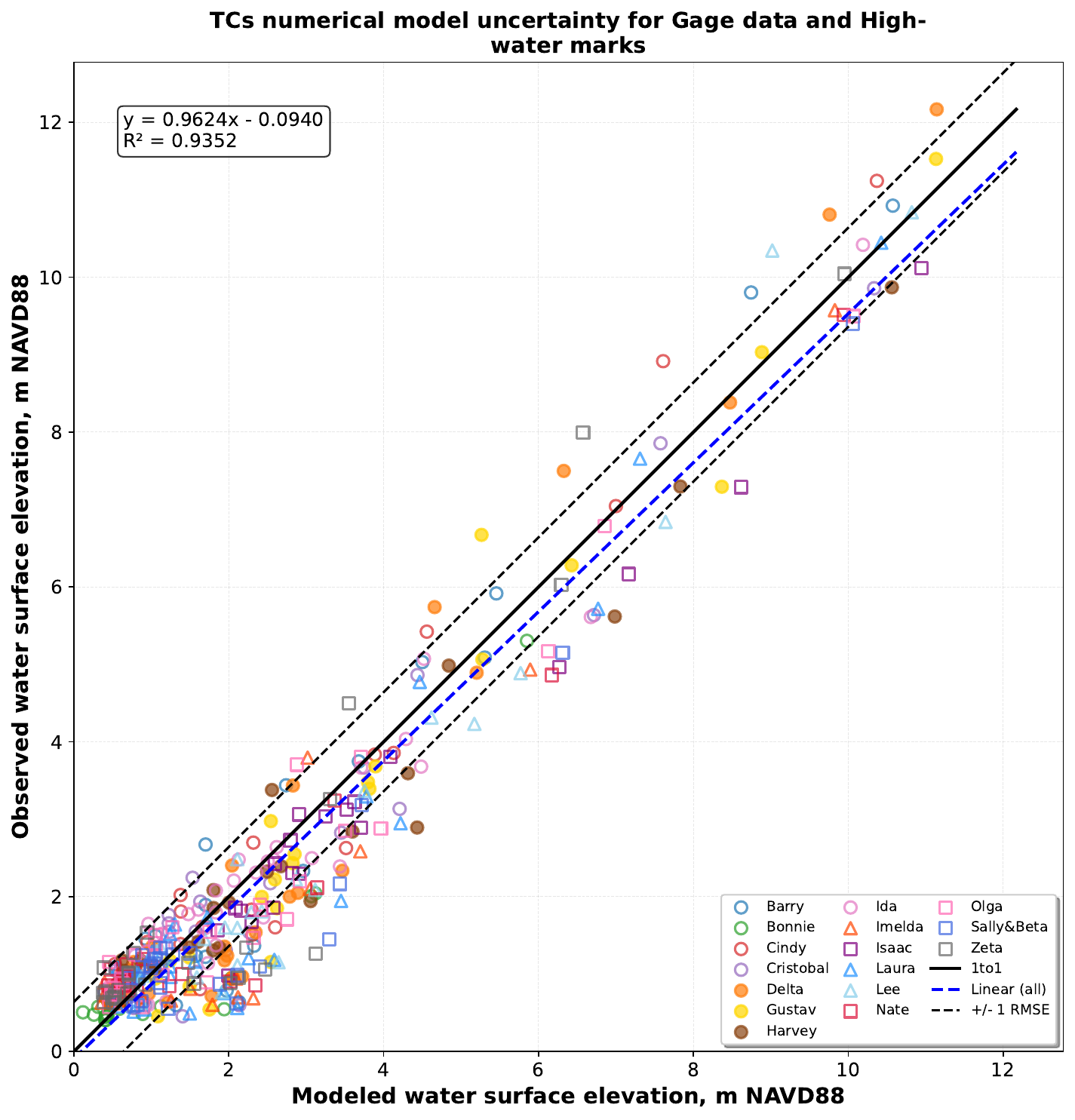}
    \caption{For TC events, comparison of modeled peak WSE against the combined gauge data WSE peaks and HWMs within the HEC-RAS domain.}
    \label{fig:rascomp}
\end{figure}

\begin{figure}
    \centering
    \includegraphics[width=6in]{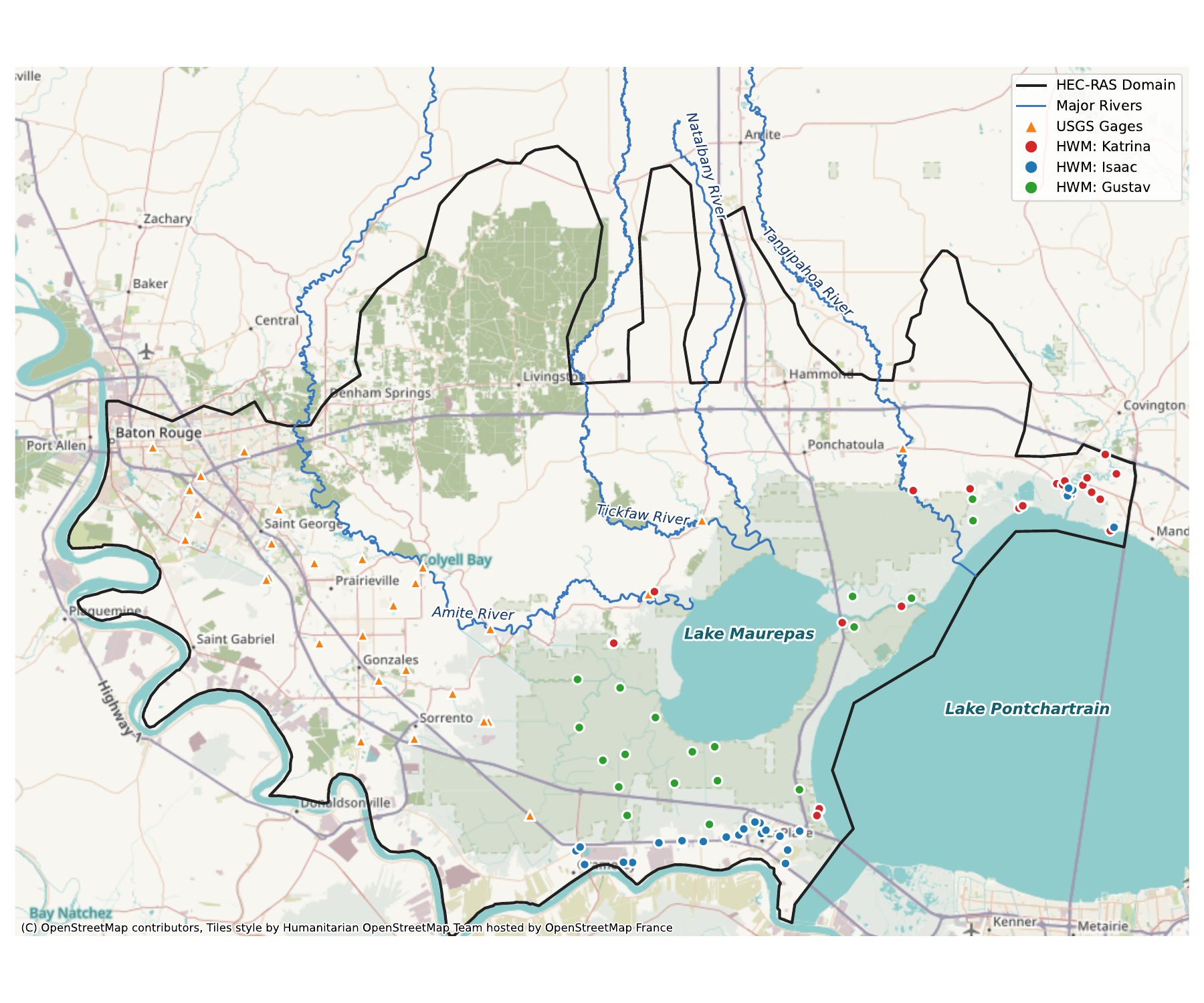}
    \caption{Location of the HWMs and USGS gages for the comparison of the observed versus modeled water surface elevations of Fig. \ref{fig:rascomp}.}
    \label{fig:location_rascomp}
\end{figure}

Calibration of the HEC-RAS model focused on the hydraulic roughness (Manning's $n$), wind-stress parameterization, and downstream boundary configuration. Historical hindcast simulations were forced using observed atmospheric conditions (OWI wind and pressure fields), observed AORC rainfall, HEC-HMS inflow hydrographs, observed antecedent watershed conditions, and downstream water levels from the calibrated Louisiana Coastal Master Plan ADCIRC model \cite{johnson2023coastal}. Model performance was evaluated by comparing simulated peak water-surface elevations with the USGS stage records and high-water marks (Table \ref{tab:gages}; Figs. \ref{fig:rascomp} and \ref{fig:location_rascomp}). Because these hindcast simulations simultaneously represent the observed rainfall-runoff response, antecedent watershed conditions, coastal storm surge, tides, wind forcing, and atmospheric pressure associated with historical compound flood events, the agreement between modeled and observed peak water-surface elevations provides a hindcast validation of the integrated compound flood modeling framework.

\section{JPM Storm Attributes PDF}
\label{sec:JPM_PDFs}
 
The joint PDF of the JPM storm attributes was represented by the following structure:
\begin{align}
p(c_p,R_{\max},v_f, \theta_l, x_l)= p(c_p|x_l)p(R_{\max}|c_p)p(v_f|\theta_l)p(\theta_l|x_l)p(x_l),
 \end{align}
where the PDF of the landfall location (along the coast), $p(x_l)$ was empirically derived, and the conditional PDFs were defined as
\begin{align}
\label{eq:pcp|xl}
p(c_p|x_l) =& \frac{\partial}{\partial x_l}\left\{\text{exp}\left[-\text{exp}\left(-\frac{c_p-a_0(x_l)}{a_l(x_l)}\right)\right]\right\} \\
p(R_{\max}|c_p) = & \frac{1}{R_{\max} \sigma(c_p)\sqrt{2\pi}}^{-\frac{(\ln [R_{\max}]-\overline{R}_{\max}(c_p))^2}{2 \sigma^2(c_p)}}\\
p(v_f|\theta_l) = &  \frac{1}{\sigma\sqrt{2\pi}}e^{-\frac{(\overline{v}_f(\theta_l)-v_f)^2}{2 \sigma^2}} \\
p(\theta_l|x_l) = &  \frac{1}{\sigma(x_l)\sqrt{2\pi}}e^{-\frac{(\overline{\theta}_l(x_l)-\theta_l)^2}{2 \sigma^2(x_l)}},
\label{eq:ptheta|xl}
\end{align}
where the PDF of the central deficit conditional on the landfall location, $p(c_p|x_l)$, was a Gumbel PDF, a log-normal distribution represented the PDF of the radius of maximum winds, $p(R_{max}|c_p)$, and normal distributions represented the PDFs of forward velocity, and storm heading at landfall, $p(\theta|x_l)$, and $p(v_f|\theta)$,  which respectively were conditional on the central pressure deficit, $c_p$, the storm landfall location, $x_l$, and the storm heading at landfall, $\theta_l$.  The landfall location PDF $p(x_l)$ represented the likelihood of a storm making landfall along the coast between $x_1 = 0$ and $x_1 = x_{l,\max}$ (see Fig. \ref{fig:1}). This empirical PDF was obtained by counting the number of times a storm makes landfall at each location \( x_l \), smoothing the counts (from  $x_1 = 0$ to $x_1 = x_{l,\max})$ with a Gaussian kernel, and normalizing this smoothed count by the total number of storms observed \cite{johnson2023coastal,nadal2020coastal,nadal2022coastal}.

In this pilot study, the PDFs in Eqs. (\ref{eq:pcp|xl})–(\ref{eq:ptheta|xl}) were fit to data from the HURricane DATa 2nd generation (HURDAT2) dataset, while the empirical PDF, $p(x_l)$, was constructed by counting how often HURDAT2 storm tracks intersected the coastline. This empirical PDF was defined at the resolution of discrete coastal segments, determined by the landfall locations of north-heading synthetic storm tracks (Fig. \ref{fig:2}), with segment edges set at equidistant points between adjacent landfall locations (Fig. \ref{fig:2}). To fit the PDFs in Eqs. (\ref{eq:pcp|xl})–(\ref{eq:ptheta|xl}), each coastline segment was associated with historical central pressure deficit data and track heading data from storms within a $\pm 150$-km distance of the segment center. The central pressure deficit, $c_p$, was de-trended to account for a linear drift, i.e., $c_p = c_{p, \text{historic}}+a_t t$ where $c_{p, \text{historic}}$ are the historic data values, $a_t$ is the drift coefficient, and $t$ is the time (in years from 1950 to 2022). This drift coefficient was found from a linear regression of the historical values, $c_{p, \text{historic}}$, over the observation period of the dataset (1950-2022). In turn, the normal PDFs $p(c_p|x_l)$ and $p(\theta|x_l)$  were fit to this data using the method of moments. The detrended $c_p$ data was discretized into bins spanning the greater of 10 mb or the minimum range needed to include at least ten historical tropical cyclones (TCs). For each $c_p$ bin, the PDF $p(R_{\max} | c_p)$ was fitted to the corresponding $R_{\max}$ data using the method of moments. Finally, for forward velocity and heading data ($v_f$ and $\theta_l$), the normal PDF $p(v_f | \theta_l)$ was fitted using a linear regression on $\theta_l$, assuming normally distributed residuals.

\section{Discretization of the continuous PDFs}

 \label{sec:PDF_discretized}

 \begin{table}
\begin{minipage}{\textwidth}
\linespread{.5}\selectfont
\caption{PDF discretization parameters \label{tab:vars_params_weights} }
\noindent
\begin{tabular}{c p{13cm}}
\hline
\noalign{\vskip 0.04in}
Symbol & Description \\
\hline
\noalign{\vskip 0.05in}
$\omega_i$                & Probability weight for the $i$-th JPM storm $\omega_i = \omega_{x_l,i}\times \omega_{c_pR_{\max}v_f\theta_{i}}$; see Eq. (\ref{eq:TC_dis})\\
$\omega_{x_l,i}$                & Probability weight for the $i$-th JPM storm for the associated segment of coastline $x_{l,i}\in[x_{l,k}, x_{l,k+1}]$; see Eqs. (\ref{eq:JPMset}), (\ref{eq:TC_dis}), and (\ref{eq:JPM-sub})\\
$\omega_{c_pR_{\max}v_f\theta_{i}}$  &  Probability weight for the $i$-th JPM storm parameters conditional on the coastline segment associated with the storm $x_{l,i}$; see Eqs. (\ref{eq:JPMset}), (\ref{eq:TC_dis}), and (\ref{eq:JPM-sub})\\
$\omega_m$  &  Probability weight for the $m$-th soil moisture condition; see   (\ref{eq:TC_dis})\\
\noalign{\vskip 0.04in}
\hline
\end{tabular}
\end{minipage}
\end{table}

This appendix gives the discretization of the continuous driver PDFs of Eqs.~(\ref{eq:pTC_PS}) and (\ref{eq:pNT_PS}) into the weighted realizations of Eq. (\ref{eq:p_discrete}), together with the construction of the associated probability weights (Table \ref{tab:vars_params_weights}).

For tropical cyclones, the continuous probability distribution $p_{TC}(\mathbf{x}_{TC})$ was discretized with respect to the JPM storm characteristics, stochastic rainfall realizations, and antecedent soil-moisture conditions. The JPM storm PDF, $p(\mathbf{x}_{JPM})$, was represented by the 645 synthetic storms developed for the Louisiana Coastal Master Plan \cite{johnson2023coastal,nadal2022coastal},
\begin{align}
S_{TC}=\{(c_{p,i},R_{\max,i},v_{f,i},\theta_i,x_{l,i})\mid i=1,\ldots,645\},
\label{eq:JPMset}
\end{align}
where each storm is uniquely defined by its central pressure deficit, radius of maximum winds, forward velocity, heading at landfall, and landfall location. The continuous probability distributions were approximated as

\begin{align}
p(\mathbf{x}_{JPM})p(\mathbf{r}(t)|\mathbf{\overline{r}}(t))p_{TC}(\mathbf{\overline{s}})
\approx
\sum_{i=1}^{645}\sum_{j=1}^{100}\sum_{m=1}^{5}
&
\overbrace{\omega_i
\delta(c_p-c_{p,i})
\delta(R_{\max}-R_{\max,i})
\delta(v_f-v_{f,i})
\delta(\theta-\theta_i)
\delta(x_l-x_{l,i})}^{\substack{\text{Discretized}\\p(\mathbf{x}_{JPM})}}
\nonumber\\
&\times
\underbrace{\frac{1}{100}
\delta(\mathbf{r}(t)-h[\mathbf{\overline{r}}_i(t)]\boldsymbol{\epsilon}_j)}_{\substack{\text{Discretized}\\p(\mathbf{r}(t)|\mathbf{\overline{r}}(t))}}
\underbrace{\omega_m
\delta(\mathbf{\overline{s}}-\mathbf{\overline{s}}_m)}_{\substack{\text{Discretized}\\p(\mathbf{\overline{s}})}},
\label{eq:TC_dis}
\end{align}
where $\omega_i$ denotes the probability weight assigned to the $i$th JPM storm, $\omega_m$ denotes the probability weight associated with the $m$th antecedent soil-moisture condition, and each rainfall realization was assigned a probability of $1/100$. The Dirac delta functions enforce the discretization by restricting each random variable to its corresponding discrete realization. The probability weights used in the discretization are summarized in Table~\ref{tab:vars_params_weights}.

The probability weight assigned to each JPM storm was decomposed as
\begin{align}
\omega_i=\omega_{x_l,i}\,\omega_{c_pR_{\max}v_f\theta,i},
\end{align}
where $\omega_{x_l,i}$ represents the probability associated with the coastline segment containing the storm landfall location, and $\omega_{c_pR_{\max}v_f\theta,i}$ represents the conditional probability of the remaining JPM storm characteristics within that segment. Coastline segments $[x_{l,k},x_{l,k+1}]$ were defined by the midpoints between adjacent north-heading synthetic storm landfall locations (Fig.~\ref{fig:2}), and the coastline-segment weights were obtained by integrating the empirical landfall-location PDF over each segment.

Within each coastline segment, the subset of storms
\begin{align}
\{(c_{p,i},R_{\max,i},v_{f,i},\theta_{l,i},x_{l,i})
\mid
x_{l,i}\in[x_{l,k},x_{l,k+1}]\}
\label{eq:JPM-sub}
\end{align}
was assigned the conditional probability weights provided by the Louisiana Coastal Master Plan. These weights were obtained by partitioning the continuous joint PDF $p(c_p,R_{\max},v_f,\theta_l)$ into Voronoi-like cells bounded midway between neighboring storm attributes and integrating the continuous PDF over each cell. The final JPM storm weight was therefore computed as $\omega_i=\omega_{x_l,i}\omega_{c_pR_{\max}v_f\theta,i}$.

The antecedent hydrologic conditions, $\overline{\mathbf{s}}_m$, were represented by five soil-moisture states corresponding to the 0.05, 0.25, 0.50, 0.75, and 0.95 quantiles across the 23 modeled watersheds. The probability weight, $\omega_m$, for each state was obtained by averaging the watershed-specific probabilities associated with the corresponding quantile, where the integration limits were selected so that each quantile represented the average soil-moisture condition over its probability interval.

For each combination of JPM storm and antecedent soil-moisture condition, 100 realizations of the spatiotemporally varying rainfall field were generated. Each rainfall field was derived from the IPET rainfall associated with the corresponding JPM storm, $\overline{r}(t)$, modified by the bias-correction function $h[\cdot]$, and scaled by the stochastic attenuation factor $\epsilon_j$ sampled through Monte Carlo simulation. Because the attenuation factor was sampled 100 times, each rainfall realization was assigned an equal probability of $1/100$. 

Unlike tropical cyclones, non-tropical storms were represented directly from the historical event record rather than through a parametric Joint Probability Method. Consequently, the historical storm catalog simultaneously defines the rainfall forcing, wind forcing, antecedent hydrologic conditions, and downstream tidal response used in the numerical simulations. The catalog consists of the 44 non-tropical storms listed in Table \ref{tab:historic_storms}, selected using the precipitation threshold described in Section \ref{sec:freq}. Specifically, the non-tropical characteristics PDF, $p_{NT}(\mathbf{x}_{NT})$, components of PDFs for the non-tidal residual peaking factor and lag time, $p(\kappa)$ and $p(\tau_l)$, as well as the wind, rainfall and soil moisture PDFs, $p(\mathbf{u}(t))p(\mathbf{\overline{r}}(t))p(\mathbf{r}(t)|\mathbf{\overline{r}}(t))p_{NT}(\mathbf{\overline{s}})$,  were discretized as follows:
\begin{align}
p(\kappa)p(\tau_l)p(\mathbf{u}(t)|\mathbf{\overline{r}}(t))p(\mathbf{\overline{r}}(t))p(\mathbf{r}(t)|\mathbf{\overline{r}}(t))p_{NT}(\mathbf{\overline{s}}) 
&\approx \sum_{i=1}^5\sum_{j=1}^{5}\sum_{m=1}^{44}  
\overbrace{\frac{1}{5}\delta(\kappa-\kappa_i)\frac{1}{5}\delta(\tau_l - \tau_{l,j})}^{\substack{\text{Discretized} \\ p(\kappa)p(\tau_l)}} \notag \\ 
&\quad\quad\quad\quad\times \underbrace{\frac{1}{44}\delta(\mathbf{u}(t)-\mathbf{u}_m(t))\delta(\mathbf{r}(t)-\mathbf{r}_m(t))\delta(\mathbf{\overline{s}}-\mathbf{\overline{s}}_m)}_{\substack{\text{Discretized} \\ p(\mathbf{u}(t)|\mathbf{\overline{r}}(t))p(\mathbf{\overline{r}}(t))p(\mathbf{r}(t)|\mathbf{\overline{r}}(t))p_{NT}(\mathbf{\overline{s}})}},
\label{eq:NT_dis}
\end{align}
where the Dirac delta functions enforce that each sampled value of the peaking factor and lag time exactly corresponds to one of the discrete values \( \kappa_i \) and \( \tau_{l,j} \), while the time-dependent rainfall and wind fields, as well as the soil moisture were assigned  respective historical values, $\mathbf{r}(t)$, $\mathbf{u}_m(t)$ and $\overline{\mathbf{s}}_m$. The quantized values of \( \kappa \) and \( \tau_l \) correspond to the 0.1, 0.3, 0.5, 0.7, and 0.9 quantiles of their respective distributions, each assigned an equal probability of \( \frac{1}{5} \). Likewise, the discrete rainfall and wind fields and soil moisture values correspond to 44 historical non-tropical storm events, each weighted by \( \frac{1}{44} \) (see Table \ref{tab:historic_storms}). These discrete values were applied to an averaged, detrended, and nondimensionalized downstream stage hydrograph that was applied across storm events.


\begin{thebibliography}{}

\bibitem [\protect \citeauthoryear {%
Ahmadisharaf%
\ \BBA {} Kalyanapu%
}{%
Ahmadisharaf%
\ \BBA {} Kalyanapu%
}{%
{\protect \APACyear {2019}}%
}]{%
ahmadisharaf2019coupled}
\APACinsertmetastar {%
ahmadisharaf2019coupled}%
\begin{APACrefauthors}%
Ahmadisharaf, E.%
\BCBT {}\ \BBA {} Kalyanapu, A\BPBI J.%
\end{APACrefauthors}%
\unskip\
\newblock
\APACrefYearMonthDay{2019}{}{}.
\newblock
{\BBOQ}\APACrefatitle {A coupled probabilistic hydrologic and hydraulic modelling framework to investigate the uncertainty of flood loss estimates} {A coupled probabilistic hydrologic and hydraulic modelling framework to investigate the uncertainty of flood loss estimates}.{\BBCQ}
\newblock
\APACjournalVolNumPages{Journal of Flood Risk Management}{12}{S2}{e12536}.
\newblock
\begin{APACrefDOI} \doi{10.1111/jfr3.12536} \end{APACrefDOI}
\PrintBackRefs{\CurrentBib}

\bibitem [\protect \citeauthoryear {%
Au%
\ \BBA {} Tam%
}{%
Au%
\ \BBA {} Tam%
}{%
{\protect \APACyear {1999}}%
}]{%
au1999transforming}
\APACinsertmetastar {%
au1999transforming}%
\begin{APACrefauthors}%
Au, C.%
\BCBT {}\ \BBA {} Tam, J.%
\end{APACrefauthors}%
\unskip\
\newblock
\APACrefYearMonthDay{1999}{}{}.
\newblock
{\BBOQ}\APACrefatitle {Transforming variables using the Dirac generalized function} {Transforming variables using the dirac generalized function}.{\BBCQ}
\newblock
\APACjournalVolNumPages{The American Statistician}{53}{3}{270--272}.
\PrintBackRefs{\CurrentBib}

\bibitem [\protect \citeauthoryear {%
Bartles%
\ \protect \BOthers {.}}{%
Bartles%
\ \protect \BOthers {.}}{%
{\protect \APACyear {2022}}%
}]{%
bartles2022hydrologic}
\APACinsertmetastar {%
bartles2022hydrologic}%
\begin{APACrefauthors}%
Bartles, M.%
, Brauer, T.%
, Ho, D.%
, Fleming, M.%
, Karlovits, G.%
, Pak, J.%
\BDBL {}Willis, J.%
\end{APACrefauthors}%
\unskip\
\newblock
\APACrefYearMonthDay{2022}{}{}.
\newblock
\APACrefbtitle {Hydrologic modeling system HEC-HMS User’s manual} {Hydrologic modeling system hec-hms user’s manual}\ \APACbVolEdTR{}{\BTR{}\ \BNUM\ CPD-74A}.
\newblock
\APACaddressInstitution{}{US Army Corps of Engineers. Davis, CA}.
\PrintBackRefs{\CurrentBib}

\bibitem [\protect \citeauthoryear {%
Bartlett%
, Cultra%
, Geldner%
\BCBL {}\ \BBA {} Porporato%
}{%
Bartlett%
, Cultra%
\BCBL {}\ \protect \BOthers {.}}{%
{\protect \APACyear {2025}}%
}]{%
bartlett2025stochasticE}
\APACinsertmetastar {%
bartlett2025stochasticE}%
\begin{APACrefauthors}%
Bartlett, M\BPBI S.%
, Cultra, E.%
, Geldner, N.%
\BCBL {}\ \BBA {} Porporato, A.%
\end{APACrefauthors}%
\unskip\
\newblock
\APACrefYearMonthDay{2025}{}{}.
\newblock
{\BBOQ}\APACrefatitle {Stochastic ecohydrological perspective on semi-distributed rainfall-runoff dynamics} {Stochastic ecohydrological perspective on semi-distributed rainfall-runoff dynamics}.{\BBCQ}
\newblock
\APACjournalVolNumPages{Water Resources Research}{61}{9}{e2025WR040606}.
\PrintBackRefs{\CurrentBib}

\bibitem [\protect \citeauthoryear {%
Bartlett%
, Daly%
, McDonnell%
, Parolari%
\BCBL {}\ \BBA {} Porporato%
}{%
Bartlett%
\ \protect \BOthers {.}}{%
{\protect \APACyear {2015}}%
}]{%
bartlett2015stochastic}
\APACinsertmetastar {%
bartlett2015stochastic}%
\begin{APACrefauthors}%
Bartlett, M\BPBI S.%
, Daly, E.%
, McDonnell, J\BPBI J.%
, Parolari, A\BPBI J.%
\BCBL {}\ \BBA {} Porporato, A.%
\end{APACrefauthors}%
\unskip\
\newblock
\APACrefYearMonthDay{2015}{}{}.
\newblock
{\BBOQ}\APACrefatitle {Stochastic rainfall-runoff model with explicit soil moisture dynamics} {Stochastic rainfall-runoff model with explicit soil moisture dynamics}.{\BBCQ}
\newblock
\APACjournalVolNumPages{Proceedings of the Royal Society A: Mathematical, Physical and Engineering Sciences}{471}{2183}{20150389}.
\PrintBackRefs{\CurrentBib}

\bibitem [\protect \citeauthoryear {%
Bartlett%
\ \protect \BOthers {.}}{%
Bartlett%
\ \protect \BOthers {.}}{%
{\protect \APACyear {2026}}%
}]{%
Bartlett2025extended-jpm}
\APACinsertmetastar {%
Bartlett2025extended-jpm}%
\begin{APACrefauthors}%
Bartlett, M\BPBI S.%
, Geldner, N.%
, Roberts, H\BPBI J.%
, Cobell, Z.%
, McMann, B.%
, Partida, L.%
\BDBL {}Narayanaswamy, M.%
\end{APACrefauthors}%
\unskip\
\newblock
\APACrefYearMonthDay{2026}{}{}.
\newblock
\APACrefbtitle {Extending the Joint Probability Method to Compound Flooding: Transition Zone Delineation, Flood Depth Attribution, and Design Event Selection [{Dataset, Computational Notebooks}].} {Extending the joint probability method to compound flooding: Transition zone delineation, flood depth attribution, and design event selection [{Dataset, Computational Notebooks}].}
\newblock
\APACaddressPublisher{}{Zenodo}.
\newblock
\begin{APACrefURL} \url{https://zenodo.org/records/21498949} \end{APACrefURL}
\newblock
\begin{APACrefDOI} \doi{10.5281/zenodo.21498948} \end{APACrefDOI}
\PrintBackRefs{\CurrentBib}

\bibitem [\protect \citeauthoryear {%
Bartlett%
, Parolari%
, McDonnell%
\BCBL {}\ \BBA {} Porporato%
}{%
Bartlett%
\ \protect \BOthers {.}}{%
{\protect \APACyear {2017}}%
}]{%
bartlett2017reply}
\APACinsertmetastar {%
bartlett2017reply}%
\begin{APACrefauthors}%
Bartlett, M\BPBI S.%
, Parolari, A\BPBI J.%
, McDonnell, J.%
\BCBL {}\ \BBA {} Porporato, A.%
\end{APACrefauthors}%
\unskip\
\newblock
\APACrefYearMonthDay{2017}{}{}.
\newblock
{\BBOQ}\APACrefatitle {Reply to comment by {F}red {L}. {O}gden et al. on ‘‘{B}eyond the {SCS-CN} method: {A} theoretical framework for spatially lumped rainfall-runoff response,’’} {Reply to comment by {F}red {L}. {O}gden et al. on ‘‘{B}eyond the {SCS-CN} method: {A} theoretical framework for spatially lumped rainfall-runoff response,’’}.{\BBCQ}
\newblock
\APACjournalVolNumPages{Water Resources Research}{53}{7}{6351-6354}.
\PrintBackRefs{\CurrentBib}

\bibitem [\protect \citeauthoryear {%
Bartlett%
, Parolari%
, McDonnell%
\BCBL {}\ \BBA {} Porporato%
}{%
Bartlett%
, Parolari%
\BCBL {}\ \protect \BOthers {.}}{%
{\protect \APACyear {2016}}%
}]{%
bartlett2015unified2}
\APACinsertmetastar {%
bartlett2015unified2}%
\begin{APACrefauthors}%
Bartlett, M\BPBI S.%
, Parolari, A\BPBI J.%
, McDonnell, J\BPBI J.%
\BCBL {}\ \BBA {} Porporato, A.%
\end{APACrefauthors}%
\unskip\
\newblock
\APACrefYearMonthDay{2016}{}{}.
\newblock
{\BBOQ}\APACrefatitle {Framework for event-based semidistributed modeling that unifies the {SCS}-{CN} method, {VIC}, {PDM}, and {TOPMODEL}} {Framework for event-based semidistributed modeling that unifies the {SCS}-{CN} method, {VIC}, {PDM}, and {TOPMODEL}}.{\BBCQ}
\newblock
\APACjournalVolNumPages{Water Resources Research}{52}{9}{7036--7052}.
\PrintBackRefs{\CurrentBib}

\bibitem [\protect \citeauthoryear {%
Bartlett%
, Rodriguez-Iturbe%
\BCBL {}\ \BBA {} Porporato%
}{%
Bartlett%
, Rodriguez-Iturbe%
\BCBL {}\ \BBA {} Porporato%
}{%
{\protect \APACyear {2016}}%
}]{%
bartlett2016mean}
\APACinsertmetastar {%
bartlett2016mean}%
\begin{APACrefauthors}%
Bartlett, M\BPBI S.%
, Rodriguez-Iturbe, I.%
\BCBL {}\ \BBA {} Porporato, A\BPBI M.%
\end{APACrefauthors}%
\unskip\
\newblock
\APACrefYearMonthDay{2016}{}{}.
\newblock
{\BBOQ}\APACrefatitle {A mean field approach to the watershed response under stochastic seasonal forcing} {A mean field approach to the watershed response under stochastic seasonal forcing}.{\BBCQ}
\newblock
\BIn{} \APACrefbtitle {{AGU Fall Meeting Abstracts}} {{AGU Fall Meeting Abstracts}}\ (\BVOL\ 2016, \BPGS\ H52A--06).
\PrintBackRefs{\CurrentBib}

\bibitem [\protect \citeauthoryear {%
Bartlett%
, Van~Blitterswyk%
\BCBL {}\ \protect \BOthers {.}}{%
Bartlett%
, Van~Blitterswyk%
\BCBL {}\ \protect \BOthers {.}}{%
{\protect \APACyear {2025}}%
}]{%
bartlett2025physically}
\APACinsertmetastar {%
bartlett2025physically}%
\begin{APACrefauthors}%
Bartlett, M\BPBI S.%
, Van~Blitterswyk, J.%
, Farella, M.%
, Li, J.%
, Smith, C.%
, Parolari, A\BPBI J.%
\BDBL {}Mrad, A.%
\end{APACrefauthors}%
\unskip\
\newblock
\APACrefYearMonthDay{2025}{}{}.
\newblock
{\BBOQ}\APACrefatitle {Physically based dimensionless features for pluvial flood mapping with machine learning} {Physically based dimensionless features for pluvial flood mapping with machine learning}.{\BBCQ}
\newblock
\APACjournalVolNumPages{Water Resources Research}{61}{4}{e2024WR039086}.
\PrintBackRefs{\CurrentBib}

\bibitem [\protect \citeauthoryear {%
Bass%
\ \BBA {} Bedient%
}{%
Bass%
\ \BBA {} Bedient%
}{%
{\protect \APACyear {2018}}%
}]{%
bass2018surrogate}
\APACinsertmetastar {%
bass2018surrogate}%
\begin{APACrefauthors}%
Bass, B.%
\BCBT {}\ \BBA {} Bedient, P.%
\end{APACrefauthors}%
\unskip\
\newblock
\APACrefYearMonthDay{2018}{}{}.
\newblock
{\BBOQ}\APACrefatitle {Surrogate modeling of joint flood risk across coastal watersheds} {Surrogate modeling of joint flood risk across coastal watersheds}.{\BBCQ}
\newblock
\APACjournalVolNumPages{Journal of Hydrology}{558}{}{159--173}.
\newblock
\begin{APACrefDOI} \doi{10.1016/j.jhydrol.2018.01.035} \end{APACrefDOI}
\PrintBackRefs{\CurrentBib}

\bibitem [\protect \citeauthoryear {%
Basso%
, Schirmer%
\BCBL {}\ \BBA {} Botter%
}{%
Basso%
\ \protect \BOthers {.}}{%
{\protect \APACyear {2015}}%
}]{%
basso2015emergence}
\APACinsertmetastar {%
basso2015emergence}%
\begin{APACrefauthors}%
Basso, S.%
, Schirmer, M.%
\BCBL {}\ \BBA {} Botter, G.%
\end{APACrefauthors}%
\unskip\
\newblock
\APACrefYearMonthDay{2015}{}{}.
\newblock
{\BBOQ}\APACrefatitle {On the emergence of heavy-tailed streamflow distributions} {On the emergence of heavy-tailed streamflow distributions}.{\BBCQ}
\newblock
\APACjournalVolNumPages{Advances in Water Resources}{82}{}{98--105}.
\PrintBackRefs{\CurrentBib}

\bibitem [\protect \citeauthoryear {%
Basso%
, Schirmer%
\BCBL {}\ \BBA {} Botter%
}{%
Basso%
\ \protect \BOthers {.}}{%
{\protect \APACyear {2016}}%
}]{%
basso2016physically}
\APACinsertmetastar {%
basso2016physically}%
\begin{APACrefauthors}%
Basso, S.%
, Schirmer, M.%
\BCBL {}\ \BBA {} Botter, G.%
\end{APACrefauthors}%
\unskip\
\newblock
\APACrefYearMonthDay{2016}{}{}.
\newblock
{\BBOQ}\APACrefatitle {A physically based analytical model of flood frequency curves} {A physically based analytical model of flood frequency curves}.{\BBCQ}
\newblock
\APACjournalVolNumPages{Geophysical Research Letters}{43}{17}{9070--9076}.
\PrintBackRefs{\CurrentBib}

\bibitem [\protect \citeauthoryear {%
Bender%
, Wahl%
, M{\"u}ller%
\BCBL {}\ \BBA {} Jensen%
}{%
Bender%
\ \protect \BOthers {.}}{%
{\protect \APACyear {2016}}%
}]{%
bender2016multivariate}
\APACinsertmetastar {%
bender2016multivariate}%
\begin{APACrefauthors}%
Bender, J.%
, Wahl, T.%
, M{\"u}ller, A.%
\BCBL {}\ \BBA {} Jensen, J.%
\end{APACrefauthors}%
\unskip\
\newblock
\APACrefYearMonthDay{2016}{}{}.
\newblock
{\BBOQ}\APACrefatitle {A multivariate design framework for river confluences} {A multivariate design framework for river confluences}.{\BBCQ}
\newblock
\APACjournalVolNumPages{Hydrological Sciences Journal}{61}{3}{471--482}.
\PrintBackRefs{\CurrentBib}

\bibitem [\protect \citeauthoryear {%
Bensi%
, Mohammadi%
, Kao%
\BCBL {}\ \BBA {} DeNeale%
}{%
Bensi%
\ \protect \BOthers {.}}{%
{\protect \APACyear {2020}}%
}]{%
bensi2020multi}
\APACinsertmetastar {%
bensi2020multi}%
\begin{APACrefauthors}%
Bensi, M.%
, Mohammadi, S.%
, Kao, S\BHBI C.%
\BCBL {}\ \BBA {} DeNeale, S\BPBI T.%
\end{APACrefauthors}%
\unskip\
\newblock
\APACrefYearMonthDay{2020}{}{}.
\newblock
\APACrefbtitle {Multi-Mechanism Flood Hazard Assessment: Critical Review of Current Practice and Approaches} {Multi-mechanism flood hazard assessment: Critical review of current practice and approaches}\ \APACbVolEdTR{}{\BTR{}\ \BNUM\ ORNL/TM-2020/1447}.
\newblock
\APACaddressInstitution{Oak Ridge, TN}{Oak Ridge National Laboratory (ORNL)}.
\newblock
\begin{APACrefURL} \url{https://www.osti.gov/biblio/1637939} \end{APACrefURL}
\PrintBackRefs{\CurrentBib}

\bibitem [\protect \citeauthoryear {%
Berg%
}{%
Berg%
}{%
{\protect \APACyear {2013}}%
}]{%
berg2013tropical}
\APACinsertmetastar {%
berg2013tropical}%
\begin{APACrefauthors}%
Berg, R.%
\end{APACrefauthors}%
\unskip\
\newblock
\APACrefYearMonthDay{2013}{}{}.
\newblock
{\BBOQ}\APACrefatitle {Tropical cyclone report hurricane isaac (al092012) 21 august--1 september 2012} {Tropical cyclone report hurricane isaac (al092012) 21 august--1 september 2012}.{\BBCQ}
\newblock
\APACjournalVolNumPages{National Hurricane Center}{}{}{1--78}.
\PrintBackRefs{\CurrentBib}

\bibitem [\protect \citeauthoryear {%
Berghuijs%
, Harrigan%
, Molnar%
, Slater%
\BCBL {}\ \BBA {} Kirchner%
}{%
Berghuijs%
\ \protect \BOthers {.}}{%
{\protect \APACyear {2019}}%
}]{%
berghuijs2019relative}
\APACinsertmetastar {%
berghuijs2019relative}%
\begin{APACrefauthors}%
Berghuijs, W\BPBI R.%
, Harrigan, S.%
, Molnar, P.%
, Slater, L\BPBI J.%
\BCBL {}\ \BBA {} Kirchner, J\BPBI W.%
\end{APACrefauthors}%
\unskip\
\newblock
\APACrefYearMonthDay{2019}{}{}.
\newblock
{\BBOQ}\APACrefatitle {The relative importance of different flood-generating mechanisms across Europe} {The relative importance of different flood-generating mechanisms across europe}.{\BBCQ}
\newblock
\APACjournalVolNumPages{Water Resources Research}{55}{6}{4582--4593}.
\PrintBackRefs{\CurrentBib}

\bibitem [\protect \citeauthoryear {%
Bevacqua%
\ \protect \BOthers {.}}{%
Bevacqua%
\ \protect \BOthers {.}}{%
{\protect \APACyear {2020}}%
}]{%
bevacqua2020more}
\APACinsertmetastar {%
bevacqua2020more}%
\begin{APACrefauthors}%
Bevacqua, E.%
, Vousdoukas, M\BPBI I.%
, Zappa, G.%
, Hodges, K.%
, Shepherd, T\BPBI G.%
, Maraun, D.%
\BDBL {}Feyen, L.%
\end{APACrefauthors}%
\unskip\
\newblock
\APACrefYearMonthDay{2020}{}{}.
\newblock
{\BBOQ}\APACrefatitle {More Meteorological Events That Drive Compound Coastal Flooding Are Projected under Climate Change} {More meteorological events that drive compound coastal flooding are projected under climate change}.{\BBCQ}
\newblock
\APACjournalVolNumPages{Communications Earth \& Environment}{1}{}{47}.
\newblock
\begin{APACrefURL} \url{https://www.nature.com/articles/s43247-020-00044-z} \end{APACrefURL}
\newblock
\begin{APACrefDOI} \doi{10.1038/s43247-020-00044-z} \end{APACrefDOI}
\PrintBackRefs{\CurrentBib}

\bibitem [\protect \citeauthoryear {%
Beven%
}{%
Beven%
}{%
{\protect \APACyear {2012}}%
}]{%
beven2012rainfall}
\APACinsertmetastar {%
beven2012rainfall}%
\begin{APACrefauthors}%
Beven, K.%
\end{APACrefauthors}%
\unskip\
\newblock
\APACrefYear{2012}.
\newblock
\APACrefbtitle {Rainfall-Runoff Modelling: The Primer} {Rainfall-runoff modelling: The primer}.
\newblock
\APACaddressPublisher{}{Wiley}.
\PrintBackRefs{\CurrentBib}

\bibitem [\protect \citeauthoryear {%
Beven%
\ \BBA {} Kirkby%
}{%
Beven%
\ \BBA {} Kirkby%
}{%
{\protect \APACyear {1979}}%
}]{%
beven1979physically}
\APACinsertmetastar {%
beven1979physically}%
\begin{APACrefauthors}%
Beven, K.%
\BCBT {}\ \BBA {} Kirkby, M.%
\end{APACrefauthors}%
\unskip\
\newblock
\APACrefYearMonthDay{1979}{}{}.
\newblock
{\BBOQ}\APACrefatitle {A physically based, variable contributing area model of basin hydrology/Un mod{\`e}le {\`a} base physique de zone d'appel variable de l'hydrologie du bassin versant} {A physically based, variable contributing area model of basin hydrology/un mod{\`e}le {\`a} base physique de zone d'appel variable de l'hydrologie du bassin versant}.{\BBCQ}
\newblock
\APACjournalVolNumPages{Hydrological Sciences Journal}{24}{1}{43--69}.
\PrintBackRefs{\CurrentBib}

\bibitem [\protect \citeauthoryear {%
Bilskie%
\ \BBA {} Hagen%
}{%
Bilskie%
\ \BBA {} Hagen%
}{%
{\protect \APACyear {2018}}%
}]{%
bilskie2018defining}
\APACinsertmetastar {%
bilskie2018defining}%
\begin{APACrefauthors}%
Bilskie, M\BPBI V.%
\BCBT {}\ \BBA {} Hagen, S\BPBI C.%
\end{APACrefauthors}%
\unskip\
\newblock
\APACrefYearMonthDay{2018}{}{}.
\newblock
{\BBOQ}\APACrefatitle {Defining flood zone transitions in low-gradient coastal regions} {Defining flood zone transitions in low-gradient coastal regions}.{\BBCQ}
\newblock
\APACjournalVolNumPages{Geophysical Research Letters}{45}{6}{2761--2770}.
\newblock
\begin{APACrefDOI} \doi{10.1002/2018GL077524} \end{APACrefDOI}
\PrintBackRefs{\CurrentBib}

\bibitem [\protect \citeauthoryear {%
Bilskie%
\ \protect \BOthers {.}}{%
Bilskie%
\ \protect \BOthers {.}}{%
{\protect \APACyear {2021}}%
}]{%
bilskie2021enhancing}
\APACinsertmetastar {%
bilskie2021enhancing}%
\begin{APACrefauthors}%
Bilskie, M\BPBI V.%
, Zhao, H.%
, Resio, D.%
, Atkinson, J.%
, Cobell, Z.%
\BCBL {}\ \BBA {} Hagen, S\BPBI C.%
\end{APACrefauthors}%
\unskip\
\newblock
\APACrefYearMonthDay{2021}{}{}.
\newblock
{\BBOQ}\APACrefatitle {Enhancing flood hazard assessments in coastal Louisiana through coupled hydrologic and surge processes} {Enhancing flood hazard assessments in coastal louisiana through coupled hydrologic and surge processes}.{\BBCQ}
\newblock
\APACjournalVolNumPages{Frontiers in Water}{3}{}{609231}.
\newblock
\begin{APACrefDOI} \doi{10.3389/frwa.2021.609231} \end{APACrefDOI}
\PrintBackRefs{\CurrentBib}

\bibitem [\protect \citeauthoryear {%
Booij%
, Ris%
\BCBL {}\ \BBA {} Holthuijsen%
}{%
Booij%
\ \protect \BOthers {.}}{%
{\protect \APACyear {1999}}%
}]{%
booij1999third}
\APACinsertmetastar {%
booij1999third}%
\begin{APACrefauthors}%
Booij, N.%
, Ris, R\BPBI C.%
\BCBL {}\ \BBA {} Holthuijsen, L\BPBI H.%
\end{APACrefauthors}%
\unskip\
\newblock
\APACrefYearMonthDay{1999}{}{}.
\newblock
{\BBOQ}\APACrefatitle {A third-generation wave model for coastal regions: 1. Model description and validation} {A third-generation wave model for coastal regions: 1. model description and validation}.{\BBCQ}
\newblock
\APACjournalVolNumPages{Journal of geophysical research: Oceans}{104}{C4}{7649--7666}.
\PrintBackRefs{\CurrentBib}

\bibitem [\protect \citeauthoryear {%
Botter%
, Peratoner%
, Porporato%
, Rodriguez-Iturbe%
\BCBL {}\ \BBA {} Rinaldo%
}{%
Botter%
, Peratoner%
\BCBL {}\ \protect \BOthers {.}}{%
{\protect \APACyear {2007}}%
}]{%
botter2007signatures}
\APACinsertmetastar {%
botter2007signatures}%
\begin{APACrefauthors}%
Botter, G.%
, Peratoner, F.%
, Porporato, A.%
, Rodriguez-Iturbe, I.%
\BCBL {}\ \BBA {} Rinaldo, A.%
\end{APACrefauthors}%
\unskip\
\newblock
\APACrefYearMonthDay{2007}{}{}.
\newblock
{\BBOQ}\APACrefatitle {Signatures of large-scale soil moisture dynamics on streamflow statistics across US climate regimes} {Signatures of large-scale soil moisture dynamics on streamflow statistics across us climate regimes}.{\BBCQ}
\newblock
\APACjournalVolNumPages{Water resources research}{43}{11}{}.
\PrintBackRefs{\CurrentBib}

\bibitem [\protect \citeauthoryear {%
Botter%
, Porporato%
, Rodriguez-Iturbe%
\BCBL {}\ \BBA {} Rinaldo%
}{%
Botter%
, Porporato%
\BCBL {}\ \protect \BOthers {.}}{%
{\protect \APACyear {2007}}%
}]{%
botter2007basin}
\APACinsertmetastar {%
botter2007basin}%
\begin{APACrefauthors}%
Botter, G.%
, Porporato, A.%
, Rodriguez-Iturbe, I.%
\BCBL {}\ \BBA {} Rinaldo, A.%
\end{APACrefauthors}%
\unskip\
\newblock
\APACrefYearMonthDay{2007}{}{}.
\newblock
{\BBOQ}\APACrefatitle {Basin-scale soil moisture dynamics and the probabilistic characterization of carrier hydrologic flows: Slow, leaching-prone components of the hydrologic response} {Basin-scale soil moisture dynamics and the probabilistic characterization of carrier hydrologic flows: Slow, leaching-prone components of the hydrologic response}.{\BBCQ}
\newblock
\APACjournalVolNumPages{Water resources research}{43}{2}{{W02}{417}}.
\PrintBackRefs{\CurrentBib}

\bibitem [\protect \citeauthoryear {%
Brunner%
}{%
Brunner%
}{%
{\protect \APACyear {2021}}%
}]{%
brunner2021hecras}
\APACinsertmetastar {%
brunner2021hecras}%
\begin{APACrefauthors}%
Brunner, G\BPBI W.%
\end{APACrefauthors}%
\unskip\
\newblock
\APACrefYearMonthDay{2021}{}{}.
\newblock
\APACrefbtitle {HEC-RAS, River Analysis System: User's Manual, Version 6.1} {Hec-ras, river analysis system: User's manual, version 6.1}\ \APACbVolEdTR{}{\BTR{}\ \BNUM\ CPD-68}.
\newblock
\APACaddressInstitution{Davis, CA}{US Army Corps of Engineers, Hydrologic Engineering Center}.
\newblock
\begin{APACrefURL} \url{https://army.mil} \end{APACrefURL}
\PrintBackRefs{\CurrentBib}

\bibitem [\protect \citeauthoryear {%
Burnash%
}{%
Burnash%
}{%
{\protect \APACyear {1973}}%
}]{%
burnash1973generalized}
\APACinsertmetastar {%
burnash1973generalized}%
\begin{APACrefauthors}%
Burnash, R\BPBI J.%
\end{APACrefauthors}%
\unskip\
\newblock
\APACrefYear{1973}.
\newblock
\APACrefbtitle {A generalized streamflow simulation system: Conceptual modeling for digital computers} {A generalized streamflow simulation system: Conceptual modeling for digital computers}.
\newblock
\APACaddressPublisher{}{US Department of Commerce, National Weather Service, and State of California~…}.
\PrintBackRefs{\CurrentBib}

\bibitem [\protect \citeauthoryear {%
Cialone%
\ \protect \BOthers {.}}{%
Cialone%
\ \protect \BOthers {.}}{%
{\protect \APACyear {2015}}%
}]{%
cialone2015north}
\APACinsertmetastar {%
cialone2015north}%
\begin{APACrefauthors}%
Cialone, M\BPBI A.%
, Massey, T\BPBI C.%
, Anderson, M\BPBI E.%
, Grzegorzewski, A\BPBI S.%
, Jensen, R\BPBI E.%
, Cialone, A.%
\BDBL {}others%
\end{APACrefauthors}%
\unskip\
\newblock
\APACrefYearMonthDay{2015}{}{}.
\newblock
\APACrefbtitle {North Atlantic Coast Comprehensive Study (NACCS) Coastal Storm Model Simulations: Waves and Water Levels} {North atlantic coast comprehensive study (naccs) coastal storm model simulations: Waves and water levels}\ \APACbVolEdTR{}{\BTR{}}.
\newblock
\APACaddressInstitution{Vicksburg, Mississippi}{U.S. Army Engineer Research and Development Center, Coastal and Hydraulics Laboratory}.
\newblock
\begin{APACrefURL} \url{https://usace.contentdm.oclc.org/digital/collection/p266001coll1/id/3681/} \end{APACrefURL}
\PrintBackRefs{\CurrentBib}

\bibitem [\protect \citeauthoryear {%
Clark%
}{%
Clark%
}{%
{\protect \APACyear {1945}}%
}]{%
clark1945storage}
\APACinsertmetastar {%
clark1945storage}%
\begin{APACrefauthors}%
Clark, C.%
\end{APACrefauthors}%
\unskip\
\newblock
\APACrefYearMonthDay{1945}{}{}.
\newblock
{\BBOQ}\APACrefatitle {Storage and the unit hydrograph} {Storage and the unit hydrograph}.{\BBCQ}
\newblock
\APACjournalVolNumPages{Transactions of the American Society of Civil Engineers}{110}{1}{1419--1446}.
\PrintBackRefs{\CurrentBib}

\bibitem [\protect \citeauthoryear {%
Cobell%
\ \BBA {} Roberts%
}{%
Cobell%
\ \BBA {} Roberts%
}{%
{\protect \APACyear {2021}}%
}]{%
cobell2023coastal}
\APACinsertmetastar {%
cobell2023coastal}%
\begin{APACrefauthors}%
Cobell, Z.%
\BCBT {}\ \BBA {} Roberts, H.%
\end{APACrefauthors}%
\unskip\
\newblock
\APACrefYearMonthDay{2021}{}{}.
\newblock
\APACrefbtitle {2023 Coastal Master Plan: Storm Surge and Waves Model Improvements} {2023 coastal master plan: Storm surge and waves model improvements}\ \APACbVolEdTR{}{\BTR{}}.
\newblock
\APACaddressInstitution{}{Coastal Protection and Restoration Authority of Louisiana}.
\newblock
\begin{APACrefURL} \url{https://coastal.la.gov/wp-content/uploads/2021/03/StormSurge_Waves_Report_Jan2021.pdf} \end{APACrefURL}
\PrintBackRefs{\CurrentBib}

\bibitem [\protect \citeauthoryear {%
Couasnon%
\ \protect \BOthers {.}}{%
Couasnon%
\ \protect \BOthers {.}}{%
{\protect \APACyear {2020}}%
}]{%
couasnon2020measuring}
\APACinsertmetastar {%
couasnon2020measuring}%
\begin{APACrefauthors}%
Couasnon, A.%
, Eilander, D.%
, Muis, S.%
, Veldkamp, T\BPBI I\BPBI E.%
, Haigh, I\BPBI D.%
, Wahl, T.%
\BDBL {}Ward, P\BPBI J.%
\end{APACrefauthors}%
\unskip\
\newblock
\APACrefYearMonthDay{2020}{}{}.
\newblock
{\BBOQ}\APACrefatitle {Measuring Compound Flood Potential from River Discharge and Storm Surge Extremes at the Global Scale} {Measuring compound flood potential from river discharge and storm surge extremes at the global scale}.{\BBCQ}
\newblock
\APACjournalVolNumPages{Natural Hazards and Earth System Sciences}{20}{2}{489--504}.
\newblock
\begin{APACrefURL} \url{https://nhess.copernicus.org/articles/20/489/2020/} \end{APACrefURL}
\newblock
\begin{APACrefDOI} \doi{10.5194/nhess-20-489-2020} \end{APACrefDOI}
\PrintBackRefs{\CurrentBib}

\bibitem [\protect \citeauthoryear {%
Danielson%
, Poppenga%
, Tyler%
, Palaseanu-Lovejoy%
\BCBL {}\ \BBA {} Gesch%
}{%
Danielson%
\ \protect \BOthers {.}}{%
{\protect \APACyear {2018}}%
}]{%
danielson2018coastal}
\APACinsertmetastar {%
danielson2018coastal}%
\begin{APACrefauthors}%
Danielson, J\BPBI J.%
, Poppenga, S\BPBI K.%
, Tyler, D\BPBI J.%
, Palaseanu-Lovejoy, M.%
\BCBL {}\ \BBA {} Gesch, D\BPBI B.%
\end{APACrefauthors}%
\unskip\
\newblock
\APACrefYearMonthDay{2018}{}{}.
\newblock
\APACrefbtitle {Coastal National Elevation Database} {Coastal national elevation database}\ \APACbVolEdTR {}{Fact Sheet\ \BNUM\ 2018-3037}.
\newblock
\APACaddressInstitution{}{U.S. Geological Survey}.
\newblock
\begin{APACrefURL} \url{https://pubs.er.usgs.gov/publication/fs20183037} \end{APACrefURL}
\newblock
\begin{APACrefDOI} \doi{10.3133/fs20183037} \end{APACrefDOI}
\PrintBackRefs{\CurrentBib}

\bibitem [\protect \citeauthoryear {%
de Moel%
\ \BBA {} Aerts%
}{%
de Moel%
\ \BBA {} Aerts%
}{%
{\protect \APACyear {2011}}%
}]{%
demoel2011effect}
\APACinsertmetastar {%
demoel2011effect}%
\begin{APACrefauthors}%
de Moel, H.%
\BCBT {}\ \BBA {} Aerts, J.%
\end{APACrefauthors}%
\unskip\
\newblock
\APACrefYearMonthDay{2011}{}{}.
\newblock
{\BBOQ}\APACrefatitle {Effect of uncertainty in land use, damage models and inundation depth on flood damage estimates} {Effect of uncertainty in land use, damage models and inundation depth on flood damage estimates}.{\BBCQ}
\newblock
\APACjournalVolNumPages{Natural Hazards}{58}{}{407--425}.
\newblock
\begin{APACrefDOI} \doi{10.1007/s11069-010-9675-6} \end{APACrefDOI}
\PrintBackRefs{\CurrentBib}

\bibitem [\protect \citeauthoryear {%
De~Moel%
\ \protect \BOthers {.}}{%
De~Moel%
\ \protect \BOthers {.}}{%
{\protect \APACyear {2015}}%
}]{%
de2015flood}
\APACinsertmetastar {%
de2015flood}%
\begin{APACrefauthors}%
De~Moel, H.%
, Jongman, B.%
, Kreibich, H.%
, Merz, B.%
, Penning-Rowsell, E.%
\BCBL {}\ \BBA {} Ward, P\BPBI J.%
\end{APACrefauthors}%
\unskip\
\newblock
\APACrefYearMonthDay{2015}{}{}.
\newblock
{\BBOQ}\APACrefatitle {Flood risk assessments at different spatial scales} {Flood risk assessments at different spatial scales}.{\BBCQ}
\newblock
\APACjournalVolNumPages{Mitigation and Adaptation Strategies for Global Change}{20}{6}{865--890}.
\PrintBackRefs{\CurrentBib}

\bibitem [\protect \citeauthoryear {%
{Dewberry Engineers Inc.}%
}{%
{Dewberry Engineers Inc.}%
}{%
{\protect \APACyear {2019}}%
}]{%
dewberry2019amite}
\APACinsertmetastar {%
dewberry2019amite}%
\begin{APACrefauthors}%
{Dewberry Engineers Inc.}%
\end{APACrefauthors}%
\unskip\
\newblock
\APACrefYearMonthDay{2019}{mar}{25}.
\newblock
\APACrefbtitle {Amite River Basin Numerical Model Project Report} {Amite river basin numerical model project report}\ \APACbVolEdTR {}{Project Report}.
\newblock
\APACaddressInstitution{Baton Rouge, LA}{Louisiana Department of Transportation \& Development (LA DOTD)}.
\PrintBackRefs{\CurrentBib}

\bibitem [\protect \citeauthoryear {%
J.~Dietrich%
\ \protect \BOthers {.}}{%
J.~Dietrich%
\ \protect \BOthers {.}}{%
{\protect \APACyear {2011}}%
}]{%
dietrich2011modeling}
\APACinsertmetastar {%
dietrich2011modeling}%
\begin{APACrefauthors}%
Dietrich, J.%
, Zijlema, M.%
, Westerink, J.%
, Holthuijsen, L.%
, Dawson, C.%
, Luettich~Jr, R.%
\BDBL {}Stone, G.%
\end{APACrefauthors}%
\unskip\
\newblock
\APACrefYearMonthDay{2011}{}{}.
\newblock
{\BBOQ}\APACrefatitle {Modeling hurricane waves and storm surge using integrally-coupled, scalable computations} {Modeling hurricane waves and storm surge using integrally-coupled, scalable computations}.{\BBCQ}
\newblock
\APACjournalVolNumPages{Coastal Engineering}{58}{1}{45--65}.
\PrintBackRefs{\CurrentBib}

\bibitem [\protect \citeauthoryear {%
J\BPBI C.~Dietrich%
\ \protect \BOthers {.}}{%
J\BPBI C.~Dietrich%
\ \protect \BOthers {.}}{%
{\protect \APACyear {2012}}%
}]{%
dietrich2012performance}
\APACinsertmetastar {%
dietrich2012performance}%
\begin{APACrefauthors}%
Dietrich, J\BPBI C.%
, Tanaka, S.%
, Westerink, J\BPBI J.%
, Dawson, C\BPBI N.%
, Luettich~Jr, R\BPBI A.%
, Zijlema, M.%
\BDBL {}Westerink, H.%
\end{APACrefauthors}%
\unskip\
\newblock
\APACrefYearMonthDay{2012}{}{}.
\newblock
{\BBOQ}\APACrefatitle {Performance of the unstructured-mesh, SWAN+ ADCIRC model in computing hurricane waves and surge} {Performance of the unstructured-mesh, swan+ adcirc model in computing hurricane waves and surge}.{\BBCQ}
\newblock
\APACjournalVolNumPages{Journal of Scientific Computing}{52}{2}{468--497}.
\PrintBackRefs{\CurrentBib}

\bibitem [\protect \citeauthoryear {%
Eagleson%
}{%
Eagleson%
}{%
{\protect \APACyear {1978}}%
}]{%
eagleson1978climate2}
\APACinsertmetastar {%
eagleson1978climate2}%
\begin{APACrefauthors}%
Eagleson, P\BPBI S.%
\end{APACrefauthors}%
\unskip\
\newblock
\APACrefYearMonthDay{1978}{}{}.
\newblock
{\BBOQ}\APACrefatitle {Climate, soil, and vegetation: 2. The distribution of annual precipitation derived from observed storm sequences} {Climate, soil, and vegetation: 2. the distribution of annual precipitation derived from observed storm sequences}.{\BBCQ}
\newblock
\APACjournalVolNumPages{Water Resources Research}{14}{5}{713--721}.
\PrintBackRefs{\CurrentBib}

\bibitem [\protect \citeauthoryear {%
Eckhardt%
}{%
Eckhardt%
}{%
{\protect \APACyear {2005}}%
}]{%
eckhardt2005construct}
\APACinsertmetastar {%
eckhardt2005construct}%
\begin{APACrefauthors}%
Eckhardt, K.%
\end{APACrefauthors}%
\unskip\
\newblock
\APACrefYearMonthDay{2005}{}{}.
\newblock
{\BBOQ}\APACrefatitle {How to construct recursive digital filters for baseflow separation} {How to construct recursive digital filters for baseflow separation}.{\BBCQ}
\newblock
\APACjournalVolNumPages{Hydrological Processes: An International Journal}{19}{2}{507--515}.
\PrintBackRefs{\CurrentBib}

\bibitem [\protect \citeauthoryear {%
Eckhardt%
}{%
Eckhardt%
}{%
{\protect \APACyear {2008}}%
}]{%
eckhardt2008comparison}
\APACinsertmetastar {%
eckhardt2008comparison}%
\begin{APACrefauthors}%
Eckhardt, K.%
\end{APACrefauthors}%
\unskip\
\newblock
\APACrefYearMonthDay{2008}{}{}.
\newblock
{\BBOQ}\APACrefatitle {A comparison of baseflow indices, which were calculated with seven different baseflow separation methods} {A comparison of baseflow indices, which were calculated with seven different baseflow separation methods}.{\BBCQ}
\newblock
\APACjournalVolNumPages{Journal of hydrology}{352}{1-2}{168--173}.
\PrintBackRefs{\CurrentBib}

\bibitem [\protect \citeauthoryear {%
Emanuel%
, Sundararajan%
\BCBL {}\ \BBA {} Williams%
}{%
Emanuel%
\ \protect \BOthers {.}}{%
{\protect \APACyear {2008}}%
}]{%
emanuel2008hurricanes}
\APACinsertmetastar {%
emanuel2008hurricanes}%
\begin{APACrefauthors}%
Emanuel, K.%
, Sundararajan, R.%
\BCBL {}\ \BBA {} Williams, J.%
\end{APACrefauthors}%
\unskip\
\newblock
\APACrefYearMonthDay{2008}{}{}.
\newblock
{\BBOQ}\APACrefatitle {Hurricanes and global warming: Results from downscaling IPCC AR4 simulations} {Hurricanes and global warming: Results from downscaling ipcc ar4 simulations}.{\BBCQ}
\newblock
\APACjournalVolNumPages{Bulletin of the American Meteorological Society}{89}{3}{347--368}.
\newblock
\begin{APACrefDOI} \doi{10.1175/BAMS-89-3-347} \end{APACrefDOI}
\PrintBackRefs{\CurrentBib}

\bibitem [\protect \citeauthoryear {%
Fall%
\ \protect \BOthers {.}}{%
Fall%
\ \protect \BOthers {.}}{%
{\protect \APACyear {2023}}%
}]{%
fall2023office}
\APACinsertmetastar {%
fall2023office}%
\begin{APACrefauthors}%
Fall, G.%
, Kitzmiller, D.%
, Pavlovic, S.%
, Zhang, Z.%
, Patrick, N.%
, St.~Laurent, M.%
\BDBL {}Miller, D.%
\end{APACrefauthors}%
\unskip\
\newblock
\APACrefYearMonthDay{2023}{}{}.
\newblock
{\BBOQ}\APACrefatitle {The Office of Water Prediction's Analysis of Record for Calibration, version 1.1: Dataset description and precipitation evaluation} {The office of water prediction's analysis of record for calibration, version 1.1: Dataset description and precipitation evaluation}.{\BBCQ}
\newblock
\APACjournalVolNumPages{JAWRA Journal of the American Water Resources Association}{59}{6}{1246--1272}.
\PrintBackRefs{\CurrentBib}

\bibitem [\protect \citeauthoryear {%
{Federal Emergency Management Agency}%
}{%
{Federal Emergency Management Agency}%
}{%
{\protect \APACyear {2023}}%
}]{%
FEMA2023Coastal}
\APACinsertmetastar {%
FEMA2023Coastal}%
\begin{APACrefauthors}%
{Federal Emergency Management Agency}.%
\end{APACrefauthors}%
\unskip\
\newblock
\APACrefYearMonthDay{2023}{November}{}.
\newblock
\APACrefbtitle {Coastal Statistical Simulation Methods} {Coastal statistical simulation methods}\ \APACbVolEdTR{}{\BTR{}}.
\newblock
\APACaddressInstitution{}{FEMA}.
\newblock
\begin{APACrefURL} \url{https://www.fema.gov/sites/default/files/documents/Coastal_Statistical_Simulation_Methods_Nov_2023.pdf} \end{APACrefURL}
\PrintBackRefs{\CurrentBib}

\bibitem [\protect \citeauthoryear {%
Feng%
\ \protect \BOthers {.}}{%
Feng%
\ \protect \BOthers {.}}{%
{\protect \APACyear {2022}}%
}]{%
feng2022investigating}
\APACinsertmetastar {%
feng2022investigating}%
\begin{APACrefauthors}%
Feng, D.%
, Tan, Z.%
, Engwirda, D.%
, Liao, C.%
, Xu, D.%
, Bisht, G.%
\BDBL {}Leung, L\BPBI R.%
\end{APACrefauthors}%
\unskip\
\newblock
\APACrefYearMonthDay{2022}{}{}.
\newblock
{\BBOQ}\APACrefatitle {Investigating coastal backwater effects and flooding in the coastal zone using a global river transport model on an unstructured mesh} {Investigating coastal backwater effects and flooding in the coastal zone using a global river transport model on an unstructured mesh}.{\BBCQ}
\newblock
\APACjournalVolNumPages{Hydrology and Earth System Sciences}{26}{21}{5473--5491}.
\PrintBackRefs{\CurrentBib}

\bibitem [\protect \citeauthoryear {%
Ganguli%
\ \BBA {} Reddy%
}{%
Ganguli%
\ \BBA {} Reddy%
}{%
{\protect \APACyear {2013}}%
}]{%
ganguli2013probabilistic}
\APACinsertmetastar {%
ganguli2013probabilistic}%
\begin{APACrefauthors}%
Ganguli, P.%
\BCBT {}\ \BBA {} Reddy, M\BPBI J.%
\end{APACrefauthors}%
\unskip\
\newblock
\APACrefYearMonthDay{2013}{}{}.
\newblock
{\BBOQ}\APACrefatitle {Probabilistic assessment of flood risks using trivariate copulas} {Probabilistic assessment of flood risks using trivariate copulas}.{\BBCQ}
\newblock
\APACjournalVolNumPages{Theoretical and Applied Climatology}{111}{1-2}{341--360}.
\newblock
\begin{APACrefDOI} \doi{10.1007/s00704-012-0664-4} \end{APACrefDOI}
\PrintBackRefs{\CurrentBib}

\bibitem [\protect \citeauthoryear {%
Ghanbari%
, Arabi%
, Kao%
, Obeysekera%
\BCBL {}\ \BBA {} Sweet%
}{%
Ghanbari%
\ \protect \BOthers {.}}{%
{\protect \APACyear {2021}}%
}]{%
ghanbari2021climate}
\APACinsertmetastar {%
ghanbari2021climate}%
\begin{APACrefauthors}%
Ghanbari, M.%
, Arabi, M.%
, Kao, S\BHBI C.%
, Obeysekera, J.%
\BCBL {}\ \BBA {} Sweet, W.%
\end{APACrefauthors}%
\unskip\
\newblock
\APACrefYearMonthDay{2021}{}{}.
\newblock
{\BBOQ}\APACrefatitle {Climate Change and Changes in Compound Coastal-Riverine Flooding Hazard Along the US Coasts} {Climate change and changes in compound coastal-riverine flooding hazard along the us coasts}.{\BBCQ}
\newblock
\APACjournalVolNumPages{Earth's Future}{9}{5}{e2021EF002055}.
\newblock
\begin{APACrefURL} \url{https://agupubs.onlinelibrary.wiley.com/doi/10.1029/2021EF002055} \end{APACrefURL}
\newblock
\begin{APACrefDOI} \doi{10.1029/2021EF002055} \end{APACrefDOI}
\PrintBackRefs{\CurrentBib}

\bibitem [\protect \citeauthoryear {%
Gonzalez%
, Nadal-Caraballo%
, Melby%
\BCBL {}\ \BBA {} Cialone%
}{%
Gonzalez%
\ \protect \BOthers {.}}{%
{\protect \APACyear {2019}}%
}]{%
gonzalez2019quantification}
\APACinsertmetastar {%
gonzalez2019quantification}%
\begin{APACrefauthors}%
Gonzalez, V\BPBI M.%
, Nadal-Caraballo, N\BPBI C.%
, Melby, J\BPBI A.%
\BCBL {}\ \BBA {} Cialone, M\BPBI A.%
\end{APACrefauthors}%
\unskip\
\newblock
\APACrefYearMonthDay{2019}{}{}.
\newblock
\APACrefbtitle {Quantification of uncertainty in probabilistic storm surge models: literature review} {Quantification of uncertainty in probabilistic storm surge models: literature review}\ \APACbVolEdTR{}{\BTR{}\ \BNUM\ ERDC-CHL SR-19-1}.
\newblock
\APACaddressInstitution{Vicksburg, MS}{U.S. Army Engineer Research and Development Center, Coastal and Hydraulics Laboratory}.
\newblock
\begin{APACrefURL} \url{https://erdc-library.erdc.dren.mil/jspui/handle/11681/32295} \end{APACrefURL}
\PrintBackRefs{\CurrentBib}

\bibitem [\protect \citeauthoryear {%
Gori%
\ \BBA {} Lin%
}{%
Gori%
\ \BBA {} Lin%
}{%
{\protect \APACyear {2022}}%
}]{%
gori2022jpmosbq}
\APACinsertmetastar {%
gori2022jpmosbq}%
\begin{APACrefauthors}%
Gori, A.%
\BCBT {}\ \BBA {} Lin, N.%
\end{APACrefauthors}%
\unskip\
\newblock
\APACrefYearMonthDay{2022}{}{}.
\newblock
{\BBOQ}\APACrefatitle {Projecting compound flood hazard under climate change with physical models and joint probability methods} {Projecting compound flood hazard under climate change with physical models and joint probability methods}.{\BBCQ}
\newblock
\APACjournalVolNumPages{Earth's Future}{10}{}{1--19}.
\newblock
\begin{APACrefDOI} \doi{10.1029/2022EF003097} \end{APACrefDOI}
\PrintBackRefs{\CurrentBib}

\bibitem [\protect \citeauthoryear {%
Gori%
, Lin%
\BCBL {}\ \BBA {} Smith%
}{%
Gori%
, Lin%
\BCBL {}\ \BBA {} Smith%
}{%
{\protect \APACyear {2020}}%
}]{%
gori2020assessing}
\APACinsertmetastar {%
gori2020assessing}%
\begin{APACrefauthors}%
Gori, A.%
, Lin, N.%
\BCBL {}\ \BBA {} Smith, J.%
\end{APACrefauthors}%
\unskip\
\newblock
\APACrefYearMonthDay{2020}{}{}.
\newblock
{\BBOQ}\APACrefatitle {Assessing compound flooding from landfalling tropical cyclones on the North Carolina coast} {Assessing compound flooding from landfalling tropical cyclones on the north carolina coast}.{\BBCQ}
\newblock
\APACjournalVolNumPages{Water Resources Research}{56}{4}{e2019WR026788}.
\PrintBackRefs{\CurrentBib}

\bibitem [\protect \citeauthoryear {%
Gori%
, Lin%
\BCBL {}\ \BBA {} Xi%
}{%
Gori%
, Lin%
\BCBL {}\ \BBA {} Xi%
}{%
{\protect \APACyear {2020}}%
}]{%
gori2020tropical}
\APACinsertmetastar {%
gori2020tropical}%
\begin{APACrefauthors}%
Gori, A.%
, Lin, N.%
\BCBL {}\ \BBA {} Xi, D.%
\end{APACrefauthors}%
\unskip\
\newblock
\APACrefYearMonthDay{2020}{}{}.
\newblock
{\BBOQ}\APACrefatitle {Tropical cyclone compound flood hazard assessment: From investigating drivers to quantifying extreme water levels} {Tropical cyclone compound flood hazard assessment: From investigating drivers to quantifying extreme water levels}.{\BBCQ}
\newblock
\APACjournalVolNumPages{Earth's Future}{8}{12}{e2020EF001660}.
\PrintBackRefs{\CurrentBib}

\bibitem [\protect \citeauthoryear {%
Gr{\"a}ler%
\ \protect \BOthers {.}}{%
Gr{\"a}ler%
\ \protect \BOthers {.}}{%
{\protect \APACyear {2013}}%
}]{%
graler2013multivariate}
\APACinsertmetastar {%
graler2013multivariate}%
\begin{APACrefauthors}%
Gr{\"a}ler, B.%
, van~den BERG, M\BPBI J.%
, Vandenberghe, S.%
, Petroselli, A.%
, Grimaldi, S.%
, De~Baets, B.%
\BCBL {}\ \BBA {} Verhoest, N.%
\end{APACrefauthors}%
\unskip\
\newblock
\APACrefYearMonthDay{2013}{}{}.
\newblock
{\BBOQ}\APACrefatitle {Multivariate return periods in hydrology: a critical and practical review focusing on synthetic design hydrograph estimation} {Multivariate return periods in hydrology: a critical and practical review focusing on synthetic design hydrograph estimation}.{\BBCQ}
\newblock
\APACjournalVolNumPages{Hydrology and Earth System Sciences}{17}{4}{1281--1296}.
\PrintBackRefs{\CurrentBib}

\bibitem [\protect \citeauthoryear {%
Green%
\ \protect \BOthers {.}}{%
Green%
\ \protect \BOthers {.}}{%
{\protect \APACyear {2025}}%
}]{%
green2025comprehensive}
\APACinsertmetastar {%
green2025comprehensive}%
\begin{APACrefauthors}%
Green, J.%
, Haigh, I\BPBI D.%
, Quinn, N.%
, Neal, J.%
, Wahl, T.%
, Wood, M.%
\BDBL {}Camus, P.%
\end{APACrefauthors}%
\unskip\
\newblock
\APACrefYearMonthDay{2025}{}{}.
\newblock
{\BBOQ}\APACrefatitle {A comprehensive review of compound flooding literature with a focus on coastal and estuarine regions} {A comprehensive review of compound flooding literature with a focus on coastal and estuarine regions}.{\BBCQ}
\newblock
\APACjournalVolNumPages{Natural Hazards and Earth System Sciences}{25}{2}{747--816}.
\PrintBackRefs{\CurrentBib}

\bibitem [\protect \citeauthoryear {%
Habel%
, Fletcher%
, Anderson%
\BCBL {}\ \BBA {} Thompson%
}{%
Habel%
\ \protect \BOthers {.}}{%
{\protect \APACyear {2020}}%
}]{%
habel2020sea}
\APACinsertmetastar {%
habel2020sea}%
\begin{APACrefauthors}%
Habel, S.%
, Fletcher, C\BPBI H.%
, Anderson, T\BPBI R.%
\BCBL {}\ \BBA {} Thompson, P\BPBI R.%
\end{APACrefauthors}%
\unskip\
\newblock
\APACrefYearMonthDay{2020}{}{}.
\newblock
{\BBOQ}\APACrefatitle {Sea-Level Rise Induced Multi-Mechanism Flooding and Contribution to Urban Infrastructure Failure} {Sea-level rise induced multi-mechanism flooding and contribution to urban infrastructure failure}.{\BBCQ}
\newblock
\APACjournalVolNumPages{Scientific Reports}{10}{1}{3796}.
\newblock
\begin{APACrefURL} \url{https://www.nature.com/articles/s41598-020-60762-4} \end{APACrefURL}
\newblock
\begin{APACrefDOI} \doi{10.1038/s41598-020-60762-4} \end{APACrefDOI}
\PrintBackRefs{\CurrentBib}

\bibitem [\protect \citeauthoryear {%
Han%
\ \BBA {} Tahvildari%
}{%
Han%
\ \BBA {} Tahvildari%
}{%
{\protect \APACyear {2024}}%
}]{%
han2024compound}
\APACinsertmetastar {%
han2024compound}%
\begin{APACrefauthors}%
Han, S.%
\BCBT {}\ \BBA {} Tahvildari, N.%
\end{APACrefauthors}%
\unskip\
\newblock
\APACrefYearMonthDay{2024}{}{}.
\newblock
{\BBOQ}\APACrefatitle {Compound flooding hazards due to storm surge and pluvial flow in a low-gradient coastal region} {Compound flooding hazards due to storm surge and pluvial flow in a low-gradient coastal region}.{\BBCQ}
\newblock
\APACjournalVolNumPages{Water Resources Research}{60}{11}{e2023WR037014}.
\PrintBackRefs{\CurrentBib}

\bibitem [\protect \citeauthoryear {%
Hinkel%
\ \protect \BOthers {.}}{%
Hinkel%
\ \protect \BOthers {.}}{%
{\protect \APACyear {2021}}%
}]{%
hinkel2021uncertainty}
\APACinsertmetastar {%
hinkel2021uncertainty}%
\begin{APACrefauthors}%
Hinkel, J.%
, Feyen, L.%
, Hemer, M.%
, Le~Cozannet, G.%
, Lincke, D.%
, Marcos, M.%
\BDBL {}Wolff, C.%
\end{APACrefauthors}%
\unskip\
\newblock
\APACrefYearMonthDay{2021}{}{}.
\newblock
{\BBOQ}\APACrefatitle {Uncertainty and bias in global to regional scale assessments of current and future coastal flood risk} {Uncertainty and bias in global to regional scale assessments of current and future coastal flood risk}.{\BBCQ}
\newblock
\APACjournalVolNumPages{Earth's Future}{9}{7}{e2020EF001882}.
\newblock
\begin{APACrefDOI} \doi{10.1029/2020EF001882} \end{APACrefDOI}
\PrintBackRefs{\CurrentBib}

\bibitem [\protect \citeauthoryear {%
Hsiao%
\ \protect \BOthers {.}}{%
Hsiao%
\ \protect \BOthers {.}}{%
{\protect \APACyear {2021}}%
}]{%
hsiao2021flood}
\APACinsertmetastar {%
hsiao2021flood}%
\begin{APACrefauthors}%
Hsiao, S\BHBI C.%
, Chiang, W\BHBI S.%
, Jang, J\BHBI H.%
, Wu, H\BHBI L.%
, Lu, W\BHBI S.%
, Chen, W\BHBI B.%
\BCBL {}\ \BBA {} Wu, Y\BHBI T.%
\end{APACrefauthors}%
\unskip\
\newblock
\APACrefYearMonthDay{2021}{}{}.
\newblock
{\BBOQ}\APACrefatitle {Flood Risk Influenced by the Compound Effect of Storm Surge and Rainfall Under Climate Change for Low-Lying Coastal Areas} {Flood risk influenced by the compound effect of storm surge and rainfall under climate change for low-lying coastal areas}.{\BBCQ}
\newblock
\APACjournalVolNumPages{Science of The Total Environment}{764}{}{144439}.
\newblock
\begin{APACrefURL} \url{https://www.sciencedirect.com/science/article/pii/S0048969720376320} \end{APACrefURL}
\newblock
\begin{APACrefDOI} \doi{10.1016/j.scitotenv.2020.144439} \end{APACrefDOI}
\PrintBackRefs{\CurrentBib}

\bibitem [\protect \citeauthoryear {%
Jane%
, Cadavid%
, Obeysekera%
\BCBL {}\ \BBA {} Wahl%
}{%
Jane%
\ \protect \BOthers {.}}{%
{\protect \APACyear {2020}}%
}]{%
jane2020multivariate}
\APACinsertmetastar {%
jane2020multivariate}%
\begin{APACrefauthors}%
Jane, R.%
, Cadavid, L.%
, Obeysekera, J.%
\BCBL {}\ \BBA {} Wahl, T.%
\end{APACrefauthors}%
\unskip\
\newblock
\APACrefYearMonthDay{2020}{}{}.
\newblock
{\BBOQ}\APACrefatitle {Multivariate statistical modelling of the drivers of compound flood events in south Florida} {Multivariate statistical modelling of the drivers of compound flood events in south florida}.{\BBCQ}
\newblock
\APACjournalVolNumPages{Natural Hazards and Earth System Sciences}{20}{10}{2681--2699}.
\PrintBackRefs{\CurrentBib}

\bibitem [\protect \citeauthoryear {%
Jane%
, Wahl%
, Santos%
, Misra%
\BCBL {}\ \BBA {} White%
}{%
Jane%
\ \protect \BOthers {.}}{%
{\protect \APACyear {2022}}%
}]{%
jane2022assessing}
\APACinsertmetastar {%
jane2022assessing}%
\begin{APACrefauthors}%
Jane, R.%
, Wahl, T.%
, Santos, V\BPBI M.%
, Misra, S\BPBI K.%
\BCBL {}\ \BBA {} White, K\BPBI D.%
\end{APACrefauthors}%
\unskip\
\newblock
\APACrefYearMonthDay{2022}{}{}.
\newblock
{\BBOQ}\APACrefatitle {Assessing the potential for compound storm surge and extreme river discharge events at the catchment scale with statistical models: sensitivity analysis and recommendations for best practice} {Assessing the potential for compound storm surge and extreme river discharge events at the catchment scale with statistical models: sensitivity analysis and recommendations for best practice}.{\BBCQ}
\newblock
\APACjournalVolNumPages{Journal of Hydrologic Engineering}{27}{3}{}.
\newblock
\begin{APACrefDOI} \doi{10.1061/(ASCE)HE.1943-5584.0002154} \end{APACrefDOI}
\PrintBackRefs{\CurrentBib}

\bibitem [\protect \citeauthoryear {%
Johnson%
\ \protect \BOthers {.}}{%
Johnson%
\ \protect \BOthers {.}}{%
{\protect \APACyear {2023}}%
}]{%
johnson2023coastal}
\APACinsertmetastar {%
johnson2023coastal}%
\begin{APACrefauthors}%
Johnson, D.%
, Fischbach, J.%
, Geldner, N.%
, Wilson, M.%
, Story, C.%
\BCBL {}\ \BBA {} Wang, J.%
\end{APACrefauthors}%
\unskip\
\newblock
\APACrefYearMonthDay{2023}{}{}.
\newblock
{\BBOQ}\APACrefatitle {Coastal Master Plan: Attachment C11: 2023 Risk Model} {Coastal master plan: Attachment c11: 2023 risk model}.{\BBCQ}
\newblock
\APACjournalVolNumPages{Version}{3}{}{31}.
\PrintBackRefs{\CurrentBib}

\bibitem [\protect \citeauthoryear {%
Kavetski%
, Kuczera%
\BCBL {}\ \BBA {} Franks%
}{%
Kavetski%
\ \protect \BOthers {.}}{%
{\protect \APACyear {2003}}%
}]{%
kavetski2003semidistributed}
\APACinsertmetastar {%
kavetski2003semidistributed}%
\begin{APACrefauthors}%
Kavetski, D.%
, Kuczera, G.%
\BCBL {}\ \BBA {} Franks, S\BPBI W.%
\end{APACrefauthors}%
\unskip\
\newblock
\APACrefYearMonthDay{2003}{}{}.
\newblock
{\BBOQ}\APACrefatitle {Semidistributed hydrological modeling: A “saturation path” perspective on TOPMODEL and VIC} {Semidistributed hydrological modeling: A “saturation path” perspective on topmodel and vic}.{\BBCQ}
\newblock
\APACjournalVolNumPages{Water resources research}{39}{9}{}.
\PrintBackRefs{\CurrentBib}

\bibitem [\protect \citeauthoryear {%
Kheradmand%
, Seidou%
, Konte%
\BCBL {}\ \BBA {} Barmou~Batoure%
}{%
Kheradmand%
\ \protect \BOthers {.}}{%
{\protect \APACyear {2018}}%
}]{%
kheradmand2018evaluation}
\APACinsertmetastar {%
kheradmand2018evaluation}%
\begin{APACrefauthors}%
Kheradmand, S.%
, Seidou, O.%
, Konte, D.%
\BCBL {}\ \BBA {} Barmou~Batoure, M\BPBI B.%
\end{APACrefauthors}%
\unskip\
\newblock
\APACrefYearMonthDay{2018}{}{}.
\newblock
{\BBOQ}\APACrefatitle {Evaluation of adaptation options to flood risk in a probabilistic framework} {Evaluation of adaptation options to flood risk in a probabilistic framework}.{\BBCQ}
\newblock
\APACjournalVolNumPages{Journal of Hydrology: Regional Studies}{19}{}{1--16}.
\newblock
\begin{APACrefDOI} \doi{10.1016/j.ejrh.2018.07.001} \end{APACrefDOI}
\PrintBackRefs{\CurrentBib}

\bibitem [\protect \citeauthoryear {%
Kim%
\ \BBA {} Villarini%
}{%
Kim%
\ \BBA {} Villarini%
}{%
{\protect \APACyear {2022}}%
}]{%
kim2022evaluation}
\APACinsertmetastar {%
kim2022evaluation}%
\begin{APACrefauthors}%
Kim, H.%
\BCBT {}\ \BBA {} Villarini, G.%
\end{APACrefauthors}%
\unskip\
\newblock
\APACrefYearMonthDay{2022}{}{}.
\newblock
{\BBOQ}\APACrefatitle {Evaluation of the Analysis of Record for Calibration (AORC) rainfall across Louisiana} {Evaluation of the analysis of record for calibration (aorc) rainfall across louisiana}.{\BBCQ}
\newblock
\APACjournalVolNumPages{Remote Sensing}{14}{14}{3284}.
\PrintBackRefs{\CurrentBib}

\bibitem [\protect \citeauthoryear {%
Kim%
\ \protect \BOthers {.}}{%
Kim%
\ \protect \BOthers {.}}{%
{\protect \APACyear {2023}}%
}]{%
kim2023generation}
\APACinsertmetastar {%
kim2023generation}%
\begin{APACrefauthors}%
Kim, H.%
, Villarini, G.%
, Jane, R.%
, Wahl, T.%
, Misra, S.%
\BCBL {}\ \BBA {} Michalek, A.%
\end{APACrefauthors}%
\unskip\
\newblock
\APACrefYearMonthDay{2023}{}{}.
\newblock
{\BBOQ}\APACrefatitle {On the generation of high-resolution probabilistic design events capturing the joint occurrence of rainfall and storm surge in coastal basins} {On the generation of high-resolution probabilistic design events capturing the joint occurrence of rainfall and storm surge in coastal basins}.{\BBCQ}
\newblock
\APACjournalVolNumPages{International Journal of climatology}{43}{2}{761--771}.
\PrintBackRefs{\CurrentBib}

\bibitem [\protect \citeauthoryear {%
Kleiber%
, Sain%
, Madaus%
\BCBL {}\ \BBA {} Harr%
}{%
Kleiber%
\ \protect \BOthers {.}}{%
{\protect \APACyear {2023}}%
}]{%
kleiber2023stochastic}
\APACinsertmetastar {%
kleiber2023stochastic}%
\begin{APACrefauthors}%
Kleiber, W.%
, Sain, S.%
, Madaus, L.%
\BCBL {}\ \BBA {} Harr, P.%
\end{APACrefauthors}%
\unskip\
\newblock
\APACrefYearMonthDay{2023}{}{}.
\newblock
{\BBOQ}\APACrefatitle {Stochastic tropical cyclone precipitation field generation} {Stochastic tropical cyclone precipitation field generation}.{\BBCQ}
\newblock
\APACjournalVolNumPages{Environmetrics}{34}{1}{e2766}.
\PrintBackRefs{\CurrentBib}

\bibitem [\protect \citeauthoryear {%
Koster%
, Reichle%
, Schubert%
\BCBL {}\ \BBA {} Mahanama%
}{%
Koster%
\ \protect \BOthers {.}}{%
{\protect \APACyear {2019}}%
}]{%
koster2019length}
\APACinsertmetastar {%
koster2019length}%
\begin{APACrefauthors}%
Koster, R\BPBI D.%
, Reichle, R\BPBI H.%
, Schubert, S\BPBI D.%
\BCBL {}\ \BBA {} Mahanama, S\BPBI P.%
\end{APACrefauthors}%
\unskip\
\newblock
\APACrefYearMonthDay{2019}{}{}.
\newblock
{\BBOQ}\APACrefatitle {Length scales of hydrological variability as inferred from SMAP soil moisture retrievals} {Length scales of hydrological variability as inferred from smap soil moisture retrievals}.{\BBCQ}
\newblock
\APACjournalVolNumPages{Journal of Hydrometeorology}{20}{11}{2129--2146}.
\PrintBackRefs{\CurrentBib}

\bibitem [\protect \citeauthoryear {%
Landsea%
\ \BBA {} Franklin%
}{%
Landsea%
\ \BBA {} Franklin%
}{%
{\protect \APACyear {2013}}%
}]{%
landsea2013atlantic}
\APACinsertmetastar {%
landsea2013atlantic}%
\begin{APACrefauthors}%
Landsea, C\BPBI W.%
\BCBT {}\ \BBA {} Franklin, J\BPBI L.%
\end{APACrefauthors}%
\unskip\
\newblock
\APACrefYearMonthDay{2013}{}{}.
\newblock
{\BBOQ}\APACrefatitle {Atlantic Hurricane Database Uncertainty and Presentation of a New Database Format} {Atlantic hurricane database uncertainty and presentation of a new database format}.{\BBCQ}
\newblock
\APACjournalVolNumPages{Monthly Weather Review}{141}{10}{3576--3592}.
\newblock
\begin{APACrefURL} \url{https://journals.ametsoc.org/view/journals/mwre/141/10/mwr-d-12-00254.1.xml} \end{APACrefURL}
\newblock
\begin{APACrefDOI} \doi{10.1175/MWR-D-12-00254.1} \end{APACrefDOI}
\PrintBackRefs{\CurrentBib}

\bibitem [\protect \citeauthoryear {%
Leopold%
\ \BBA {} Maddock%
}{%
Leopold%
\ \BBA {} Maddock%
}{%
{\protect \APACyear {1953}}%
}]{%
leopold1953hydraulic}
\APACinsertmetastar {%
leopold1953hydraulic}%
\begin{APACrefauthors}%
Leopold, L\BPBI B.%
\BCBT {}\ \BBA {} Maddock, T.%
\end{APACrefauthors}%
\unskip\
\newblock
\APACrefYear{1953}.
\newblock
\APACrefbtitle {The hydraulic geometry of stream channels and some physiographic implications} {The hydraulic geometry of stream channels and some physiographic implications}\ (\BVOL~252).
\newblock
\APACaddressPublisher{}{US Government Printing Office}.
\PrintBackRefs{\CurrentBib}

\bibitem [\protect \citeauthoryear {%
Liang%
, Lettenmaier%
, Wood%
\BCBL {}\ \BBA {} Burges%
}{%
Liang%
\ \protect \BOthers {.}}{%
{\protect \APACyear {1994}}%
}]{%
liang1994simple}
\APACinsertmetastar {%
liang1994simple}%
\begin{APACrefauthors}%
Liang, X.%
, Lettenmaier, D\BPBI P.%
, Wood, E\BPBI F.%
\BCBL {}\ \BBA {} Burges, S\BPBI J.%
\end{APACrefauthors}%
\unskip\
\newblock
\APACrefYearMonthDay{1994}{}{}.
\newblock
{\BBOQ}\APACrefatitle {A simple hydrologically based model of land surface water and energy fluxes for general circulation models} {A simple hydrologically based model of land surface water and energy fluxes for general circulation models}.{\BBCQ}
\newblock
\APACjournalVolNumPages{Journal of Geophysical Research: Atmospheres (1984--2012)}{99}{D7}{14415--14428}.
\PrintBackRefs{\CurrentBib}

\bibitem [\protect \citeauthoryear {%
{Louisiana Watershed Initiative}%
}{%
{Louisiana Watershed Initiative}%
}{%
{\protect \APACyear {2026}}%
}]{%
lwi_endmc_2025}
\APACinsertmetastar {%
lwi_endmc_2025}%
\begin{APACrefauthors}%
{Louisiana Watershed Initiative}.%
\end{APACrefauthors}%
\unskip\
\newblock
\APACrefYearMonthDay{2026}{}{}.
\newblock
\APACrefbtitle {Louisiana {W}atershed {I}nitiative model data [{Data Catalog}].} {Louisiana {W}atershed {I}nitiative model data [{Data Catalog}].}
\newblock
\APAChowpublished {EnDMC: Environmental Data and Model Catalog, The Water Institute}.
\newblock
\begin{APACrefURL} \url{https://lwi.endmc.org/} \end{APACrefURL}
\PrintBackRefs{\CurrentBib}

\bibitem [\protect \citeauthoryear {%
Luettich~Jr%
\ \BBA {} Westerink%
}{%
Luettich~Jr%
\ \BBA {} Westerink%
}{%
{\protect \APACyear {1991}}%
}]{%
luettich1991solution}
\APACinsertmetastar {%
luettich1991solution}%
\begin{APACrefauthors}%
Luettich~Jr, R\BPBI A.%
\BCBT {}\ \BBA {} Westerink, J\BPBI J.%
\end{APACrefauthors}%
\unskip\
\newblock
\APACrefYearMonthDay{1991}{}{}.
\newblock
{\BBOQ}\APACrefatitle {A solution for the vertical variation of stress, rather than velocity, in a three-dimensional circulation model} {A solution for the vertical variation of stress, rather than velocity, in a three-dimensional circulation model}.{\BBCQ}
\newblock
\APACjournalVolNumPages{International Journal for Numerical Methods in Fluids}{12}{10}{911--928}.
\PrintBackRefs{\CurrentBib}

\bibitem [\protect \citeauthoryear {%
Maranzoni%
, D'Oria%
\BCBL {}\ \BBA {} Rizzo%
}{%
Maranzoni%
\ \protect \BOthers {.}}{%
{\protect \APACyear {2023}}%
}]{%
maranzoni2023quantitative}
\APACinsertmetastar {%
maranzoni2023quantitative}%
\begin{APACrefauthors}%
Maranzoni, A.%
, D'Oria, M.%
\BCBL {}\ \BBA {} Rizzo, C.%
\end{APACrefauthors}%
\unskip\
\newblock
\APACrefYearMonthDay{2023}{}{}.
\newblock
{\BBOQ}\APACrefatitle {Quantitative flood hazard assessment methods: A review} {Quantitative flood hazard assessment methods: A review}.{\BBCQ}
\newblock
\APACjournalVolNumPages{Journal of Flood Risk Management}{16}{1}{e12855}.
\PrintBackRefs{\CurrentBib}

\bibitem [\protect \citeauthoryear {%
Marijnissen%
, Kok%
, Kroeze%
\BCBL {}\ \BBA {} van Loon-Steensma%
}{%
Marijnissen%
\ \protect \BOthers {.}}{%
{\protect \APACyear {2019}}%
}]{%
marijnissen2019re}
\APACinsertmetastar {%
marijnissen2019re}%
\begin{APACrefauthors}%
Marijnissen, R.%
, Kok, M.%
, Kroeze, C.%
\BCBL {}\ \BBA {} van Loon-Steensma, J.%
\end{APACrefauthors}%
\unskip\
\newblock
\APACrefYearMonthDay{2019}{}{}.
\newblock
{\BBOQ}\APACrefatitle {Re-evaluating safety risks of multifunctional dikes with a probabilistic risk framework} {Re-evaluating safety risks of multifunctional dikes with a probabilistic risk framework}.{\BBCQ}
\newblock
\APACjournalVolNumPages{Natural Hazards and Earth System Sciences}{19}{4}{737--756}.
\newblock
\begin{APACrefDOI} \doi{10.5194/nhess-19-737-2019} \end{APACrefDOI}
\PrintBackRefs{\CurrentBib}

\bibitem [\protect \citeauthoryear {%
Moftakhari%
, Schubert%
, AghaKouchak%
, Matthew%
\BCBL {}\ \BBA {} Sanders%
}{%
Moftakhari%
\ \protect \BOthers {.}}{%
{\protect \APACyear {2019}}%
}]{%
moftakhari2019linking}
\APACinsertmetastar {%
moftakhari2019linking}%
\begin{APACrefauthors}%
Moftakhari, H.%
, Schubert, J\BPBI E.%
, AghaKouchak, A.%
, Matthew, R\BPBI A.%
\BCBL {}\ \BBA {} Sanders, B\BPBI F.%
\end{APACrefauthors}%
\unskip\
\newblock
\APACrefYearMonthDay{2019}{}{}.
\newblock
{\BBOQ}\APACrefatitle {Linking statistical and hydrodynamic modeling for compound flood hazard assessment in tidal channels and estuaries} {Linking statistical and hydrodynamic modeling for compound flood hazard assessment in tidal channels and estuaries}.{\BBCQ}
\newblock
\APACjournalVolNumPages{Advances in Water Resources}{128}{}{28--38}.
\PrintBackRefs{\CurrentBib}

\bibitem [\protect \citeauthoryear {%
Nadal-Caraballo%
\ \protect \BOthers {.}}{%
Nadal-Caraballo%
\ \protect \BOthers {.}}{%
{\protect \APACyear {2020}}%
}]{%
nadal2020coastal}
\APACinsertmetastar {%
nadal2020coastal}%
\begin{APACrefauthors}%
Nadal-Caraballo, N\BPBI C.%
, Campbell, M\BPBI O.%
, Gonzalez, V\BPBI M.%
, Torres, M\BPBI J.%
, Melby, J\BPBI A.%
\BCBL {}\ \BBA {} Taflanidis, A\BPBI A.%
\end{APACrefauthors}%
\unskip\
\newblock
\APACrefYearMonthDay{2020}{}{}.
\newblock
{\BBOQ}\APACrefatitle {Coastal hazards system: a probabilistic coastal hazard analysis framework} {Coastal hazards system: a probabilistic coastal hazard analysis framework}.{\BBCQ}
\newblock
\APACjournalVolNumPages{Journal of Coastal Research}{95}{SI}{1211--1216}.
\PrintBackRefs{\CurrentBib}

\bibitem [\protect \citeauthoryear {%
Nadal-Caraballo%
, Melby%
\BCBL {}\ \BBA {} Gonzalez%
}{%
Nadal-Caraballo%
\ \protect \BOthers {.}}{%
{\protect \APACyear {2016}}%
}]{%
nadal2016statistical}
\APACinsertmetastar {%
nadal2016statistical}%
\begin{APACrefauthors}%
Nadal-Caraballo, N\BPBI C.%
, Melby, J\BPBI A.%
\BCBL {}\ \BBA {} Gonzalez, V\BPBI M.%
\end{APACrefauthors}%
\unskip\
\newblock
\APACrefYearMonthDay{2016}{}{}.
\newblock
{\BBOQ}\APACrefatitle {Statistical analysis of historical extreme water levels for the US North Atlantic coast using Monte Carlo life-cycle simulation} {Statistical analysis of historical extreme water levels for the us north atlantic coast using monte carlo life-cycle simulation}.{\BBCQ}
\newblock
\APACjournalVolNumPages{Journal of Coastal Research}{32}{1}{35--45}.
\newblock
\begin{APACrefDOI} \doi{10.2112/JCOASTRES-D-15-00031.1} \end{APACrefDOI}
\PrintBackRefs{\CurrentBib}

\bibitem [\protect \citeauthoryear {%
Nadal-Caraballo%
\ \protect \BOthers {.}}{%
Nadal-Caraballo%
\ \protect \BOthers {.}}{%
{\protect \APACyear {2022}}%
}]{%
nadal2022coastal}
\APACinsertmetastar {%
nadal2022coastal}%
\begin{APACrefauthors}%
Nadal-Caraballo, N\BPBI C.%
, Yawn, M\BPBI C.%
, Aucoin, L\BPBI A.%
, Carr, M\BPBI L.%
, Melby, J\BPBI A.%
, Ramos-Santiago, E.%
\BDBL {}others%
\end{APACrefauthors}%
\unskip\
\newblock
\APACrefYearMonthDay{2022}{}{}.
\newblock
\APACrefbtitle {Coastal Hazards System--Louisiana (CHS-LA)} {Coastal hazards system--louisiana (chs-la)}\ \APACbVolEdTR{}{\BTR{}}.
\newblock
\APACaddressInstitution{}{Engineer Research and Development Center (US)}.
\PrintBackRefs{\CurrentBib}

\bibitem [\protect \citeauthoryear {%
Nasr%
, Wahl%
, Rashid%
, Camus%
\BCBL {}\ \BBA {} Haigh%
}{%
Nasr%
\ \protect \BOthers {.}}{%
{\protect \APACyear {2021}}%
}]{%
nasr2021assessing}
\APACinsertmetastar {%
nasr2021assessing}%
\begin{APACrefauthors}%
Nasr, A\BPBI A.%
, Wahl, T.%
, Rashid, M\BPBI M.%
, Camus, P.%
\BCBL {}\ \BBA {} Haigh, I\BPBI D.%
\end{APACrefauthors}%
\unskip\
\newblock
\APACrefYearMonthDay{2021}{}{}.
\newblock
{\BBOQ}\APACrefatitle {Assessing the Dependence Structure Between Oceanographic, Fluvial, and Pluvial Flooding Drivers Along the United States Coastline} {Assessing the dependence structure between oceanographic, fluvial, and pluvial flooding drivers along the united states coastline}.{\BBCQ}
\newblock
\APACjournalVolNumPages{Hydrology and Earth System Sciences}{25}{12}{6203--6222}.
\newblock
\begin{APACrefURL} \url{https://hess.copernicus.org/articles/25/6203/2021/} \end{APACrefURL}
\newblock
\begin{APACrefDOI} \doi{10.5194/hess-25-6203-2021} \end{APACrefDOI}
\PrintBackRefs{\CurrentBib}

\bibitem [\protect \citeauthoryear {%
Nofal%
\ \BBA {} Van De~Lindt%
}{%
Nofal%
\ \BBA {} Van De~Lindt%
}{%
{\protect \APACyear {2022}}%
}]{%
nofal2022understanding}
\APACinsertmetastar {%
nofal2022understanding}%
\begin{APACrefauthors}%
Nofal, O\BPBI M.%
\BCBT {}\ \BBA {} Van De~Lindt, J\BPBI W.%
\end{APACrefauthors}%
\unskip\
\newblock
\APACrefYearMonthDay{2022}{}{}.
\newblock
{\BBOQ}\APACrefatitle {Understanding flood risk in the context of community resilience modeling for the built environment: Research needs and trends} {Understanding flood risk in the context of community resilience modeling for the built environment: Research needs and trends}.{\BBCQ}
\newblock
\APACjournalVolNumPages{Sustainable and Resilient Infrastructure}{7}{3}{171--187}.
\PrintBackRefs{\CurrentBib}

\bibitem [\protect \citeauthoryear {%
Peña%
\ \protect \BOthers {.}}{%
Peña%
\ \protect \BOthers {.}}{%
{\protect \APACyear {2022}}%
}]{%
pena2022}
\APACinsertmetastar {%
pena2022}%
\begin{APACrefauthors}%
Peña, F.%
, Nardi, F.%
, Melesse, A.%
, Obeysekera, J.%
, Castelli, F.%
, Price, R\BPBI M.%
\BDBL {}Gonzalez-Ramirez, N.%
\end{APACrefauthors}%
\unskip\
\newblock
\APACrefYearMonthDay{2022}{}{}.
\newblock
{\BBOQ}\APACrefatitle {Compound flood modeling framework for surface–subsurface water interactions} {Compound flood modeling framework for surface–subsurface water interactions}.{\BBCQ}
\newblock
\APACjournalVolNumPages{Nat. Hazards Earth Syst. Sci.}{22}{}{775-793}.
\newblock
\begin{APACrefDOI} \doi{10.5194/nhess-22-775-2022} \end{APACrefDOI}
\PrintBackRefs{\CurrentBib}

\bibitem [\protect \citeauthoryear {%
Ponce%
}{%
Ponce%
}{%
{\protect \APACyear {1986}}%
}]{%
ponce1986diffusion}
\APACinsertmetastar {%
ponce1986diffusion}%
\begin{APACrefauthors}%
Ponce, V\BPBI M.%
\end{APACrefauthors}%
\unskip\
\newblock
\APACrefYearMonthDay{1986}{}{}.
\newblock
{\BBOQ}\APACrefatitle {Diffusion wave modeling of catchment dynamics} {Diffusion wave modeling of catchment dynamics}.{\BBCQ}
\newblock
\APACjournalVolNumPages{Journal of hydraulic engineering}{112}{8}{716--727}.
\PrintBackRefs{\CurrentBib}

\bibitem [\protect \citeauthoryear {%
Ponce%
\ \BBA {} Hawkins%
}{%
Ponce%
\ \BBA {} Hawkins%
}{%
{\protect \APACyear {1996}}%
}]{%
ponce1996runoff}
\APACinsertmetastar {%
ponce1996runoff}%
\begin{APACrefauthors}%
Ponce, V\BPBI M.%
\BCBT {}\ \BBA {} Hawkins, R\BPBI H.%
\end{APACrefauthors}%
\unskip\
\newblock
\APACrefYearMonthDay{1996}{}{}.
\newblock
{\BBOQ}\APACrefatitle {Runoff curve number: Has it reached maturity?} {Runoff curve number: Has it reached maturity?}{\BBCQ}
\newblock
\APACjournalVolNumPages{Journal of hydrologic engineering}{1}{1}{11--19}.
\PrintBackRefs{\CurrentBib}

\bibitem [\protect \citeauthoryear {%
Ponce%
\ \BBA {} Yevjevich%
}{%
Ponce%
\ \BBA {} Yevjevich%
}{%
{\protect \APACyear {1978}}%
}]{%
ponce1978muskingum}
\APACinsertmetastar {%
ponce1978muskingum}%
\begin{APACrefauthors}%
Ponce, V\BPBI M.%
\BCBT {}\ \BBA {} Yevjevich, V.%
\end{APACrefauthors}%
\unskip\
\newblock
\APACrefYearMonthDay{1978}{}{}.
\newblock
{\BBOQ}\APACrefatitle {Muskingum-Cunge method with variable parameters} {Muskingum-cunge method with variable parameters}.{\BBCQ}
\newblock
\APACjournalVolNumPages{Journal of the Hydraulics Division}{104}{12}{1663--1667}.
\PrintBackRefs{\CurrentBib}

\bibitem [\protect \citeauthoryear {%
Porporato%
\ \BBA {} Yin%
}{%
Porporato%
\ \BBA {} Yin%
}{%
{\protect \APACyear {2022}}%
}]{%
porporato2022ecohydrology}
\APACinsertmetastar {%
porporato2022ecohydrology}%
\begin{APACrefauthors}%
Porporato, A.%
\BCBT {}\ \BBA {} Yin, J.%
\end{APACrefauthors}%
\unskip\
\newblock
\APACrefYear{2022}.
\newblock
\APACrefbtitle {Ecohydrology: Dynamics of life and water in the critical zone} {Ecohydrology: Dynamics of life and water in the critical zone}.
\newblock
\APACaddressPublisher{}{Cambridge University Press}.
\PrintBackRefs{\CurrentBib}

\bibitem [\protect \citeauthoryear {%
Rahman%
\ \protect \BOthers {.}}{%
Rahman%
\ \protect \BOthers {.}}{%
{\protect \APACyear {2021}}%
}]{%
rahman2021investigation}
\APACinsertmetastar {%
rahman2021investigation}%
\begin{APACrefauthors}%
Rahman, M\BPBI A.%
, Sadeghi~Tabas, S.%
, House, E.%
, Storey, A.%
, Meselhe, E.%
, Ferreira, C.%
\BCBL {}\ \BBA {} Hu, K.%
\end{APACrefauthors}%
\unskip\
\newblock
\APACrefYearMonthDay{2021}{}{}.
\newblock
{\BBOQ}\APACrefatitle {Investigation of Compound Flooding in the Coastal Transition Zone within Lake Maurepas Watershed and Southeastern Coastal Louisiana} {Investigation of compound flooding in the coastal transition zone within lake maurepas watershed and southeastern coastal louisiana}.{\BBCQ}
\newblock
\BIn{} \APACrefbtitle {AGU Fall Meeting Abstracts} {Agu fall meeting abstracts}\ (\BVOL\ 2021, \BPGS\ H45O--1342).
\PrintBackRefs{\CurrentBib}

\bibitem [\protect \citeauthoryear {%
Resio%
\ \protect \BOthers {.}}{%
Resio%
\ \protect \BOthers {.}}{%
{\protect \APACyear {2007}}%
}]{%
resio2007white}
\APACinsertmetastar {%
resio2007white}%
\begin{APACrefauthors}%
Resio, D\BPBI T.%
, Boc, S\BPBI J.%
, Borgman, L\BPBI E.%
, Cardone, V\BPBI J.%
, Dean, R\BPBI G.%
, Ratcliff, J\BPBI J.%
\BDBL {}Westerink, J\BPBI J.%
\end{APACrefauthors}%
\unskip\
\newblock
\APACrefYearMonthDay{2007}{}{}.
\newblock
\APACrefbtitle {White Paper on Estimating Hurricane Inundation Probabilities} {White paper on estimating hurricane inundation probabilities}\ \APACbVolEdTR{}{\BTR{}}.
\newblock
\APACaddressInstitution{Vicksburg, Mississippi}{U.S. Army Engineer Research and Development Center, Coastal and Hydraulics Laboratory}.
\newblock
\begin{APACrefURL} \url{https://erdc-library.erdc.dren.mil/jspui/handle/11681/22643} \end{APACrefURL}
\PrintBackRefs{\CurrentBib}

\bibitem [\protect \citeauthoryear {%
Resio%
, Irish%
\BCBL {}\ \BBA {} Cialone%
}{%
Resio%
\ \protect \BOthers {.}}{%
{\protect \APACyear {2009}}%
}]{%
resio2009surge}
\APACinsertmetastar {%
resio2009surge}%
\begin{APACrefauthors}%
Resio, D\BPBI T.%
, Irish, J.%
\BCBL {}\ \BBA {} Cialone, M.%
\end{APACrefauthors}%
\unskip\
\newblock
\APACrefYearMonthDay{2009}{}{}.
\newblock
{\BBOQ}\APACrefatitle {A surge response function approach to coastal hazard assessment--part 1: basic concepts} {A surge response function approach to coastal hazard assessment--part 1: basic concepts}.{\BBCQ}
\newblock
\APACjournalVolNumPages{Natural hazards}{51}{}{163--182}.
\PrintBackRefs{\CurrentBib}

\bibitem [\protect \citeauthoryear {%
Rigby%
\ \BBA {} Porporato%
}{%
Rigby%
\ \BBA {} Porporato%
}{%
{\protect \APACyear {2006}}%
}]{%
rigby2006simplified}
\APACinsertmetastar {%
rigby2006simplified}%
\begin{APACrefauthors}%
Rigby, J.%
\BCBT {}\ \BBA {} Porporato, A.%
\end{APACrefauthors}%
\unskip\
\newblock
\APACrefYearMonthDay{2006}{}{}.
\newblock
{\BBOQ}\APACrefatitle {Simplified stochastic soil-moisture models: a look at infiltration} {Simplified stochastic soil-moisture models: a look at infiltration}.{\BBCQ}
\newblock
\APACjournalVolNumPages{Hydrology and Earth System Sciences}{10}{}{861--871}.
\PrintBackRefs{\CurrentBib}

\bibitem [\protect \citeauthoryear {%
Salvadori%
\ \BBA {} De~Michele%
}{%
Salvadori%
\ \BBA {} De~Michele%
}{%
{\protect \APACyear {2004}}%
}]{%
salvadori2004frequency}
\APACinsertmetastar {%
salvadori2004frequency}%
\begin{APACrefauthors}%
Salvadori, G.%
\BCBT {}\ \BBA {} De~Michele, C.%
\end{APACrefauthors}%
\unskip\
\newblock
\APACrefYearMonthDay{2004}{}{}.
\newblock
{\BBOQ}\APACrefatitle {Frequency analysis via copulas: Theoretical aspects and applications to hydrological events} {Frequency analysis via copulas: Theoretical aspects and applications to hydrological events}.{\BBCQ}
\newblock
\APACjournalVolNumPages{Water resources research}{40}{12}{}.
\PrintBackRefs{\CurrentBib}

\bibitem [\protect \citeauthoryear {%
Salvadori%
, De~Michele%
\BCBL {}\ \BBA {} Durante%
}{%
Salvadori%
\ \protect \BOthers {.}}{%
{\protect \APACyear {2011}}%
}]{%
salvadori2011return}
\APACinsertmetastar {%
salvadori2011return}%
\begin{APACrefauthors}%
Salvadori, G.%
, De~Michele, C.%
\BCBL {}\ \BBA {} Durante, F.%
\end{APACrefauthors}%
\unskip\
\newblock
\APACrefYearMonthDay{2011}{}{}.
\newblock
{\BBOQ}\APACrefatitle {On the return period and design in a multivariate framework} {On the return period and design in a multivariate framework}.{\BBCQ}
\newblock
\APACjournalVolNumPages{Hydrology and Earth System Sciences}{15}{11}{3293--3305}.
\PrintBackRefs{\CurrentBib}

\bibitem [\protect \citeauthoryear {%
Salvadori%
, Durante%
\BCBL {}\ \BBA {} De~Michele%
}{%
Salvadori%
\ \protect \BOthers {.}}{%
{\protect \APACyear {2013}}%
}]{%
salvadori2013multivariate}
\APACinsertmetastar {%
salvadori2013multivariate}%
\begin{APACrefauthors}%
Salvadori, G.%
, Durante, F.%
\BCBL {}\ \BBA {} De~Michele, C.%
\end{APACrefauthors}%
\unskip\
\newblock
\APACrefYearMonthDay{2013}{}{}.
\newblock
{\BBOQ}\APACrefatitle {Multivariate return period calculation via survival functions} {Multivariate return period calculation via survival functions}.{\BBCQ}
\newblock
\APACjournalVolNumPages{Water Resources Research}{49}{4}{2308--2311}.
\PrintBackRefs{\CurrentBib}

\bibitem [\protect \citeauthoryear {%
Santamaria-Aguilar%
, Maduwantha%
, Enriquez%
\BCBL {}\ \BBA {} Wahl%
}{%
Santamaria-Aguilar%
\ \protect \BOthers {.}}{%
{\protect \APACyear {2026}}%
}]{%
santamaria2026large}
\APACinsertmetastar {%
santamaria2026large}%
\begin{APACrefauthors}%
Santamaria-Aguilar, S.%
, Maduwantha, P.%
, Enriquez, A\BPBI R.%
\BCBL {}\ \BBA {} Wahl, T.%
\end{APACrefauthors}%
\unskip\
\newblock
\APACrefYearMonthDay{2026}{}{}.
\newblock
{\BBOQ}\APACrefatitle {Large discrepancies between event-and response-based compound flood hazard estimates} {Large discrepancies between event-and response-based compound flood hazard estimates}.{\BBCQ}
\newblock
\APACjournalVolNumPages{Natural Hazards and Earth System Sciences}{26}{1}{571--586}.
\PrintBackRefs{\CurrentBib}

\bibitem [\protect \citeauthoryear {%
Santiago-Collazo%
, Bilskie%
, Bacopoulos%
\BCBL {}\ \BBA {} Hagen%
}{%
Santiago-Collazo%
\ \protect \BOthers {.}}{%
{\protect \APACyear {2021}}%
}]{%
santiago2021examination}
\APACinsertmetastar {%
santiago2021examination}%
\begin{APACrefauthors}%
Santiago-Collazo, F\BPBI L.%
, Bilskie, M\BPBI V.%
, Bacopoulos, P.%
\BCBL {}\ \BBA {} Hagen, S\BPBI C.%
\end{APACrefauthors}%
\unskip\
\newblock
\APACrefYearMonthDay{2021}{}{}.
\newblock
{\BBOQ}\APACrefatitle {An examination of compound flood hazard zones for past, present, and future low-gradient coastal land-margins} {An examination of compound flood hazard zones for past, present, and future low-gradient coastal land-margins}.{\BBCQ}
\newblock
\APACjournalVolNumPages{Frontiers in Climate}{3}{}{684035}.
\PrintBackRefs{\CurrentBib}

\bibitem [\protect \citeauthoryear {%
Santiago-Collazo%
, Bilskie%
\BCBL {}\ \BBA {} Hagen%
}{%
Santiago-Collazo%
\ \protect \BOthers {.}}{%
{\protect \APACyear {2019}}%
}]{%
santiago2019comprehensive}
\APACinsertmetastar {%
santiago2019comprehensive}%
\begin{APACrefauthors}%
Santiago-Collazo, F\BPBI L.%
, Bilskie, M\BPBI V.%
\BCBL {}\ \BBA {} Hagen, S\BPBI C.%
\end{APACrefauthors}%
\unskip\
\newblock
\APACrefYearMonthDay{2019}{}{}.
\newblock
{\BBOQ}\APACrefatitle {A comprehensive review of compound inundation models in low-gradient coastal watersheds} {A comprehensive review of compound inundation models in low-gradient coastal watersheds}.{\BBCQ}
\newblock
\APACjournalVolNumPages{Environmental Modelling \& Software}{119}{}{166--181}.
\newblock
\begin{APACrefDOI} \doi{10.1016/j.envsoft.2019.06.002} \end{APACrefDOI}
\PrintBackRefs{\CurrentBib}

\bibitem [\protect \citeauthoryear {%
Serafin%
, Ruggiero%
, Parker%
\BCBL {}\ \BBA {} Hill%
}{%
Serafin%
\ \protect \BOthers {.}}{%
{\protect \APACyear {2019}}%
}]{%
serafin2019s}
\APACinsertmetastar {%
serafin2019s}%
\begin{APACrefauthors}%
Serafin, K\BPBI A.%
, Ruggiero, P.%
, Parker, K.%
\BCBL {}\ \BBA {} Hill, D\BPBI F.%
\end{APACrefauthors}%
\unskip\
\newblock
\APACrefYearMonthDay{2019}{}{}.
\newblock
{\BBOQ}\APACrefatitle {What's streamflow got to do with it? A probabilistic simulation of the competing oceanographic and fluvial processes driving extreme along-river water levels} {What's streamflow got to do with it? a probabilistic simulation of the competing oceanographic and fluvial processes driving extreme along-river water levels}.{\BBCQ}
\newblock
\APACjournalVolNumPages{Natural Hazards and Earth System Sciences}{19}{7}{1415--1431}.
\newblock
\begin{APACrefDOI} \doi{10.5194/nhess-19-1415-2019} \end{APACrefDOI}
\PrintBackRefs{\CurrentBib}

\bibitem [\protect \citeauthoryear {%
Serinaldi%
}{%
Serinaldi%
}{%
{\protect \APACyear {2015}}%
}]{%
serinaldi2015dismissing}
\APACinsertmetastar {%
serinaldi2015dismissing}%
\begin{APACrefauthors}%
Serinaldi, F.%
\end{APACrefauthors}%
\unskip\
\newblock
\APACrefYearMonthDay{2015}{}{}.
\newblock
{\BBOQ}\APACrefatitle {Dismissing return periods!} {Dismissing return periods!}{\BBCQ}
\newblock
\APACjournalVolNumPages{Stochastic environmental research and risk assessment}{29}{4}{1179--1189}.
\PrintBackRefs{\CurrentBib}

\bibitem [\protect \citeauthoryear {%
Shen%
, Morsy%
, Huxley%
, Tahvildari%
\BCBL {}\ \BBA {} Goodall%
}{%
Shen%
\ \protect \BOthers {.}}{%
{\protect \APACyear {2019}}%
}]{%
shen2019flood}
\APACinsertmetastar {%
shen2019flood}%
\begin{APACrefauthors}%
Shen, Y.%
, Morsy, M\BPBI M.%
, Huxley, C.%
, Tahvildari, N.%
\BCBL {}\ \BBA {} Goodall, J\BPBI L.%
\end{APACrefauthors}%
\unskip\
\newblock
\APACrefYearMonthDay{2019}{}{}.
\newblock
{\BBOQ}\APACrefatitle {Flood risk assessment and increased resilience for coastal urban watersheds under the combined impact of storm tide and heavy rainfall} {Flood risk assessment and increased resilience for coastal urban watersheds under the combined impact of storm tide and heavy rainfall}.{\BBCQ}
\newblock
\APACjournalVolNumPages{Journal of Hydrology}{579}{}{124159}.
\PrintBackRefs{\CurrentBib}

\bibitem [\protect \citeauthoryear {%
Tanim%
\ \BBA {} Goharian%
}{%
Tanim%
\ \BBA {} Goharian%
}{%
{\protect \APACyear {2021}}%
}]{%
tanim2021developing}
\APACinsertmetastar {%
tanim2021developing}%
\begin{APACrefauthors}%
Tanim, A\BPBI H.%
\BCBT {}\ \BBA {} Goharian, E.%
\end{APACrefauthors}%
\unskip\
\newblock
\APACrefYearMonthDay{2021}{}{}.
\newblock
{\BBOQ}\APACrefatitle {Developing a hybrid modeling and multivariate analysis framework for storm surge and runoff interactions in urban coastal flooding} {Developing a hybrid modeling and multivariate analysis framework for storm surge and runoff interactions in urban coastal flooding}.{\BBCQ}
\newblock
\APACjournalVolNumPages{Journal of Hydrology}{595}{}{125670}.
\PrintBackRefs{\CurrentBib}

\bibitem [\protect \citeauthoryear {%
Thompson%
\ \BBA {} Frazier%
}{%
Thompson%
\ \BBA {} Frazier%
}{%
{\protect \APACyear {2014}}%
}]{%
thompson2014deterministic}
\APACinsertmetastar {%
thompson2014deterministic}%
\begin{APACrefauthors}%
Thompson, C\BPBI M.%
\BCBT {}\ \BBA {} Frazier, T\BPBI G.%
\end{APACrefauthors}%
\unskip\
\newblock
\APACrefYearMonthDay{2014}{}{}.
\newblock
{\BBOQ}\APACrefatitle {Deterministic and probabilistic flood modeling for contemporary and future coastal and inland precipitation inundation} {Deterministic and probabilistic flood modeling for contemporary and future coastal and inland precipitation inundation}.{\BBCQ}
\newblock
\APACjournalVolNumPages{Applied Geography}{50}{}{1--14}.
\newblock
\begin{APACrefDOI} \doi{10.1016/j.apgeog.2014.01.013} \end{APACrefDOI}
\PrintBackRefs{\CurrentBib}

\bibitem [\protect \citeauthoryear {%
Toro%
}{%
Toro%
}{%
{\protect \APACyear {2008}}%
}]{%
toro2008joint}
\APACinsertmetastar {%
toro2008joint}%
\begin{APACrefauthors}%
Toro, G\BPBI R.%
\end{APACrefauthors}%
\unskip\
\newblock
\APACrefYearMonthDay{2008}{October}{}.
\newblock
\APACrefbtitle {Joint probability analysis of hurricane flood hazards for Mississippi} {Joint probability analysis of hurricane flood hazards for mississippi}\ \APACbVolEdTR{}{\BTR{}}.
\newblock
\APACaddressInstitution{Acton, MA}{Risk Engineering, Inc.}
\PrintBackRefs{\CurrentBib}

\bibitem [\protect \citeauthoryear {%
Toro%
, Resio%
, Divoky%
, Niedoroda%
\BCBL {}\ \BBA {} Reed%
}{%
Toro%
\ \protect \BOthers {.}}{%
{\protect \APACyear {2010}}%
}]{%
toro2010efficient}
\APACinsertmetastar {%
toro2010efficient}%
\begin{APACrefauthors}%
Toro, G\BPBI R.%
, Resio, D\BPBI T.%
, Divoky, D.%
, Niedoroda, A\BPBI W.%
\BCBL {}\ \BBA {} Reed, C.%
\end{APACrefauthors}%
\unskip\
\newblock
\APACrefYearMonthDay{2010}{}{}.
\newblock
{\BBOQ}\APACrefatitle {Efficient joint-probability methods for hurricane surge frequency analysis} {Efficient joint-probability methods for hurricane surge frequency analysis}.{\BBCQ}
\newblock
\APACjournalVolNumPages{Ocean Engineering}{37}{1}{125--134}.
\PrintBackRefs{\CurrentBib}

\bibitem [\protect \citeauthoryear {%
Troy%
, Wood%
\BCBL {}\ \BBA {} Sheffield%
}{%
Troy%
\ \protect \BOthers {.}}{%
{\protect \APACyear {2008}}%
}]{%
troy2008efficient}
\APACinsertmetastar {%
troy2008efficient}%
\begin{APACrefauthors}%
Troy, T\BPBI J.%
, Wood, E\BPBI F.%
\BCBL {}\ \BBA {} Sheffield, J.%
\end{APACrefauthors}%
\unskip\
\newblock
\APACrefYearMonthDay{2008}{}{}.
\newblock
{\BBOQ}\APACrefatitle {An efficient calibration method for continental-scale land surface modeling} {An efficient calibration method for continental-scale land surface modeling}.{\BBCQ}
\newblock
\APACjournalVolNumPages{Water Resources Research}{44}{9}{}.
\PrintBackRefs{\CurrentBib}

\bibitem [\protect \citeauthoryear {%
{USDA Natural Resources Conservation Service}%
}{%
{USDA Natural Resources Conservation Service}%
}{%
{\protect \APACyear {2009}}%
}]{%
NRCS2009HSG}
\APACinsertmetastar {%
NRCS2009HSG}%
\begin{APACrefauthors}%
{USDA Natural Resources Conservation Service}.%
\end{APACrefauthors}%
\unskip\
\newblock
\APACrefYearMonthDay{2009}{}{}.
\newblock
{\BBOQ}\APACrefatitle {National Engineering Handbook, Part 630: Hydrology, Chapter 7 -- Hydrologic Soil Groups} {National engineering handbook, part 630: Hydrology, chapter 7 -- hydrologic soil groups}{\BBCQ}\ [\bibcomputersoftwaremanual].
\newblock
\begin{APACrefURL} \url{https://usda.gov} \end{APACrefURL}
\PrintBackRefs{\CurrentBib}

\bibitem [\protect \citeauthoryear {%
Villarini%
\ \protect \BOthers {.}}{%
Villarini%
\ \protect \BOthers {.}}{%
{\protect \APACyear {2022}}%
}]{%
villarini2022probabilistic}
\APACinsertmetastar {%
villarini2022probabilistic}%
\begin{APACrefauthors}%
Villarini, G.%
, Zhang, W.%
, Miller, P.%
, Johnson, D\BPBI R.%
, Grimley, L\BPBI E.%
\BCBL {}\ \BBA {} Roberts, H\BPBI J.%
\end{APACrefauthors}%
\unskip\
\newblock
\APACrefYearMonthDay{2022}{}{}.
\newblock
{\BBOQ}\APACrefatitle {Probabilistic Rainfall Generator for Tropical Cyclones Affecting Louisiana} {Probabilistic rainfall generator for tropical cyclones affecting louisiana}.{\BBCQ}
\newblock
\APACjournalVolNumPages{International Journal of Climatology}{42}{3}{1789--1802}.
\newblock
\begin{APACrefURL} \url{https://doi.org/10.1002/joc.7335} \end{APACrefURL}
\newblock
\begin{APACrefDOI} \doi{10.1002/joc.7335} \end{APACrefDOI}
\PrintBackRefs{\CurrentBib}

\bibitem [\protect \citeauthoryear {%
Volpi%
\ \BBA {} Fiori%
}{%
Volpi%
\ \BBA {} Fiori%
}{%
{\protect \APACyear {2014}}%
}]{%
volpi2014hydraulic}
\APACinsertmetastar {%
volpi2014hydraulic}%
\begin{APACrefauthors}%
Volpi, E.%
\BCBT {}\ \BBA {} Fiori, A.%
\end{APACrefauthors}%
\unskip\
\newblock
\APACrefYearMonthDay{2014}{}{}.
\newblock
{\BBOQ}\APACrefatitle {Hydraulic structures subject to bivariate hydrological loads: Return period, design, and risk assessment} {Hydraulic structures subject to bivariate hydrological loads: Return period, design, and risk assessment}.{\BBCQ}
\newblock
\APACjournalVolNumPages{Water Resources Research}{50}{2}{885--897}.
\PrintBackRefs{\CurrentBib}

\bibitem [\protect \citeauthoryear {%
Voortman%
, Van~Gelder%
\BCBL {}\ \BBA {} Vrijling%
}{%
Voortman%
\ \protect \BOthers {.}}{%
{\protect \APACyear {2003}}%
}]{%
voortman2003risk}
\APACinsertmetastar {%
voortman2003risk}%
\begin{APACrefauthors}%
Voortman, H\BPBI G.%
, Van~Gelder, P.%
\BCBL {}\ \BBA {} Vrijling, J.%
\end{APACrefauthors}%
\unskip\
\newblock
\APACrefYearMonthDay{2003}{}{}.
\newblock
{\BBOQ}\APACrefatitle {Risk-based design of large-scale flood defence systems} {Risk-based design of large-scale flood defence systems}.{\BBCQ}
\newblock
\BIn{} J\BPBI M.~Smith\ (\BED), \APACrefbtitle {Coastal Engineering 2002: Solving Coastal Conundrums} {Coastal engineering 2002: Solving coastal conundrums}\ (\BVOL~2, \BPGS\ 2373--2385).
\newblock
\APACaddressPublisher{}{World Scientific}.
\newblock
\begin{APACrefDOI} \doi{10.1142/9789812791306_0199} \end{APACrefDOI}
\PrintBackRefs{\CurrentBib}

\bibitem [\protect \citeauthoryear {%
Vousdoukas%
\ \protect \BOthers {.}}{%
Vousdoukas%
\ \protect \BOthers {.}}{%
{\protect \APACyear {2018}}%
}]{%
vousdoukas2018understanding}
\APACinsertmetastar {%
vousdoukas2018understanding}%
\begin{APACrefauthors}%
Vousdoukas, M\BPBI I.%
, Bouziotas, D.%
, Giardino, A.%
, Bouwer, L\BPBI M.%
, Mentaschi, L.%
, Voukouvalas, E.%
\BCBL {}\ \BBA {} Feyen, L.%
\end{APACrefauthors}%
\unskip\
\newblock
\APACrefYearMonthDay{2018}{}{}.
\newblock
{\BBOQ}\APACrefatitle {Understanding epistemic uncertainty in large-scale coastal flood risk assessment for present and future climates} {Understanding epistemic uncertainty in large-scale coastal flood risk assessment for present and future climates}.{\BBCQ}
\newblock
\APACjournalVolNumPages{Natural Hazards and Earth System Sciences}{18}{8}{2127--2142}.
\newblock
\begin{APACrefDOI} \doi{10.5194/nhess-18-2127-2018} \end{APACrefDOI}
\PrintBackRefs{\CurrentBib}

\bibitem [\protect \citeauthoryear {%
Wahl%
, Jain%
, Bender%
, Meyers%
\BCBL {}\ \BBA {} Luther%
}{%
Wahl%
\ \protect \BOthers {.}}{%
{\protect \APACyear {2015}}%
}]{%
wahl2015increasing}
\APACinsertmetastar {%
wahl2015increasing}%
\begin{APACrefauthors}%
Wahl, T.%
, Jain, S.%
, Bender, J.%
, Meyers, S\BPBI D.%
\BCBL {}\ \BBA {} Luther, M\BPBI E.%
\end{APACrefauthors}%
\unskip\
\newblock
\APACrefYearMonthDay{2015}{}{}.
\newblock
{\BBOQ}\APACrefatitle {Increasing risk of compound flooding from storm surge and rainfall for major US cities} {Increasing risk of compound flooding from storm surge and rainfall for major us cities}.{\BBCQ}
\newblock
\APACjournalVolNumPages{Nature Climate Change}{5}{}{1093--1097}.
\newblock
\begin{APACrefDOI} \doi{10.1038/nclimate2736} \end{APACrefDOI}
\PrintBackRefs{\CurrentBib}

\bibitem [\protect \citeauthoryear {%
Watson%
, Storm%
, Breaker%
\BCBL {}\ \BBA {} Rose%
}{%
Watson%
\ \protect \BOthers {.}}{%
{\protect \APACyear {2017}}%
}]{%
watson2017characterization}
\APACinsertmetastar {%
watson2017characterization}%
\begin{APACrefauthors}%
Watson, K\BPBI M.%
, Storm, J\BPBI B.%
, Breaker, B\BPBI K.%
\BCBL {}\ \BBA {} Rose, C\BPBI E.%
\end{APACrefauthors}%
\unskip\
\newblock
\APACrefYearMonthDay{2017}{}{}.
\newblock
\APACrefbtitle {Characterization of peak streamflows and flood inundation of selected areas in Louisiana from the August 2016 flood} {Characterization of peak streamflows and flood inundation of selected areas in louisiana from the august 2016 flood}\ \APACbVolEdTR{}{\BTR{}}.
\newblock
\APACaddressInstitution{}{US Geological Survey}.
\PrintBackRefs{\CurrentBib}

\bibitem [\protect \citeauthoryear {%
Westerink%
\ \protect \BOthers {.}}{%
Westerink%
\ \protect \BOthers {.}}{%
{\protect \APACyear {1994}}%
}]{%
westerink1994adcirc}
\APACinsertmetastar {%
westerink1994adcirc}%
\begin{APACrefauthors}%
Westerink, J\BPBI J.%
, Blain, C\BPBI A.%
, Luettich, R\BPBI A.%
, Mark, D\BPBI J.%
, Scheffner, N\BPBI W.%
\BCBL {}\ \BOthersPeriod {.}\end{APACrefauthors}%
\unskip\
\newblock
\APACrefYearMonthDay{1994}{}{}.
\newblock
\APACrefbtitle {ADCIRC: an advanced three-dimensional circulation model for shelves, coasts, and estuaries. Report 5, A tropical storm database for the east and Gulf of Mexico coasts of the United States} {Adcirc: an advanced three-dimensional circulation model for shelves, coasts, and estuaries. report 5, a tropical storm database for the east and gulf of mexico coasts of the united states}\ \APACbVolEdTR{}{\BTR{}}.
\newblock
\APACaddressInstitution{}{US Army Engineer Waterways Experiment Station}.
\PrintBackRefs{\CurrentBib}

\bibitem [\protect \citeauthoryear {%
Wilkerson%
\ \BBA {} Parker%
}{%
Wilkerson%
\ \BBA {} Parker%
}{%
{\protect \APACyear {2011}}%
}]{%
wilkerson2011physical}
\APACinsertmetastar {%
wilkerson2011physical}%
\begin{APACrefauthors}%
Wilkerson, G\BPBI V.%
\BCBT {}\ \BBA {} Parker, G.%
\end{APACrefauthors}%
\unskip\
\newblock
\APACrefYearMonthDay{2011}{}{}.
\newblock
{\BBOQ}\APACrefatitle {Physical basis for quasi-universal relationships describing bankfull hydraulic geometry of sand-bed rivers} {Physical basis for quasi-universal relationships describing bankfull hydraulic geometry of sand-bed rivers}.{\BBCQ}
\newblock
\APACjournalVolNumPages{Journal of Hydraulic Engineering}{137}{7}{739--753}.
\PrintBackRefs{\CurrentBib}

\bibitem [\protect \citeauthoryear {%
Yang%
, Paramygin%
\BCBL {}\ \BBA {} Sheng%
}{%
Yang%
\ \protect \BOthers {.}}{%
{\protect \APACyear {2019}}%
}]{%
yang2019objective}
\APACinsertmetastar {%
yang2019objective}%
\begin{APACrefauthors}%
Yang, K.%
, Paramygin, V.%
\BCBL {}\ \BBA {} Sheng, Y\BPBI P.%
\end{APACrefauthors}%
\unskip\
\newblock
\APACrefYearMonthDay{2019}{}{}.
\newblock
{\BBOQ}\APACrefatitle {An objective and efficient method for estimating probabilistic coastal inundation hazards} {An objective and efficient method for estimating probabilistic coastal inundation hazards}.{\BBCQ}
\newblock
\APACjournalVolNumPages{Natural Hazards}{99}{2}{1105--1130}.
\newblock
\begin{APACrefDOI} \doi{10.1007/s11069-019-03807-w} \end{APACrefDOI}
\PrintBackRefs{\CurrentBib}

\bibitem [\protect \citeauthoryear {%
Zscheischler%
\ \protect \BOthers {.}}{%
Zscheischler%
\ \protect \BOthers {.}}{%
{\protect \APACyear {2018}}%
}]{%
zscheischler2018future}
\APACinsertmetastar {%
zscheischler2018future}%
\begin{APACrefauthors}%
Zscheischler, J.%
, Westra, S.%
, van~den Hurk, B\BPBI J\BPBI J\BPBI M.%
, Seneviratne, S\BPBI I.%
, Ward, P\BPBI J.%
, Pitman, A.%
\BDBL {}Zhang, X.%
\end{APACrefauthors}%
\unskip\
\newblock
\APACrefYearMonthDay{2018}{}{}.
\newblock
{\BBOQ}\APACrefatitle {Future Climate Risk from Compound Events} {Future climate risk from compound events}.{\BBCQ}
\newblock
\APACjournalVolNumPages{Nature Climate Change}{8}{6}{469--477}.
\newblock
\begin{APACrefURL} \url{https://www.nature.com/articles/s41558-018-0156-3} \end{APACrefURL}
\newblock
\begin{APACrefDOI} \doi{10.1038/s41558-018-0156-3} \end{APACrefDOI}
\PrintBackRefs{\CurrentBib}

\end{thebibliography}
\end{document}